\documentclass[twocolumn]{aastex631}

\submitjournal{ApJ}

\usepackage{graphicx}	% Including figure files
\usepackage{amsmath}	% Advanced maths commands
\usepackage{amssymb}	% Extra maths symbols
\usepackage{color}
\usepackage{algorithm}
\usepackage{booktabs}
\usepackage[noend]{algpseudocode}
\usepackage{pifont}% http://ctan.org/pkg/pifont
\usepackage{soul}
\usepackage{float}
\usepackage{multibib}
\newcites{Appendix}{Appendix References}
\usepackage{natbib}
\usepackage[flushleft]{threeparttablex}
\usepackage{mathrsfs}
\usepackage{tikz}
\usetikzlibrary{positioning}
\usepackage{esint}
\usepackage{lipsum}
\usepackage{longtable}
\usepackage{pifont}% http://ctan.org/pkg/pifont
\usepackage{savesym}
\savesymbol{tablenum}
\usepackage{siunitx}
\restoresymbol{SIX}{tablenum}
\usepackage{placeins}
\usepackage{bm}

\newcommand{\appropto}{\mathrel{\vcenter{
  \offinterlineskip\halign{\hfil$##$\cr
    \propto\cr\noalign{\kern2pt}\sim\cr\noalign{\kern-2pt}}}}}

\newcommand{\Exp}[2]{\left\langle{#1}\right\rangle_{#2}}

\renewcommand{\d}[1]{\ensuremath{\operatorname{d}\!{#1}}}

\DeclareMathOperator\D{\mathcal{D}}
\DeclareMathOperator\V{\mathcal{V}}
\newcommand{\M}{\mathcal{M}}

\DeclareMathOperator\s{\text{s}}
\DeclareMathOperator\f{\text{f}}

\DeclareMathAlphabet\mathbfcal{OMS}{cmsy}{b}{n}

\renewcommand{\Re}{\text{Re}}

\renewcommand{\v}{\mathit{v}}
\renewcommand{\b}{\mathit{b}}

\renewcommand{\u}{\bm{u}}
\renewcommand{\k}{\bm{k}}

\newcommand{\x}{\bm{x}}
\newcommand{\K}{\bm{K}}

\renewcommand{\P}{\mathcal{P}_{f}}
\newcommand{\bom}{\bm{\omega}}

\newcommand{\It}{\mathbb{I}}
\newcommand{\T}{\mathcal{T}}

\newcommand{\flash}{\textsc{flash} }
\newcommand{\Ptot}{P_{\rm{tot}}}
\renewcommand{\Pi}{P_{i}}
\newcommand{\Pe}{P_{e}}
\renewcommand{\Pr}{P_{\rm{rad}}}
\newcommand{\Etot}{E_{\rm{tot}}}
\newcommand{\Ei}{E_{i}}
\newcommand{\Ee}{E_{e}}
\newcommand{\Er}{E_{\rm{rad}}}

\newcommand{\Qheat}{Q_{\rm{heat}}}
\newcommand{\Qabs}{Q_{\rm{abs}}}
\newcommand{\Qemis}{Q_{\rm{emis}}}
\newcommand{\q}{\bm{q}}

\newcommand{\Ti}{T_{i}}
\newcommand{\Te}{T_{e}}
\newcommand{\Tr}{T_{\rm{rad}}}

\newcommand{\bnab}{\bm{\nabla}}
\newcommand{\baro}{(\bnab\rho \times \bnab \Ptot)/\rho^2}

\newcommand{\cve}{c_{v,e}}

\begin{document}
    \correspondingauthor{\\
    $^{\dagger}$Stefano Merlini: \href{mailto:s.merlini19@imperial.ac.uk}{s.merlini19@imperial.ac.uk}\\
    $^{\ddagger}$James R. Beattie: \href{mailto:james.beattie@princeton.edu}{james.beattie@princeton.edu}\\
    $^\star$ these authors made equal contributions and should be deemed as joint first authors for the publication.}

    \title{Numerical simulations of shock-driven, supersonic interstellar turbulence \\ 
    in colliding three-temperature laboratory plasmas}
    
    \author[0000-0002-7128-8895]{Stefano Merlini$^{\dagger,\star}$}
    \affiliation{Blackett Laboratory, Imperial College London, London, SW7 2BW, UK}
    \author[0000-0001-9199-7771]{James R. Beattie$^{\ddagger,\star}$}
    \affiliation{Department of Astrophysical Sciences, Princeton University, Princeton, 08540, NJ, USA}
    \affiliation{Canadian Institute for Theoretical Astrophysics, University of Toronto, Toronto, M5S3H8, ON, Canada}
    \author[0000-0001-6772-1441]{Vicente Valenzuela-Villaseca}
    \affiliation{Department of Astrophysical Sciences, Princeton University, Princeton, 08540, NJ, USA}
    \affiliation{Lawrence Livermore National Laboratory, Livermore, California 94550, USA.}
    \affiliation{Plasma Science and Fusion Center, Massachusetts Institute of Technology, Cambridge, Massachusetts 02139, USA.}

    \begin{abstract}
    Shock-driven turbulence is central to the interstellar medium of our Galaxy, where explosions and compressive driving inject energy through shocks rather than steady stirring. We present three-dimensional, three-temperature (ion, electron, and radiation) radiation-hydrodynamic simulations of a new, constructed laboratory platform in which two offset CH mesh targets are irradiated by a $30\,\rm ns$ X-ray pulse, providing a new window into the study of interstellar turbulence. Mesh ablation launches counter-streaming supersonic flows whose vorticity is seeded baroclinically at mesh-cell corners, advected into collimated channels over $\sim15\,\rm ns$, and injected into the outgoing streams before collision. The flows first collide at $t\simeq75\,\rm ns$, forming a shocked turbulent mixing layer that persists for at least $300\,\rm ns$, reaches $\ell_0\simeq4.5\,\rm mm$, and evolves toward an effectively isothermal equation of state with $\gamma_{\rm eff}\simeq1.1$. After stagnation, $u_0(t)\propto t^{-1.1}$ while $t_0/t_{c_s}\simeq0.2$ remains nearly fixed. Vortex compression and stretching dominate the vorticity budget, and the velocity field relaxes toward a kinetic-energy partition of approximately $70\%$ solenoidal and $30\%$ compressive. The Reynolds stress is strongly anisotropic at the outer scale and remains measurably anisotropic over much of the resolved inertial interval, indicating directional memory of the collision axis and mesh geometry across many scales. The density-gradient spectrum, an important statistic for interstellar scintillation, is controlled by the compressible mode spectrum, which evolves independently from the incompressible mode spectrum, at odds with the conventional wisdom. These results establish the double-mesh platform as a controlled laboratory realisation of sustained shock-driven turbulence and open a new route for probing interstellar turbulence, providing a quantitative baseline for future high-energy-density laboratory astrophysics experiments that connect shock--turbulence interactions to refractive density structure.
    \end{abstract}

    
    \section{Introduction}\label{sec:intro}
    Turbulence in many astrophysical plasmas is driven not by steady stirring but by explosive and shock-producing events. Supernova blast waves in the interstellar medium (ISM) \citep{Draine1993_ISM_shocks,Korpi1999_supernova_regulated_ISM,Padoan2016_supernova_driving,Bacchini2020_supernova_drives_turb,Lu2020_supernova_driving,Gent2021_supernova_turbulence_and_dynamo,Beattie2025_so_long_kolmogorov,Beattie2025_small_scale_instabilities,Connor2026_sn_turb}, shocks in cold star-forming molecular clouds \citep{Draine1993_ISM_shocks,Klessen2000,MacLow2004,McKee2007,Federrath2018,Mandal2020}, and feedback from central engines embedded deep within a galaxy \citep{Mohapatra2019_turbulent_heat_flux_ICM,Mohapatra2020,Mohapatra2021,Schmidt2021_CGM_turbulence,Chen2023_QSO_CGM_turbulence} all inject energy compressively. Even in the absence of a central engine, galaxy interactions, groups, and cluster mergers likewise convert gravitational energy into shocked bulk motions and turbulence, from multiphase intergalactic structures such as Stephan's Quintet to the hot intracluster medium on scales of hundreds of kiloparsecs \citep{meinecke_developed_2015,Appleton2023_Stephans_Quintet,Zhang2026_Perseus_merger_turbulence}. Shock interaction with inhomogeneous media can also drive post-shock turbulence and magnetic amplification \citep{Hu2022_shock_amplification,Hew2023_lagrangian_stats}. Such driving naturally produces shocks, large density gradients, and vorticity through baroclinic misalignment \citep{Beattie2025_so_long_kolmogorov,Beattie2025_small_scale_instabilities}. Understanding how this kind of shock-driven turbulence is generated, evolves, and transfers energy across scales is therefore a central problem in astrophysics, which can be probed in the laboratory.

    The balance between compressive and solenoidal motions is one of the key diagnostics of this problem. In supersonic isothermal turbulence, compressive forcing produces stronger density enhancements, larger voids, and density probability distribution functions (PDFs) with standard deviations about three times larger than solenoidal forcing at fixed rms Mach number \citep{Federrath2009,Federrath2010_solendoidal_versus_compressive}. Similar driving-mode comparisons show that the intermittency and non-Gaussian structure of density PDFs increase with more compressive forcing in highly magnetized, molecular-cloud-like turbulence \citep{Beattie2022_spdf}. The same distinction is used to interpret turbulent driving in molecular clouds, with more compressive forcing associated with swept-up shells \citep{Federrath2010_solendoidal_versus_compressive}. Ionizing feedback provides a concrete example: simulations of expanding H~II regions and ALMA observations of Carina pillars both indicate that radiation-driven compression can shape the turbulence and enhance the star-forming state of the gas \citep{Menon2020b,Menon2020}. A closely related Helmholtz diagnostic is the compressive ratio, $\langle u_c^2\rangle/\langle u_s^2\rangle$, where $u_c$ and $u_s$ denote the compressive and solenoidal velocity components; this ratio is used to quantify cloud driving modes in galaxy simulations \citep{Jin2017}. Molecular-cloud calculations show that the effective driving mode can vary in both time and space during cloud formation and evolution \citep{Kortgen2017}, while supernova- and shock-driven calculations use the same compressive--solenoidal partition to characterize feedback-driven and shock-driven turbulence \citep{Padoan2016_supernova_driving,Dhawalikar2022_shock_driving_parameter}. Observationally, the turbulence-driving mixture has been inferred in nearby extragalactic star-forming gas and spatially mapped across the Small Magellanic Cloud (SMC), making the mode mixture an empirical diagnostic rather than only a simulation quantity \citep{Sharda2022_driving_parameter,Gerrard2023_driving_parameter}. It also affects scale-by-scale velocity statistics: compressively driven turbulence has a steeper density-weighted velocity spectrum than solenoidally driven turbulence \citep{Federrath2013_universality}.

    \begin{figure*}
        \centering
        \includegraphics[width=1\linewidth]{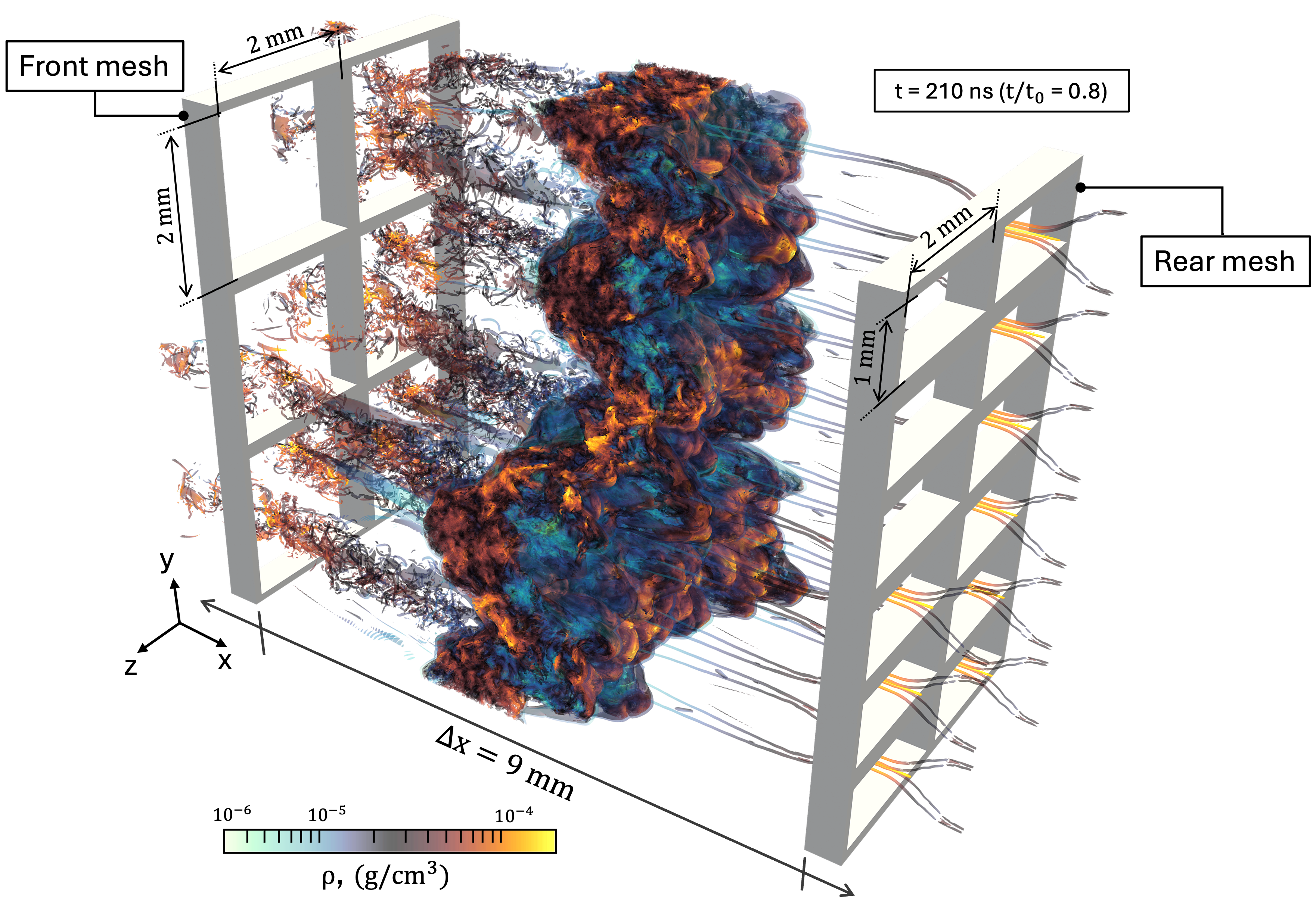}
        \caption{Full-domain view of the simulated double-mesh target irradiated by a single X-ray source. The two CH meshes are separated by $9\,\rm{mm}$ and have distinct cell geometries: $2\times2\,\rm{mm}$ square apertures in the front mesh and $2\times1\,\rm{mm}$ end-slot features in the rear mesh. The front mesh is directly exposed to the radiation drive. The iso-surfaces show the Q-criterion structures that develop within the central shocked mixing layer formed by the collision of the counter-streaming supersonic ablation flows (see \autoref{sec:GenVorticity}). The iso-surfaces are coloured by mass density, $\rho$ ($\rm{g\,cm^{-3}}$), at $t=210\,\rm{ns}$ after the start of the X-ray pulse.}
        \label{fig:ExampleFullcScaleSim}
    \end{figure*}

    Directly diagnosing the internal structure of astrophysical shock-driven turbulence remains difficult because radiative emission, projection, and line-of-sight integration obscure the underlying flow \citep{Draine1993_ISM_shocks,Gaesnsler_2011_trans_ISM}. High-energy-density laboratory astrophysics (HEDLA) experimental analogues offer a complementary route by generating shock-driven turbulence under controlled, diagnosable conditions. Previous HEDLA platforms (experiments) have established important aspects of blast-wave interaction and turbulence generation, including laser-driven blast waves \citep{tzeferacos_laboratory_2018,casner_turbulent_2018,collins_role_2020,Tzeferacos2018_dynamo_in_the_lab,Bott2021_time_resolved_dynamo,bott_insensitivity_2022} and shock tubes filled with low-density foams \citep{davidovits_turbulence_2022,Dhawalikar2022_shock_driving_parameter,Hew2023_lagrangian_stats}. What remains comparatively unexplored is a platform that produces a sustained, shock-rich turbulent mixing layer whose mode composition, large-scale anisotropy, and multiscale structure can be quantified. Before interpreting what future diagnostics may measure in such systems, one must first establish the underlying turbulent state itself: how the turbulence is seeded, how compressive and solenoidal motions coexist, whether the large scales relax toward isotropy or retain memory of the driving geometry, how rapidly that directional memory weakens with decreasing scale, and what range of scales is dynamically resolved.

    Here we study such a platform: two offset, co-aligned CH mesh targets irradiated by a short-duration X-ray pulse, extending the class of pulsed-power-driven radiative plasma platforms discussed by \citet{halliday_investigating_2022}. \autoref{fig:ExampleFullcScaleSim} provides an overview of the plasma dynamics in the system. The ablation launches counter-streaming supersonic flows whose structure is set by the mesh geometry already during the drive phase. Vorticity is seeded baroclinically near the mesh-cell corners, advected into collimated channels, and injected into the outgoing streams before the two flows collide. Their interaction forms a shocked, close-to-isothermal turbulent mixing layer, highly relevant to the cold phase ISM \citep{Ferriere2020_reynolds_numbers_for_ism}, that persists for several hundred nanoseconds (longer than an outer-scale eddy turnover time) and supports a resolved turbulent cascade with both compressive and solenoidal components. This makes the platform well suited to establishing the physics of shock-driven turbulence and providing the baseline needed to interpret future laboratory diagnostics of interstellar flow analogues.

    \begin{table*}[htb] 
    \begin{threeparttable}
    \caption{List of simulations and their associated numerical parameters.}
    \begin{tabular*}{\textwidth}{@{\extracolsep{\fill}} lccccccc}
    \hline
    Sim. ID & Front Mesh  & Rear Mesh & Domain Size & $N_x \times N_y \times N_z$  & $\delta$ & Hydro & Riemann \\
     & Design & Design & ($\rm{cm}$) & & ($\mu\rm{m}$) & Order & Solver \\
    (1) & (2) & (3) & (4) & (5) & (6) & (7) & (8) \\
    \hline\hline
    yz32o3$\delta$25      &  6 square $2\times2$ mm    &   12 end-slot $1\times2$ mm   & $1.2 \times 0.75 \times 0.5 $&    $480\times300\times200$      &     25        &     3   &  Hybrid \\
    yz32o3$\delta$12.5    &  6 square $2\times2$ mm    &   12 end-slot $1\times2$ mm   & $1.2 \times 0.75 \times 0.5 $&     $960\times600\times400$     &    12.5       &     3   &  Hybrid \\
    yz32o2$\delta$10      &  6 square $2\times2$ mm    &   12 end-slot $1\times2$ mm   & $1.2 \times 0.75 \times 0.5 $&      $1200\times750\times500$   &    10         &     3   &  HLLC   \\
    yz32o2$\delta$12.5    &  6 square $2\times2$ mm    &   12 end-slot $1\times2$ mm   & $1.2 \times 0.75 \times 0.5 $&      $960\times600\times400$    &    12.5       &     2   &  Hybrid \\
    yz32o1$\delta$12.5    &  6 square $2\times2$ mm    &   12 end-slot $1\times2$ mm   & $1.2 \times 0.75 \times 0.5 $&      $960\times600\times400$    &    12.5       &     1   &  Hybrid \\
    \hline\hline        
	\end{tabular*}
    \begin{tablenotes}[flushleft]
    \item \textit{\textbf{Notes.}} \textbf{Column (1):} simulation identifier. \textbf{Columns (2)-(3):} front and rear mesh designs, including the number and geometry of the apertures. \textbf{Column (4):} physical domain size in cm, ordered as $L_x \times L_y \times L_z$. \textbf{Column (5):} grid resolution, $N_x \times N_y \times N_z$. \textbf{Column (6):} the size of a single grid cell, $\delta$, in $\mu$m. \textbf{Column (7):} flux reconstruction order. \textbf{Column (8):} Riemann solver used in the \flash hydrodynamic step.
    \end{tablenotes}
    \label{tab:SimulationsList}
    \end{threeparttable}
\end{table*}

    In this paper, we use three-dimensional, three-temperature (3T) radiation-hydrodynamic simulations with \flash to quantify the dynamics of this platform and the structure of the resulting shock-driven interstellar-type turbulence. We first describe the numerical model and target configuration (\autoref{sec:methods}), then follow the formation of the mixing layer, its integral measures, and the evolution of its outer-scale anisotropy (\autoref{sec:time_evol}). We next show that the interaction region evolves toward an effectively close-to-isothermal thermodynamic closure (\autoref{sec:eos}) and that its vorticity is seeded baroclinically at the mesh, advected into collimated channels, and subsequently amplified by compression and stretching after collision (\autoref{sec:GenVorticity}). We then characterise the turbulence through a number of statistics regularly used in studies of interstellar turbulence: Fourier spectra, Helmholtz mode decomposition, scale-dependent Reynolds-stress anisotropy, and an effective Reynolds-number estimate based on the resolved strain spectrum (\autoref{sec:PowerSpectrum}). Finally, we examine the density and density-gradient spectra, relevant to interstellar scintillation observations \citep[e.g.,]{Armstrong1995_power_law}, showing through the hierarchy in \autoref{eq:density_gradient_hierarchy} that density structure in the layer remains tied to compressive dynamics rather than the solenoidal cascade (\autoref{sec:DensityGrad}). Taken together, these results provide a physical interpretation of how shock-driven turbulence is generated in this platform and a quantitative baseline for the design and interpretation of future HEDLA experiments that are able to probe astrophysical turbulence analogues.
    
    \section{Numerical model and methods}\label{sec:methods}
    We model the radiation-driven flows and their subsequent interaction using the publicly available simulation code, \flash \citep{Fryxell2000, tzeferacos_FLASHMHDSimulations_2015}. \flash is a multi-physics finite-volume code for Eulerian hydrodynamics (HD) and magnetohydrodynamics (MHD), with support for multi-temperature equations of state and multi-group opacities. Here we use a three-temperature (3T) model, which evolves separate internal energy equations for the electron, ion, and radiation components of the plasma, together with multi-group flux-limited radiation diffusion.
    
    Our aim is to isolate how radiation-driven counter-streaming plasma outflows generate turbulence in the hydrodynamic limit. We therefore omit external magnetic fields and Biermann-battery effects, which are not expected to play a leading role in the experimental counterpart considered here because resistive diffusion is significant in the experiment. Accordingly, we use the 3T unsplit hydrodynamics solver in \flash, i.e. the unmagnetized limit of the more general HED plasma models widely used for laboratory-astrophysics simulations \citep{tzeferacos_numerical_2017, tzeferacos_laboratory_2018}.
    
    \subsection{3T hydrodynamic fluid model}
        Estimates for the ISM place the turbulent cascade, from the driving scale down to the dissipation scale, in the collisional regime, making a hydrodynamic treatment a well-justified first description of the interstellar counterpart \citep{Ferriere2020_reynolds_numbers_for_ism}\citep[see Fig.~11 of][]{Howes2024_fundamental_parameters}. The governing equations for an ideal, 3T, unmagnetized fluid plasma can be written in conservative form as
        \begin{align}
                \frac{\partial \rho}{\partial t} + \bnab \cdot (\rho \u) &= 0, \label{eq:ConservationMassE}\\
            \frac{\partial\rho \u}{ \partial t} + \bnab\cdot ( \rho \u\otimes\u + \Ptot\It) &= 0,\label{eq:ConservationMomentumE}\\ 
            \frac{\partial\rho \Etot}{\partial t} + \bnab \cdot [(\rho \Etot + \Ptot)\u] &= - \bnab \cdot \q, \label{eq:ConservationEnergyE}
        \end{align}
        where $\rho$ is the total mass density and $\u$ is the fluid velocity. The tensor product $\u \otimes \u$ is defined component-wise as $\u \otimes \u \equiv u_i u_j$, $\Ptot = \Pi + \Pe + \Pr$, and $\Etot = \Ei + \Ee + \Er + u^2/2$. Here $\Ei$, $\Ee$, and $\Er$ are the ion, electron, and radiation specific internal energies, $\It \equiv \delta_{ij}$ is the identity tensor, and $\q = \q_{\rm e} + \q_{\rm rad}$ is the total heat flux. In the 3T model, the individual component energies are evolved according to
        \begin{align}
            \label{eq:EnergyIon}
            \frac{\partial (\rho \Ei)}{\partial t} + \bnab \cdot (\rho \Ei \u) + \Pi \bnab \cdot \u &= \Qheat, \\
            \label{eq:EnergyEle}
            \frac{\partial (\rho \Ee)}{\partial t} + \bnab \cdot (\rho \Ee \u) + \Pe \bnab\cdot \u &= -\bnab \cdot \q_{e} \nonumber \\
                - \Qheat + \Qabs - \Qemis&,  \\  
            \label{eq:EnergyRad}
            \frac{\partial (\rho \Er)}{\partial t} + \bnab \cdot (\rho \Er \u) + \Pr \bnab \cdot \u &= -\bnab \cdot \q_{\rm{rad}} \nonumber \\
            - \Qabs& + \Qemis, 
        \end{align}
        where $\Qheat$ is the electron--ion thermal exchange term,
        \begin{align}
            \Qheat = \rho \frac{\cve}{\tau_{ei}}(\Te - \Ti)
        \end{align}
        with $\cve$ the electron specific heat and $\tau_{ei}$ the ion--electron equilibration time. The radiative coupling terms $\Qabs$ and $\Qemis$ are
        \begin{align}
            \Qabs &= c \, \kappa_{\rm abs} \rho \Er, \\[4pt]
            \Qemis &= c \, \kappa_{\rm abs} \rho a \Te^4,
        \end{align}
    where $\kappa_{\rm abs}$ is the absorption opacity, $a$ is the radiation constant, and $c$ is the speed of light. 
    
\begin{figure}[t]
    \centering
    \includegraphics[width=\linewidth]{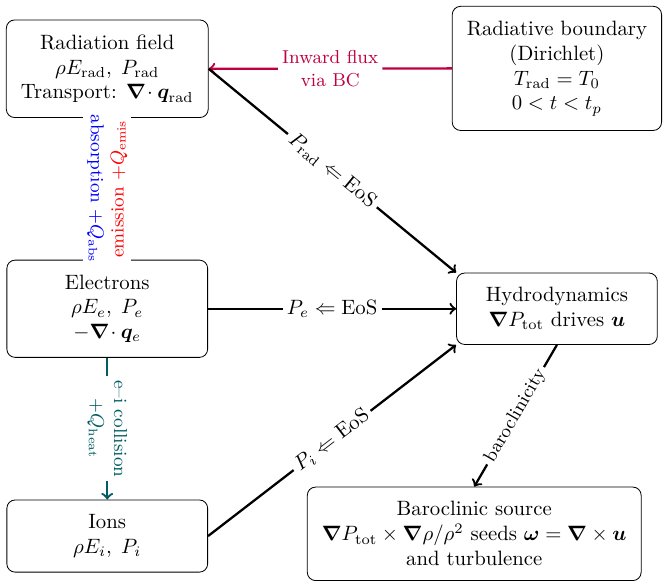}
    \caption{Schematic of the energy flow in the 3T radiation-hydrodynamics model driven by the radiative Dirichlet boundary described in \autoref{sec:BCs}. Energy enters the domain exclusively through the radiation field, $\Er$, at the driven boundary and is transported inward by $\q_{\rm rad}$. In dense material, $\Qabs$ couples $\Er$ to $\Ee$, $\Qemis$ returns energy from $\Ee$ to $\Er$, and $\Qheat$ transfers energy from $\Ee$ to $\Ei$. The 3T EoS then maps $(\Er,\Ee,\Ei)$ to $(\Pr,\Pe,\Pi)$, whose sum defines $\Ptot$. Gradients of $\Ptot$ launch the ablation flows from the mesh walls and ultimately seed the turbulence injected into the mixing layer through baroclinic vorticity generation.}
    \label{fig:energy_flow_3T}
    \end{figure}

    \autoref{fig:energy_flow_3T} summarises the energy flow in the model, including the Dirichlet radiative boundary described in \autoref{sec:BCs}. The external X-ray drive enters the system exclusively through this boundary as $\Er$; no volumetric external source term is included in the matter equations. The net injected power is given by the inward radiative flux through the driving boundary,
    \begin{equation}
        \dot{E}_{{\rm inj}}(t) = -\int_{\mathcal{S}}
        \q_{\rm rad} \cdot \hat{\bm{n}} \, dA
        \label{eqs:RadiationFlux}
    \end{equation}    
    As radiation propagates inward, $\Qabs$ transfers energy from $\Er$ to $\Ee$. This coupling is strongly localised to dense regions, i.e. the solid mesh walls, and is negligible in the low-density vacuum. Electrons can return energy to the radiation field through $\Qemis$, while $\Qheat$ transfers energy from $\Ee$ to $\Ei$.

    The system is closed with tabulated 3T equations of state and opacities from \textsc{ProPacEOS} \citep{macfarlane_HELIOSCR1DRadiation_2006}, which relate $(\Er,\Ee,\Ei)$ to $(\Pr,\Pe,\Pi)$. Their sum defines $\Ptot$, whose gradients drive the fluid motion, launch the ablation flows from the mesh walls, and, as we show in \autoref{sec:GenVorticity}, seed the turbulence.
    
    A practical challenge in solving the coupled system of \autoref{eq:ConservationMassE}, \autoref{eq:ConservationMomentumE}, and \autoref{eq:ConservationEnergyE} is that the component-energy equations contain source terms proportional to $\bnab\cdot\u$ in \autoref{eq:EnergyEle}, \autoref{eq:EnergyIon}, and \autoref{eq:EnergyRad}. Across shock discontinuities, $\bnab\cdot\u$ is undefined, so these terms cannot be evaluated directly. \flash implements two standard strategies to handle this difficulty: the entropy-advection approach and the so-called RAGE-like approach, named after its implementation in the radiation-hydrodynamics code \textsc{RAGE} \citep{gittings_RAGE_Code_2008}.

    In the simulations presented here, we adopt the RAGE-like approach implemented in \flash. In this method, shock heating is not assigned exclusively to $\Ei$; instead, it is partitioned among $\Ei$, $\Ee$, and $\Er$ in proportion to their partial pressures. This treatment is formally consistent with the smooth-flow equations, but it is not strictly correct at a shock, where the dissipative heating should enter only the ion energy equation.

    \subsection{Initial conditions and geometry of the setup}\label{sec:ICs}
    \begin{figure}
        \centering
        \includegraphics[width=1\linewidth]{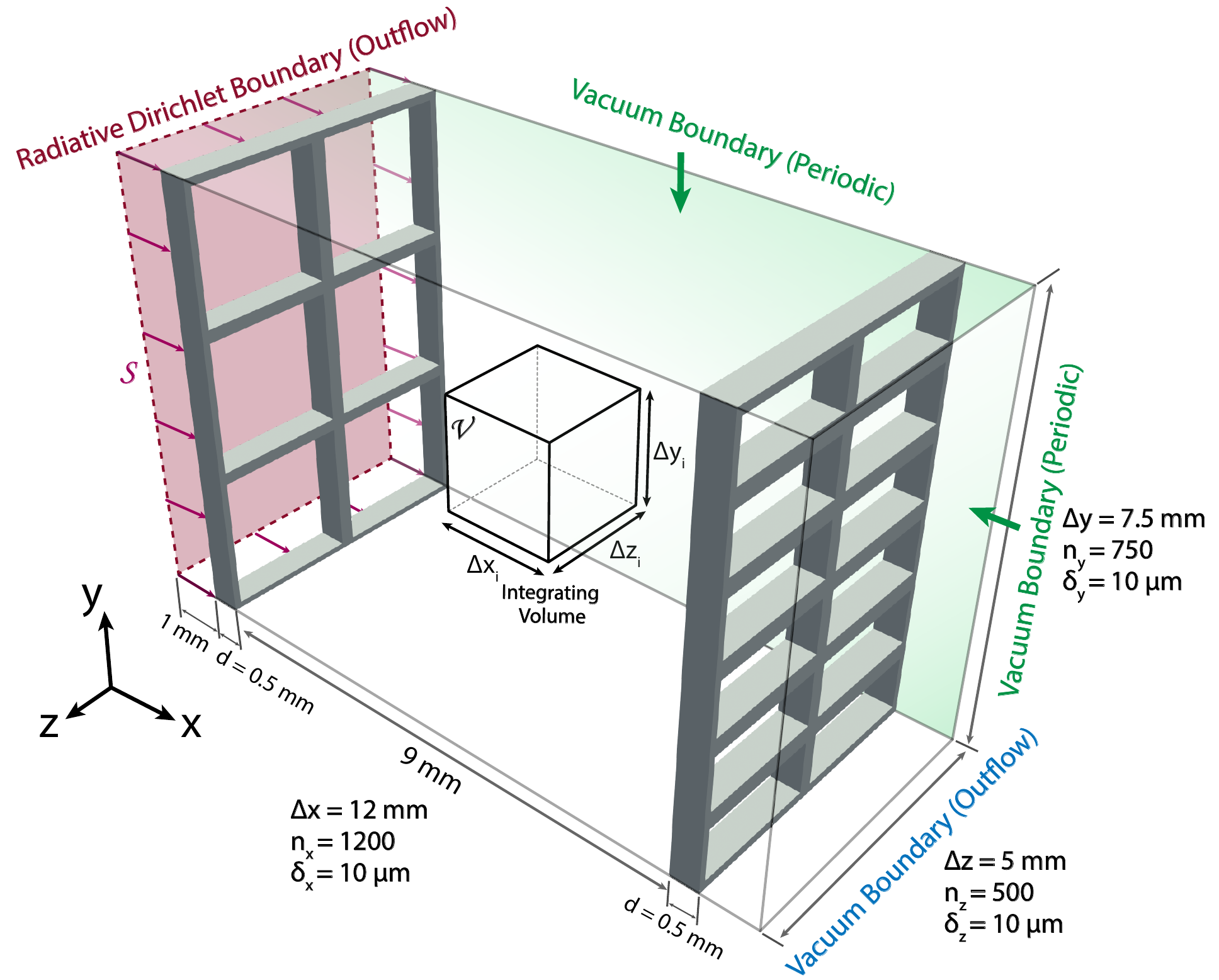}
        \caption{Geometry and boundary conditions of the computational domain. Two co-aligned CH meshes are separated by $9\,\rm{mm}$ and irradiated from the left by a radiative Dirichlet boundary imposed on the $yz$ surface, $\mathcal{S}$. Outflow boundary conditions are applied to the hydrodynamic variables on the $yz$ planes, while periodic boundary conditions are imposed on the $xy$ and $xz$ faces, emulating an extended target in the transverse directions. The front mesh is located $1\,\rm{mm}$ downstream of the emitting surface. The highlighted integrating volume defines the region over which time-resolved hydrodynamic quantities are computed, spanning $\Delta x_i \times \Delta y_i \times \Delta z_i = 400 \times 400 \times 400$ computational cells.}
        \label{fig:Setup}
    \end{figure}
        
    The \flash simulations are initialised to replicate a target design intended as a laboratory platform for studying plasma turbulence driven by soft-X-ray irradiation from pulsed-power sources \citep{Lebedev_2001_wire_array, Bland_2006_diagnostics, halliday_investigating_2022, Merlini_2023, Marrow_2026}. The target system consists of two co-aligned CH meshes normal to the $x$ direction and facing the incident X-ray source, as illustrated in \autoref{fig:Setup}. The front mesh has high transparency (82\% open area) with $2\times2\,\rm{mm}$ square apertures and is located $1\,\rm{mm}$ from the emitting $yz$ surface. The rear mesh has lower transparency (62\% open area) with $2\times1\,\rm{mm}$ end-slot features and is placed $9\,\rm{mm}$ behind the front mesh. Both meshes are initialised as plastic (CH) with density $\rho = 1.05 \, \rm{g \, cm^{-3}}$ at room temperature, with $T = T_i = T_e = T_{\rm rad} = 0.025 \, \rm eV$ ($290\, \rm{K}$). Although we employ CH tabulated EoS and opacity tables, we treat the ablated plasma as an effective single-species fluid, taking carbon as the representative constituent.
    
    To reduce computational cost while maintaining high resolution, we simulate only a subsection of the full target system and use periodic transverse boundaries to emulate its lateral repetition. The surrounding vacuum regions are initialised with the same material properties as the meshes but at a much lower density, $\rho_{\rm vac} = 10^{-8}\,\rm{g\,cm^{-3}}$, and with temperature $T_{\rm vac} = 0.025\,\rm{eV}$ ($290\,\rm{K}$). The principal plasma, transport, and dimensionless parameters derived from two temporal windows in the same control volume are summarised in \autoref{tab:plasma_inflows_params}.

    \begin{table*}[htb] 
\begin{threeparttable}
\centering 
\caption{Plasma flow parameters for the two head--on inflows and the central mixing layer, computed as averages over the control volume $\mathcal{V}$ at two evolutionary stages.}
\label{tab:plasma_inflows_params} 
\setlength{\tabcolsep}{2pt}
\begin{tabular*}{\textwidth}{@{\extracolsep{\fill}} lcccc} 
    \hline
    \textbf{Quantity} & \textbf{Symbol} & \textbf{Formula} & \textbf{Inflow Stage (A)} & \textbf{Mixing Stage (B)} \\
    \hline\hline 
    \multicolumn{5}{c}{\textbf{Plasma parameters}}\\
    \hline 
    Atomic Mass Number & $A$ & – & 12 & 12 \\  
    Mass density (g/$\text{cm}^{3}$) & $ \langle \rho \rangle_{\V}$ & $\displaystyle \frac{1}{\V}\int_{\V} \rho\,\d{\V}$ & $ (5.0 \pm 1.5) \times 10^{-6}$  & $ (2.5 \pm 1.0) \times 10^{-6}$ \\ 
    Mass density perturbations & $ \langle \rho_{\rm{rms}} \rangle_{\V}$ & $\left[ \frac{1}{\V}\int_{\V} \left(\rho-\langle\rho\rangle\right)^2 \,\d{\V} \right]^{1/2}$ & $ (1.8 \pm 0.3) \times10^{-5}$ & $ (1.0 \pm 0.7) \times 10^{-6}$\\  
    Electron density (cm$^{-3}$) & $\langle n_e\rangle_{\V}$ & $ \displaystyle \frac{1}{\V}\int_{\V} n_e\,\d{\V}$ & $(4 \pm 1) \times 10^{17}$ &  $(3.0 \pm 1.5) \times 10^{17}$ \\ 
    Electron density fluctuations (cm$^{-3}$) & $ \langle n_{e,\rm rms} \rangle_{\V} $ & $\left[ \frac{1}{\V}\int_{\V} \left(n_e-\langle n_e \rangle\right)^2 \,\d{\V} \right]^{1/2}$ & $ (1.5 \pm 0.2) \times 10^{18}$ & $ (2.0 \pm 0.6) \times 10^{17} $ \\ 
    Velocity (km/s) & $  \langle \u \rangle_{\V} $ & $ \displaystyle \frac{1}{\V} \int_{\V} \u \,\d{\V}$ & $14 \pm 4$  & $2.5 \pm 3$ \\  
    Velocity fluctuations (km/s) & $ u_0$ & $ \left[ \frac{1}{\V}\int_{\V} \left(\u-\langle \u \rangle\right)^2 \,\d{\V} \right]^{1/2} $ & $2.0 \pm 0.2 $ & $25 \pm 4$ \\  
    Ion temperature (eV) & $ \langle \Ti \rangle_{\V} $ & $\displaystyle \frac{1}{\V} \int_{\V} \Ti \,\d{\V}$ & $4.0 \pm 0.2$ & $24 \pm 5$ \\  
    Electron temperature (eV) & $ \langle \Te \rangle_{\V} $ & $\displaystyle \frac{1}{\V} \int_{\V} \Te \,\d{\V}$ & $4.0 \pm 0.2$ & $10.0 \pm 1.3$ \\  
    Radiation temperature (eV) & $ \langle \Tr \rangle_{\V} $ & $\displaystyle \frac{1}{\V} \int_{\V} \Tr\,\d{\V}$ & $4.0 \pm 0.2$ & $3.0 \pm 0.1$ \\  
    Average ionization & $\bar{Z}$ & $ \frac{1}{\V} \displaystyle \int_{\V} n_e / n_i {\rm d}\V $ & $1.7 \pm 0.1$ & $2.3 \pm 0.1$ \\  
    Ion-acoustic sound speed (km/s) &  $ \langle c_{s} \rangle_{\V}$ & $\displaystyle \frac{1}{\V} \int_{\V} \sqrt{\Ptot / \rho} \,\d{\V} $ & $9.0 \pm 0.4$ & $18.5 \pm 1.0$ \\  
    Kinematic viscosity ($\rm cm^{2}/s$) ($\nu_{\rm bulk} \simeq 0$)  & $ \langle \nu \rangle_{\V} $ & $  \displaystyle \frac{1}{\V} \int_{\V} v_{th,i} \lambda_{ii} \,\d{\V}$ &  $25 \pm 6$  & $ (5 \pm 4) \times 10^{2}$  \\  
    Thermal diffusivity ($\rm cm^{2}/s$) & $ \langle \chi \rangle_{\V} $ & $ 3.3 \times 10^{-3} \displaystyle \frac{A \langle T \rangle_{\V}^{5/2}}{[\Lambda \bar{Z} (\bar{Z}+1) \langle \rho \rangle_{\V}]} $ &  $ (1.0 \pm 0.6) \times 10^{4}$  & $ (7.5 \pm 7.0) \times 10^{5}$ \\ 
    \hline
    \multicolumn{5}{c}{\textbf{Dimensionless parameters}}\\
    \hline 
    Sonic Mach number & $\M$ & $ u_0 \Big/ \left(\displaystyle \frac{1}{\V} \int_{\V}  c_{s} \,\d{\V} \right)$ & $ 2.4 \pm 0.2$ & $ 1.6 \pm 0.5 $  \\  
    Reynolds number & $\rm{Re}$ & $ \displaystyle \int_{\V} \u_0\ell_0/ \langle \nu \rangle_{\V} \,\d{\V} $ & $ (7.0 \pm 0.3) \times 10^{4}$ & $(1.0 \pm 0.2) \times 10^{4}$ \\  
    Reynolds number fluctuations & $\rm{Re_0}$  &  $ \left[ \frac{1}{\V}\int_{\V} \left(\rm{Re}-\langle \rm{Re} \rangle\right)^2 \,\d{\V} \right]^{1/2}$  & $ (8.0 \pm 0.2) \times 10^{4}$ & $(1.50 \pm 0.25) \times 10^{4}$ \\  
    Péclet number  & $\rm{Pe}$ & $ \u_0\ell_0/ \langle \chi \rangle_{\V}$ & $ (1.50 \pm 0.15) \times 10^{2}$ & $ 10.0 \pm 1.5$ \\ 
    Péclet number fluctuations  & $\rm{Pe_0}$ & $\left[ \frac{1}{\V}\int_{\V}\left(\rm{Pe}-\langle \rm{Pe} \rangle\right)^2 \,\d{\V} \right]^{1/2}$ & $ (2 \pm 0.2) \times 10^{2} $ & $16 \pm 4$ \\ 
    \hline
    \multicolumn{5}{c}{\textbf{Characteristic timescales}} \\
    \hline 
    Outer-scale eddy turnover time (ns) & $ t_{0}(t) $ & $ \ell_0(t) / u_0 $ & $97 \pm 6$ & $ 120 \pm 8$\\ 
    Longest Outer-scale eddy turnover time (ns) & $ t_{0} $ & - & \multicolumn{2}{c}{261} \\  
    Sound crossing time (ns) & $t_{c_s}(t)$ & $L / \langle c_s \rangle_{\V}$ & $610 \pm 47$ & $675 \pm 13$ \\  
    Radiation drive time (ns)  & $t_{\rm rad}$ & - & \multicolumn{2}{c}{30} \\ 
    \hline
    \multicolumn{5}{c}{\textbf{Characteristic length scales}}\\
    \hline
    Outer-scale of the turbulence (cm) & $\ell_0(t)$ & $        \frac{\displaystyle\int_0^\infty E(k,t)\,k^{-1}\,\d{k}}{\displaystyle\int_0^\infty E(k,t)\,\d{k}}$ & \multicolumn{2}{c}{$0.45 \pm 0.04$} \\  
    System scale (cm) & $L$ &  $\displaystyle \sqrt{\Delta x^2 + \Delta y^2 + \Delta z^2}$ & \multicolumn{2}{c}{0.7} \\  
    \hline\hline
\end{tabular*}
    \begin{tablenotes}[flushleft]
\item\textit{\textbf{Notes.}} $\mathcal{V}$ denotes the control volume used in each average. Columns A and B refer to two temporal averaging windows in the same control volume: A from $t/t_0 = 0.2$ (55 ns) to $t/t_0 = 0.3$ (75 ns), when the sampled material is dominated by the incoming flows, and B from $t/t_0 = 0.3$ (75 ns) to $t/t_0 = 0.4$ (95 ns), after the central mixing layer has formed. Here $t_0 = 261$ ns is the eddy turnover time defined in \autoref{sec:time_evol}. The two inflows are comparable in their plasma conditions, therefore we report only one for clarity. Quoted uncertainties correspond to the temporal standard deviation within each averaging interval. In computing the Reynolds number, we adopt the shear-viscosity contribution to the kinematic viscosity following \citet{Ryutov_1999}, neglecting the bulk viscosity $\nu_{\rm bulk}$, which is expected to be small for a monoatomic plasma. In the absence of an ambient magnetic field (hydrodynamic regime), the thermal diffusivity is taken to be isotropic and is evaluated using the classical \citet{Braginskii_1965} formulation.
    \end{tablenotes}
\end{threeparttable}
\end{table*}

    \subsection{Boundary conditions and radiative forcing}\label{sec:BCs}
    Outflow boundary conditions are applied to the hydrodynamic variables on the $yz$ planes at the domain edges, allowing mass and energy to leave the domain along the streamwise direction. Periodic boundary conditions are imposed on the $xy$ and $xz$ faces, giving the mesh geometry an effectively infinite transverse extent without edge artefacts. This setup is illustrated in \autoref{fig:Setup}.

    For radiation transport and diffusion, Neumann boundary conditions (zero gradient) are enforced everywhere except at the driven radiative boundary. On the left $yz$ face, identified as $\mathcal{S}$ in \autoref{fig:Setup}, a vacuum Dirichlet radiative boundary is imposed by prescribing the radiation energy density through an equivalent radiation temperature. This boundary condition sets the temporal profile of the drive to a top-hat pulse with peak radiation temperature $T_0$ and duration $t_{\rm rad} = 30 \, \rm{ns}$, thereby directly controlling the inward radiative flux entering the domain,
    \begin{equation}
        T_{\rm rad}\big|_{\mathcal{S}} =
        \begin{cases}
            T_0, & 0 < t < t_{\rm{rad}} , \\[4pt]
            0, & t \ge t_{\rm{rad}} .
        \end{cases}  
    \end{equation}
    Through the relation $E_{\rm rad} = aT_{\rm rad}^4$, this imposed boundary temperature determines the incoming radiative energy density and hence the net injected power via the radiative flux in \autoref{eqs:RadiationFlux}. The radiation temperature is chosen to match the expected flux incident on a target positioned $4\,\rm{cm}$ from an imploding wire array driven by a $1\,\rm{MA}$ peak current, 240 ns rise-time pulsed-power electrical discharge \citep{Lebedev_2001_wire_array, Bland_2006_diagnostics,halliday_investigating_2022}, yielding $T_{\rm rad}\big|_{\mathcal{S}} = T_0$ in the range $8$--$10\,\rm{eV}$ ($93,\!000$--$116,\!000\,\rm{K}$) over the front mesh area. For simplicity, we neglect directional shadowing and partial self-absorption, and assume isotropic illumination, which is a reasonable approximation given the high transparency of the front mesh.
    
    \subsection{Discretisation and numerical diffusion}
    The computational domain is resolved with a uniform 3D Cartesian grid of $1,\!200 \times 750 \times 500$ cells, covering a physical volume of $1.2\,\rm{cm} \times 0.75\,\rm{cm} \times 0.5\,\rm{cm}$. This corresponds to a uniform cell size of $\Delta x = \Delta y = \Delta z = \Delta = 10 \,\rm{\mu m}$ in all directions (see \autoref{fig:Setup}). At this resolution we capture the large-scale flow structure together with a resolved solenoidal scale separation of about $\ell_0/\ell_{\nu,\rm s} \sim 50$, even though the grid does not reach the true laboratory dissipation scale (see \autoref{sec:estimating_Reynolds}).

    No explicit sub-grid turbulence model, such as RANS or BHR, is evolved \citep[e.g.,][]{Grete2015_supersonic_closures,Semenov2025_subgrid_turbulence}. Instead, the simulation operates in the implicit large-eddy simulation (ILES) regime, in which the numerical discretisation acts as an effective dissipation operator. In this sense, the resolved turbulent dynamics can be interpreted as a direct numerical simulation over the dynamically accessible range, with dissipation supplied by the numerical discretisation rather than by an explicit microphysical viscosity \citep{Grete2023_as_a_matter_of_dynamical_range,Teissier2024_higher_order_schemes,Shivakumar2025_numerical_diffusion,Grehan2025_numerical_reconnection}.

    As detailed in \autoref{sec:estimating_Reynolds}, we estimate the local dissipation (strain) scale and the effective Reynolds number by analysing the rate-of-strain tensor. For our third-order runs, this yields a solenoidal strain scale of $\ell_{\nu,\rm s}\simeq 92\,\rm{\mu m}\simeq 9\Delta$, a resolved scale separation $\ell_0/\ell_{\nu,\rm s} \simeq 49$, and hence an effective numerical Reynolds number of order $\Re\sim 2\times10^2$ under standard Kolmogorov scaling. To verify that the resolved scales adequately capture the essential turbulent statistics, we perform a reconstruction-order study (see \autoref{app:convergence}). A summary of all simulations used in this study is given in \autoref{tab:SimulationsList}.
    
    \section{Time evolution of the system and integral quantities}\label{sec:time_evol}

    \begin{figure*}
        \centering
        \includegraphics[width=1\linewidth]{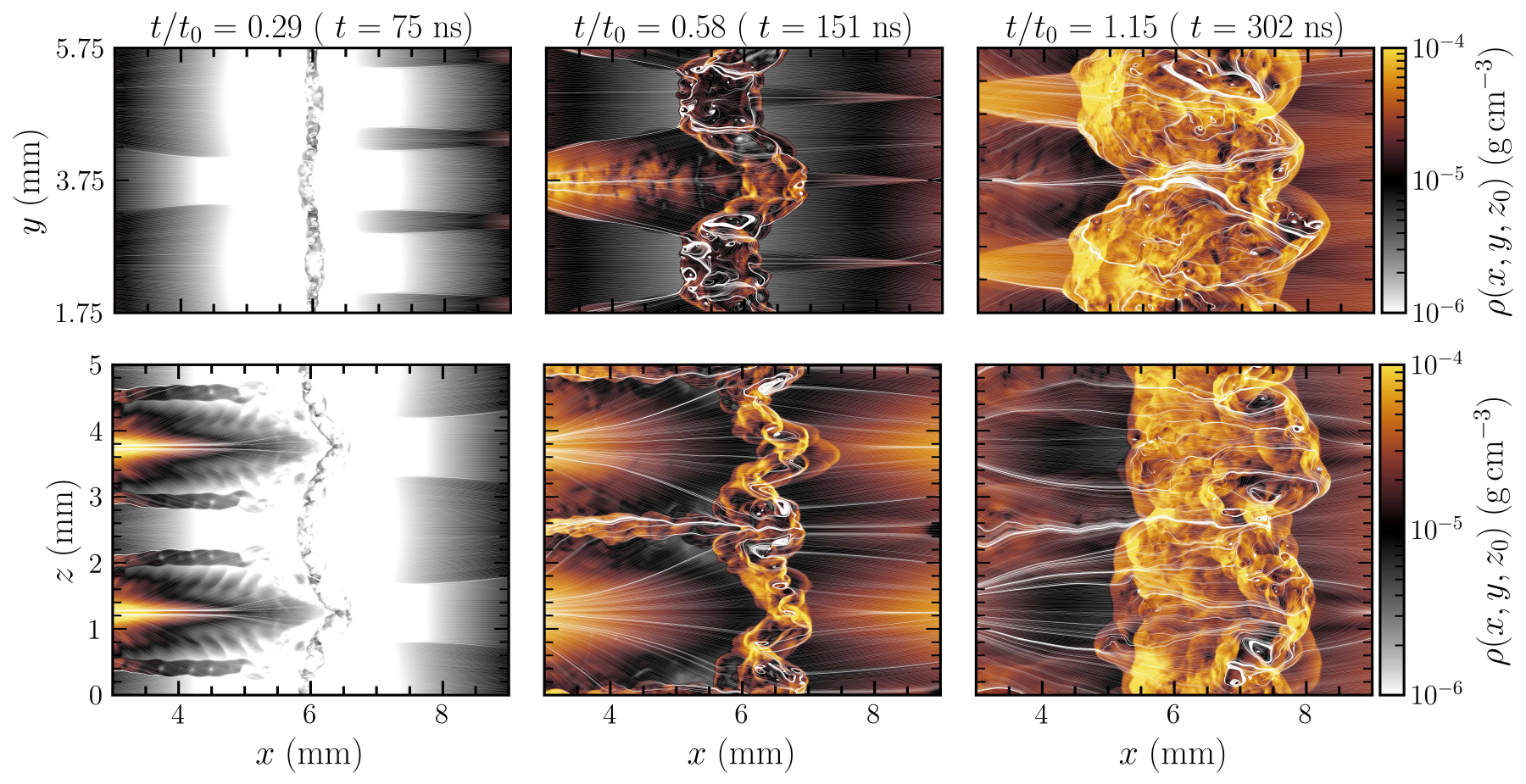}
        \caption{Time evolution of the central mixing layer at three representative stages following the onset of the X-ray drive. The top row shows density slices in the $xy$ plane at fixed $z$, while the bottom row shows complementary slices in the $xz$ plane at fixed $y$. Columns correspond to $t/t_0 = 0.29$ ($t=75\,\rm{ns}$), $t/t_0 = 0.58$ ($t=151\,\rm{ns}$), and $t/t_0 = 1.15$ ($t=302\,\rm{ns}$). The colour scale gives the mass density, $\rho$, in $\rm{g\,cm^{-3}}$, and the overlaid streamlines trace the in-plane velocity field in each slice. At $t=75\,\rm{ns}$, the counter-streaming outflows first meet near the mid-plane at $x\simeq 6\,\rm{mm}$ and the mixing layer begins to form. By $t=151\,\rm{ns}$, the interaction region has broadened and become strongly corrugated, with vortical motions and shear structures visible in both projections. By $t=302\,\rm{ns}$, the layer has expanded into a fully developed, shock-rich turbulent region extending over several millimetres. The slightly different morphologies seen in the two rows reflect the anisotropic imprint of the front and rear mesh geometries on the flow.}
        \label{fig:TimeResolvedLayer}
    \end{figure*}
    
    \subsection{Global evolution of the mixing layer}

    We first summarise the large-scale evolution of the system, following it from the ablation-driven launch of the mesh outflows to the emergence and growth of the central mixing layer, as illustrated in \autoref{fig:ExampleFullcScaleSim}.
    
    \autoref{fig:TimeResolvedLayer} shows the evolution from the early target-ablation phase to the formation of the turbulent mixing layer. The plasma is driven by a 30-ns duration X-ray pulse, which ionizes the surface of the mesh targets. As ablation occurs, and as a result of strong pressure gradients, the superficial plasma is accelerated outwards, forming two counter-streaming supersonic outflows that propagate toward the mid-plane at $x\simeq 6\,\rm{mm}$. Each outflow results from the collective merging of many adjacent aperture-scale streams as the individual mesh holes fill with ablated material during the drive. Vorticity, $\bom = \bnab\times\u$, seeded by the mesh geometry, is advected with these outflows and inherited by the central mixing layer; its origin is discussed in detail in \autoref{sec:GenVorticity}.

    \begin{figure}
        \centering
        \includegraphics[width=0.9\linewidth]{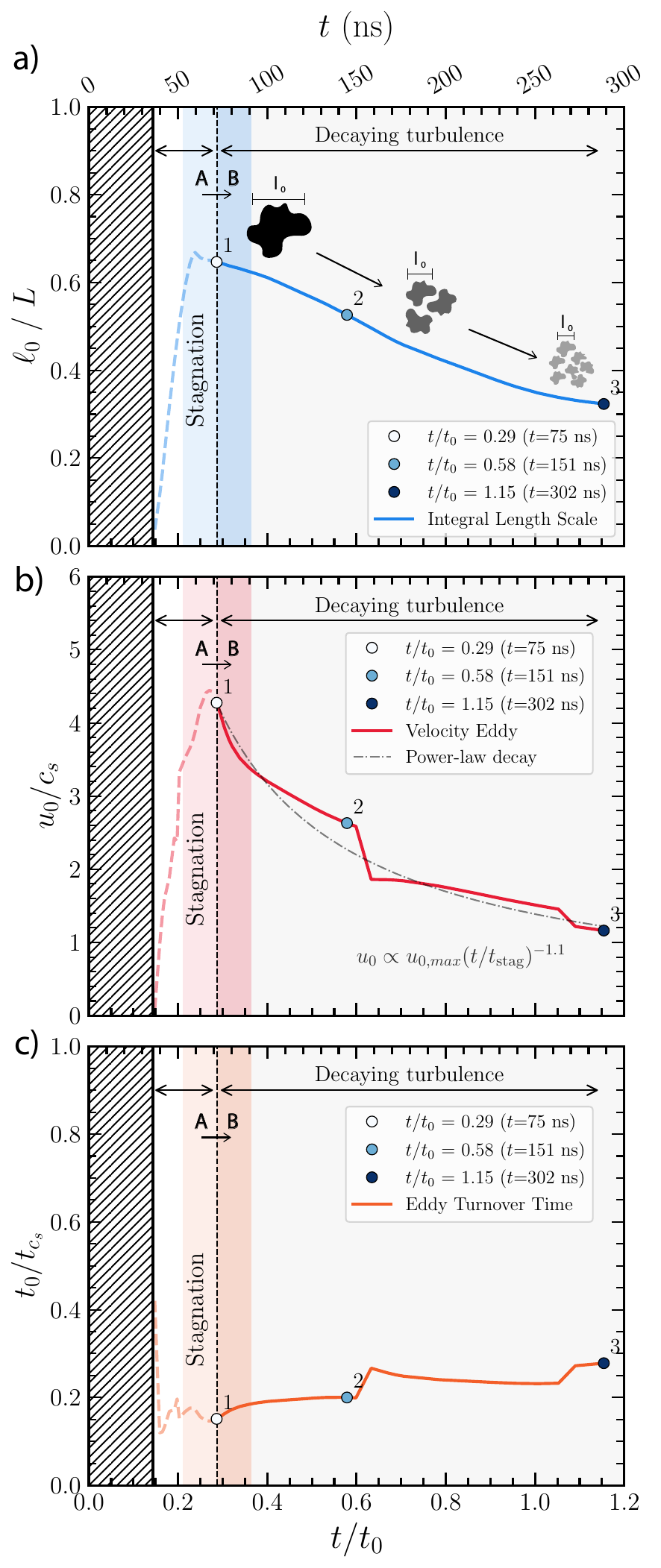}
        \caption{Time evolution of the volume-averaged integral measures within the control volume shown in \autoref{fig:Setup}. Panel (a) shows the normalised integral scale, $\ell_0/L$, panel (b) the turbulent Mach number, $u_0/c_s$, and panel (c) the timescale ratio, $t_0/t_{c_s}$, where $t_0=\ell_0/u_0$. The hatched region marks the radiative drive phase, the vertical dashed line marks the onset of stagnation at $t/t_0 \simeq 0.29$ ($t=75\,\rm ns$), and the arrows indicate the subsequent decaying-turbulence regime. The intervals labelled A and B denote the inflow and mixing-layer averaging windows used in \autoref{tab:plasma_inflows_params}. The figure shows that the largest structures are set near stagnation, after which the turbulence decays while remaining supersonic and maintaining a nearly constant ratio of nonlinear to acoustic timescales.}
        \label{fig:TimeResolvedEddy}
    \end{figure}
    
    The outflows propagate toward the mid-plane with a characteristic bulk speed of order $40\,\rm{km\,s^{-1}}$, while their volume-averaged plasma properties within the inflow control volume are summarised in \autoref{tab:plasma_inflows_params}. These averages give a characteristic mass density $\langle\rho\rangle_{\V}\simeq 5\times10^{-6}\,\rm{g\,cm^{-3}}$, temperatures $\Ti\simeq\Te\simeq 4\,\rm{eV}$, and an ion-acoustic sound speed $\langle c_s\rangle_{\V}\simeq 9\,\rm{km\,s^{-1}}$. The near equality of $\Ti$ and $\Te$ is consistent with rapid electron--ion equilibration at these densities: $\tau_{\rm eq}\simeq 2\,\rm{ns}$, much shorter than the relevant dynamical timescales in the flow ($t_{\rm rad}\sim 30\,\rm{ns}$, $t_0\sim 10^2\,\rm{ns}$, and $t_{c_s}\sim 700\,\rm{ns}$). This ordering justifies treating the inflows as close to thermal equilibrium when interpreting their bulk properties.
    
    Combining the bulk propagation speed with the inflow sound speed implies a bulk sonic Mach number $\M \approx 4.5$. This should be distinguished from the turbulent Mach number based on $u_0/c_s$ discussed below. Uncertainties in the equation-of-state and opacity tables propagate directly into $\Ptot$ and therefore into $c_s$, so the bulk $\M$ quoted here should be regarded as an upper estimate.
    
    \autoref{tab:plasma_inflows_params} summarises the principal hydrodynamic and transport quantities obtained by volume-averaging over the control volume $\mathcal{V}$ during the two temporal windows A and B shown in \autoref{fig:TimeResolvedEddy}. Quoted uncertainties correspond to the temporal standard deviation within these windows. Despite the one-sided boundary forcing, the two inflows remain closely matched in their mean hydrodynamic properties, consistent with a near-symmetric collision and only weak bias in the subsequent stagnation dynamics. The mesh-target geometry breaks symmetry between the $xy$ and $xz$ planes, so the inflows and mixing layer show different projected morphologies (see \autoref{fig:TimeResolvedLayer}). The anisotropy diagnostics in \autoref{ssec:large_scale_anisotropy} and \autoref{fig:Anisotropy_reynolds_outer_scale} quantify how much of this directional imprint survives after the layer forms.

    The collision begins at $t = 75\,\rm{ns}$ (i.e., $t/t_0 = 0.3$), measured from the start of the X-ray pulse, and persists for several hundred nanoseconds, reaching peak stagnation densities of order $10^{-4}~\mathrm{g\,cm^{-3}}$. Immediately after first contact, the plasma heats strongly: the ions rise from $\Ti \simeq 4\,\rm{eV}$ to $\Ti \simeq 24\,\rm{eV}$, while the electrons increase to $\Te \simeq 10\,\rm{eV}$. This heating and corresponding jump in density are, however, less than would be expected for a simple planar, perpendicular post-shock state. In other words, the presence of turbulence decreases the compression ratio compared to the equivalent laminar 1-D case, whilst also redistributing energy and decreasing the amount of ion heating downstream of the shock. The thermodynamic interpretation of this distributed, non-planar interaction is discussed further in \autoref{sec:eos}.
    
    \subsection{Integral measures of the turbulence}

    Further insight into the turbulence and its dynamical state is obtained from the volume-averaged integral measures shown in \autoref{fig:TimeResolvedEddy}: the normalised integral length scale $\ell_0/L$, the turbulent Mach number $u_0/c_s$, and the eddy turnover time $t_0/t_{c_s}$. All quantities are evaluated at each time step within the control volume $\Delta x_i \times \Delta y_i \times \Delta z_i$ highlighted in \autoref{fig:Setup}, corresponding to a spherical sampling radius, $L=0.7~\mathrm{cm}$. Taken together, these diagnostics show that the collision first sets the largest turbulent structures at stagnation, after which the flow enters a self-similarly decaying, but still supersonic, regime rather than a statistically stationary state.
        
    For a scalar field $f$, we define the three-dimensional spectral density
    \begin{align}
        \mathcal{P}_f(\k) = \Tilde{f}(\k)\,\Tilde{f}^{\dagger}(\k),
    \end{align}
    where $\Tilde{f}(\k)$ is the Fourier transform of $f$ and $\square^{\dagger}$ denotes complex conjugation. We isotropise by integrating over shells in $k=|\k|$,
    \begin{align}
         \mathcal{P}_f(k) = \int_{\Omega_k} \mathcal{P}_f(\k)\, k^2\, \d{\Omega_k},
    \end{align}
    with $\d{\Omega_k}$ the solid-angle element on the shell. The normalisation is
    \begin{align}
        \int_0^\infty \mathcal{P}_f(k)\,\d{k} = \Exp{f^2}{\V},
    \end{align}
    where $\Exp{\hdots}{\V}$ is a volume average over $\V$. We also use the shell power $f_k^2 \equiv k \mathcal{P}_f(k)$. We take $\mathcal{P}_u(k)\propto k^{-5/3}$ as the incompressible reference \citep{Kolmogorov1941} throughout our study. 
    
    Turbulence is intrinsically multiscale: kinetic energy is contained at large scales, transferred across an inertial range by nonlinear interactions, and ultimately dissipated at small scales. The largest energy-containing eddies are characterised by the integral length scale $\ell_0$, which provides a robust measure of the outer scale of the turbulent motions in the control volume. We define the integral scale as the energy-weighted mean wavelength, $2\pi/k$,
    \begin{equation}
        \frac{\ell_0(t)}{L}=\frac{\displaystyle\int_0^\infty \mathcal{P}_u(k,t)\,(2\pi/k)\,\d{k}}{\displaystyle\int_0^\infty \mathcal{P}_u(k,t)\,\d{k}}.
    \end{equation}
    Tracking $\ell_0/L$ therefore quantifies how much of the control volume is occupied by coherent, energy-containing structures. In our case, $\ell_0/L$ rises rapidly as the inflows enter the control volume and reaches a maximum value $\ell_0/L \simeq 0.65$, corresponding to $\ell_0 \simeq 4.5~\rm mm$, at the onset of stagnation. After the layer forms, the outer scale decreases steadily in time, as shown in \autoref{fig:TimeResolvedEddy}(a). This post-stagnation decline indicates that the largest structures are set at first contact and then progressively broken down as the flow transitions into a decaying turbulent regime.

    The rate at which these outer-scale eddies strain and feed the cascade is set by the characteristic turbulent velocity $u_0$, defined from the cumulative kinetic energy at wavenumbers $k \gtrsim k_0$, where $k_0\equiv 2\pi/\ell_0$,
    \begin{equation}
        u_0(t)\;\equiv\;
        \left[
        \int_{k_0(t)}^{\infty} \mathcal{P}_u(k,t)\, \mathrm{d}k
        \right]^{1/2}
        \qquad
        \label{eq:u0_def_int}
    \end{equation}
    This definition makes explicit that $u_0$ is an energy-containing velocity scale rather than the mean bulk flow speed $|\langle \u \rangle|$. The corresponding turbulent $\M$ is shown in \autoref{fig:TimeResolvedEddy}~(b) and directly diagnoses the acoustic response compared to the turbulent fluctuations \citep{Federrath2013_universality,beattie_taking_2025}. At the onset of collision and during the early interaction, we find $u_0/c_s > 4$, indicating a strongly supersonic turbulent regime in which compressive motions and shocklets coexist with vortical eddies. After stagnation, however, $u_0/c_s$ decays approximately as a power law, $u_0 \propto t^{-1.1}$, consistent with the secular decay expected in freely decaying turbulence once the outer-scale driving has ceased.

    In classical homogeneous turbulence, the canonical Saffman and Batchelor/Loitsyansky decay classes predict power-law energy decay rates set by the large-scale invariant and infrared spectrum, with $u^2\propto t^{-6/5}$ for $E(k\rightarrow 0)\sim k^2$ and $u^2\propto t^{-10/7}$ for $E(k\rightarrow 0)\sim k^4$ \citep{Ishida2006_decay_isotropic,KrogstadDavidson2010_grid,GorceFalcon2024_saffman,Hosking2021_reconnection_controlled_decay}. Although our mixing layer is neither incompressible nor isotropic at the injection scale, compressible and supersonic simulations likewise show rapid power-law decay once forcing ceases, with dissipation mediated by shocks and shocklets \citep{MacLow1998_decay_supersonic,Samtaney2001_decaying_compressible}. The decay observed here is therefore best interpreted as part of that broader class of freely decaying compressible turbulence, rather than as the approach to a stationary state. Importantly, $u_0/c_s \gtrsim 1$ for longer than $t_0$, implying that the interaction sustains supersonic turbulence over multiple nonlinear times rather than rapidly transitioning to a weakly compressible, $\M < 1$ state.

    Two timescales are particularly relevant in our simulation: the acoustic crossing time $t_{c_s}=L/c_s$, which governs how rapidly pressure disturbances communicate across the control volume, and the eddy turnover time $t_0$, which is the slowest nonlinear timescale for stretching and mixing. On smaller scales, the nonlinear time decreases with scale: for a Kolmogorov cascade, $t_\ell \sim \ell^{2/3}$ \citep{Kolmogorov1941}, whereas for a Burgers-like shock-dominated cascade, $t_\ell \sim \ell^{1/2}$. We define
    \begin{equation}
        t_0(t)=\frac{\ell_0(t)}{u_0(t)},
    \end{equation}
    We find that $t_0/t_{c_s} \sim (c_s / L)(\ell_0 / u_0 ) \sim (\ell_0 / L) (1 / \M)$ remains approximately constant throughout the layer evolution, with a typical value $t_0/t_{c_s} \simeq 0.2$ (see \autoref{fig:TimeResolvedEddy}~(c)). This indicates an approximately self-similar adjustment of the integral measures: as $\ell_0$ decreases in time, $u_0$ decreases proportionally, so that $\ell_0/\M$ remains nearly fixed throughout the evolution. In that sense, the post-stagnation flow decays self-similarly at the level of the integral diagnostics. It is not statistically steady, however, because both the outer scale and the turbulent velocity continue to evolve secularly. Rather, the figure shows a non-stationary but organised decaying regime in which nonlinear and acoustic timescales on $\ell_0$  remain in roughly fixed proportion.

    \subsection{Integral scale Reynolds stress anisotropy}\label{ssec:large_scale_anisotropy}

    \begin{figure}
        \centering
        \includegraphics[width=1\linewidth]{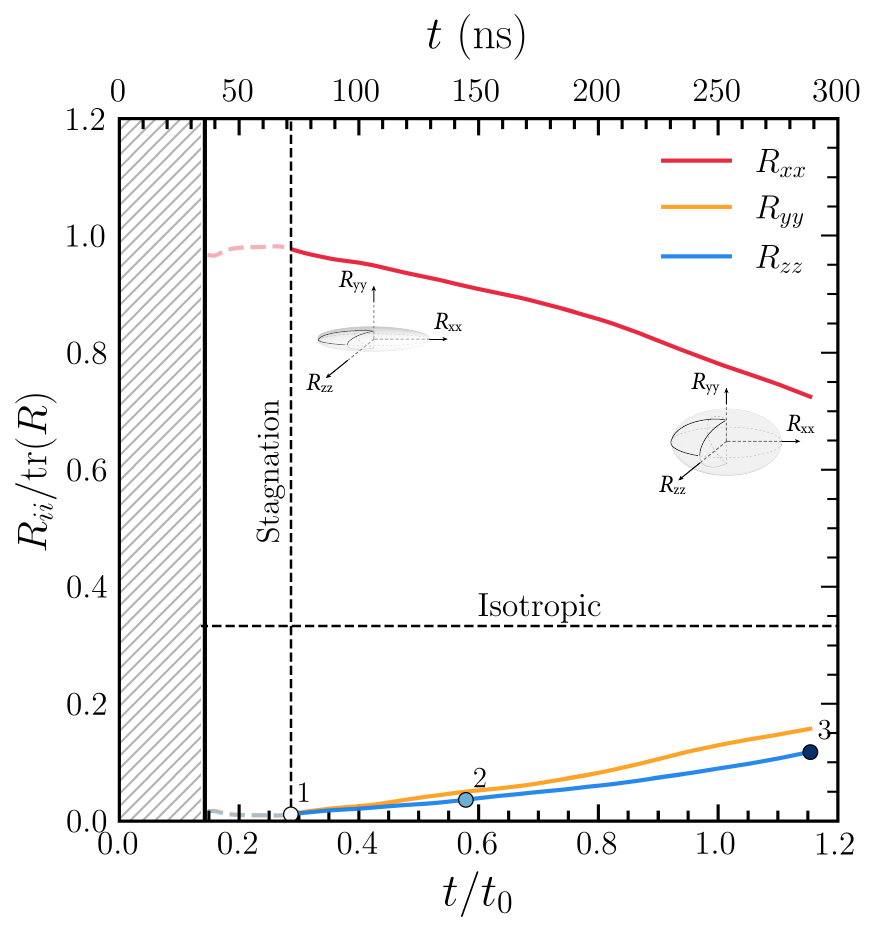}
        \caption{Time evolution of the normalised diagonal Reynolds stresses, $R_{ii}/\mathrm{tr}(R)$, within the interaction control volume, where $\mathrm{tr}(R)=R_{xx}+R_{yy}+R_{zz}$ is the trace. The hatched region marks the radiative drive phase and the vertical dashed line marks the onset of stagnation at $t/t_0\simeq 0.3$ ($t\simeq 75\,\rm ns$). The horizontal dashed line gives the isotropic value, $R_{ii}/\mathrm{tr}(R)=1/3$. At stagnation the fluctuating flow is almost entirely streamwise, with $R_{xx}/\mathrm{tr}(R)\approx 1$ and negligible transverse support. As the mixing layer develops, $R_{yy}/\mathrm{tr}(R)$ and $R_{zz}/\mathrm{tr}(R)$ grow while $R_{xx}/\mathrm{tr}(R)$ declines, but by $t/t_0=1.15$ the flow remains anisotropic, with $R_{xx}/\mathrm{tr}(R)\approx 0.72$, $R_{yy}/\mathrm{tr}(R)\approx 0.16$, and $R_{zz}/\mathrm{tr}(R)\approx 0.12$. The insets sketch the corresponding Reynolds-stress ellipsoids at the three labelled times.}
        \label{fig:Anisotropy_reynolds_outer_scale}
    \end{figure}

    To quantify how strongly the large-scale flow retains memory of the collision geometry, we examine the Reynolds-stress tensor of the fluctuating velocity field,
    \begin{equation}
        R_{ij}(t)
        \equiv
        \Exp{\left(u_i-\Exp{u_i}{\V}\right)\left(u_j-\Exp{u_j}{\V}\right)}{\V}.
    \end{equation}
    The figure itself shows the diagonal contribution of each component normalised by the trace, $R_{ii}/\mathrm{tr}(R)$, so that the isotropic limit appears at $1/3$ for each component. \autoref{fig:Anisotropy_reynolds_outer_scale} shows that this is not the case in the double-mesh layer. At the onset of stagnation, the fluctuating stress is overwhelmingly streamwise, with $R_{xx}/\mathrm{tr}(R)\simeq 0.98$, while the transverse components remain close to zero. This reflects the fact that first contact is still dominated by the head-on collision of the counter-streaming outflows, so the largest motions initially preserve the compression axis rather than immediately becoming fully three-dimensional.

    As the interaction layer develops, transverse turbulent motions grow and the stress tensor becomes less extreme: $R_{xx}/\mathrm{tr}(R)$ decreases monotonically, while both $R_{yy}/\mathrm{tr}(R)$ and $R_{zz}/\mathrm{tr}(R)$ increase. The approach toward isotropy is only partial, however. By the latest time shown, $t/t_0=1.15$, the outer-scale tensor still lies far from the isotropic limit, with $R_{xx}/\mathrm{tr}(R)\simeq 0.72$, $R_{yy}/\mathrm{tr}(R)\simeq 0.16$, and $R_{zz}/\mathrm{tr}(R)\simeq 0.12$. The ordering $R_{yy}>R_{zz}$ further shows that the two transverse directions are not equivalent, consistent with the asymmetric imprint of the front and rear mesh geometries already visible in \autoref{fig:TimeResolvedLayer}. This persistent anisotropy is also likely reinforced by the fact that the inflows do not arrive as featureless compressive streams: as already suggested by the coherent structures visible in \autoref{fig:ExampleFullcScaleSim} and discussed in detail in \autoref{sec:GenVorticity}, the mesh apertures inject narrow, highly collimated vortical jets into the collision. The large scales therefore retain a clear memory of both the collision axis and the collimation, even while the mixing layer broadens and the smaller-scale turbulence becomes progressively more intermixed.

    This has a direct implication for experimental interpretation. Proton-radiography and proton-tomography analyses of stochastic fields commonly reconstruct path-integrated magnetic structure and then infer three-dimensional magnetic spectra or $B_{\rm rms}$ under statistical homogeneity and isotropy assumptions \citep{Bott2017_proton_imaging_stochastic_fields,tzeferacos_laboratory_2018,Bott2021_time_resolved_dynamo}. Our Reynolds-stress diagnostic measures hydrodynamic velocity anisotropy, not magnetic-field anisotropy, so it does not by itself invalidate those magnetic inversions. It does show, however, that isotropy is not an innocuous outer-scale prior in gridded, collimated-ablation platforms. Any isotropic inversion should therefore be validated for the specific field and scale range being inferred; in the present simulations, the scale-dependent analysis in \autoref{sec:PowerSpectrum} suggests that such assumptions are more defensible over the intermediate resolved cascade than on the injection-scale modes that retain the strongest memory of the target geometry.

    \section{The effective equation of state}\label{sec:eos}

    \begin{figure}
        \centering
        \includegraphics[width=1.0\linewidth]{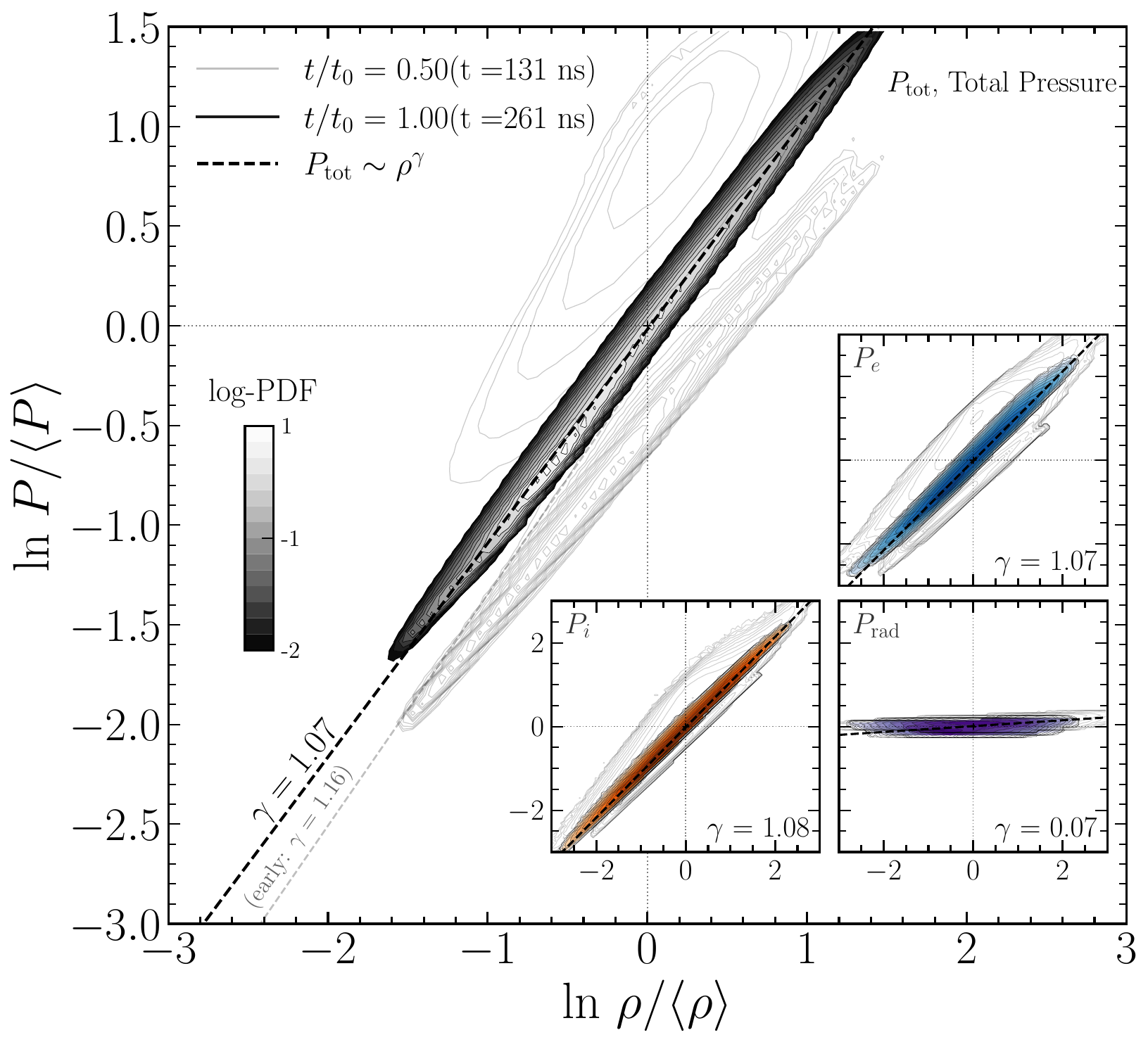}
        \caption{Joint PDFs of density and pressure in the interaction region, shown in terms of the logarithmic contrasts $\ln(\rho/\langle\rho\rangle)$ and $\ln(P/\langle P\rangle)$. The main panel shows the total pressure, $\Ptot$, while the insets show the electron, ion, and radiation partial pressures from the 3T equation of state. Contours are plotted at two representative times, $t/t_0=0.50$ ($t=131\,\rm{ns}$; grey) and $t/t_0=1.00$ ($t=261\,\rm{ns}$; black), and the dashed lines show power-law fits of the form $P\propto \rho^{\gamma_{\rm eff}}$. The total, electron, and ion pressures all lie on narrow ridges with $\gamma_{\rm eff}\simeq 1.1$, indicating that the mixing layer is thermodynamically close to isothermal. The fitted slopes decrease slightly in time, from $\gamma_{\rm eff}\simeq 1.15$--$1.17$ at $t/t_0=0.50$ to $\gamma_{\rm eff}\simeq 1.07$--$1.08$ at $t/t_0=1.00$, showing that the layer relaxes further toward an isothermal state as it develops. By contrast, the radiation pressure is nearly independent of density, with $\gamma_{\rm eff}\ll 1$, demonstrating that the local thermodynamic closure inside the mixing layer is controlled primarily by the coupled matter pressure rather than by $\Pr$. This emergent close-to-isothermal response is the laboratory analogue of the rapidly cooling thermodynamic closures commonly used for cold ISM turbulence \citep{Wolfire1995_isothermal_ISM,Ferriere2020_reynolds_numbers_for_ism,beattie_taking_2025}.}
        \label{fig:PressureDensityCorrelation}
    \end{figure}
    In the 3T model introduced in \autoref{sec:methods}, the equation of state partitions the internal energies of the radiation, electron, and ion components into the corresponding partial pressures, $\Pr$, $\Pe$, and $\Pi$, whose sum defines the total pressure, $\Ptot$ (see \autoref{eq:ConservationMomentumE}). The turbulent mixing layer therefore does not a priori obey a single ideal-gas closure. Instead, its bulk thermodynamic response can be characterised empirically from pressure--density correlations measured directly in the interaction region.

    This effective closure is astrophysically important because the cold, radiatively regulated phases of the ISM are controlled by thermal balance between heating and cooling \citep{Field1969,Wolfire1995_isothermal_ISM}. In this regime, cooling and thermal equilibration can be fast compared with the dynamical evolution, so shocks generate strong density structure without producing a purely adiabatic post-shock response \citep{Draine1993_ISM_shocks,Ferriere2020_reynolds_numbers_for_ism, MerliniRadiativeCoolingEffects_2023, beattie_taking_2025}. Nearly isothermal shock-bounded layers are therefore a standard idealisation for colliding supersonic flows in astrophysical gas \citep{Folini_2006}. Here the near-isothermal behaviour is not imposed as an equation of state; it emerges from the 3T radiation-hydrodynamic evolution of the shocked layer.

    \autoref{fig:PressureDensityCorrelation} shows joint PDFs of $\ln(\rho/\langle\rho\rangle)$ against $\ln(P/\langle P\rangle)$ for the total pressure and for the three partial pressures at two representative times, $t/t_0=0.50$ ($t=131\,\rm{ns}$) and $t/t_0=1.00$ ($t=261\,\rm{ns}$). If the flow is described locally by an effective polytropic relation, $P\propto \rho^{\gamma_{\rm eff}}$, then the slope of the ridge in each PDF defines an effective polytropic index $\gamma_{\rm eff}$. For the total pressure, the distribution remains tightly concentrated about a nearly linear relation in log-space, with $\gamma_{\rm eff}\simeq 1.16$ at the earlier time and $\gamma_{\rm eff}\simeq 1.07$ by one outer-scale eddy-turnover time.

    The same behaviour is seen in the matter pressures. The electron and ion components follow density with nearly identical slopes, evolving from $\gamma_{\rm eff}\simeq 1.17$ and $1.15$ at $t/t_0=0.50$ to $\gamma_{\rm eff}\simeq 1.07$ and $1.08$ at $t/t_0=1.00$, respectively. This close correspondence is consistent with the rapid electron--ion equilibration discussed in \autoref{sec:methods} and shows that the density-dependent part of the thermodynamic response is carried primarily by the matter pressure. By contrast, the radiation pressure is nearly independent of density, with $\gamma_{\rm eff}\simeq 0.02$ at the earlier time and $\gamma_{\rm eff}\simeq 0.07$ at the later time. This also helps explain why the interaction region discussed in \autoref{sec:time_evol} does not resemble a simple planar post-shock slab. Using the volume-averaged values in \autoref{tab:plasma_inflows_params}, the total pressure estimate $\Ptot \simeq \rho c_s^2$ gives $P_{\rm in}\simeq 4\times10^{5}\,\rm Pa$ for the inflows and $P_{\rm mix}\simeq 8.6\times10^{5}\,\rm Pa$ for the mixing layer, so the pressure contrast is only of order unity even though the post-collision plasma is substantially hotter. The interaction region is therefore better interpreted as a distributed, non-1D collision zone shaped by oblique shocks and turbulent mixing than as a single compressed post-shock state.

    Taken together, these correlations show that the interaction layer evolves toward an almost-isothermal effective closure, with $\gamma_{\rm eff}\rightarrow 1$, despite remaining shock-rich and compressible. In other words, radiation drives the system globally through the boundary forcing and the 3T energy exchange, but within the mixing layer itself the thermodynamic response of the plasma is controlled mainly by the coupled electron--ion pressure, which tracks density much more closely than an adiabatic gas would. This is also important dynamically: an effectively barotropic closure tends to align $\bnab \Ptot$ with $\bnab \rho$, thereby suppressing local baroclinic vorticity generation within the layer. That does not mean the non-adiabatic physics is irrelevant to the turbulence, however. Rather, in our platform the radiative drive and 3T energy exchange seed the turbulence earlier, during the ablation and hole-closure phase at the mesh, where the geometry and transient heating produce strong local misalignment between $\bnab \Ptot$ and $\bnab \rho$, as we show in the next section.
    
    \section{The generation of vorticity and the onset of turbulence} \label{sec:GenVorticity}

    \subsection{Baroclinic seeding at the mesh}

    The evolution of the vorticity field in the inviscid limit is governed by
    \begin{align}
        \label{eq:vorticity}
        \frac{\d{\bom}}{\d{t}}
        =
        \overbrace{-\bom(\bnab\cdot\u)}^{\text{compression}}
        +
        \underbrace{\bom\cdot\bnab\otimes\u}_{\text{stretching}}
        +
        \overbrace{\frac{1}{\rho^2} \bnab \rho \times \bnab \Ptot}^{\text{baroclinicity}}.
    \end{align}
    Here, $\d{}/\d{t} \equiv \partial/\partial t + \u\cdot\bnab$ denotes the material derivative. The vorticity is the natural quantity to follow because it directly traces the rotational, or solenoidal, part of the velocity field. In Fourier space, $\Tilde{\bom}(\k) = i \k \times \Tilde{\u}(\k)$, so only the solenoidal velocity component contributes and $|\Tilde{\bom}(\k)| \sim k |\Tilde{\u}_{s}(\k)|$ \citep[see Figure~11 in ][]{Beattie2025_small_scale_instabilities}. Probing vorticity generation is therefore equivalent to probing the generation of incompressible velocity modes, which are the modes that sustain the classical \citet{Kolmogorov1941}-type turbulent cascade.

    Astrophysically, this conversion from compressive shock driving into solenoidal motion is central to how blast waves and feedback maintain turbulent cascades in multiphase gas. Shocks propagating through nonuniform media generate post-shock velocity perturbations and turbulence through the Richtmyer--Meshkov response to density structure \citep{davidovits_turbulence_2022}, while mode-decomposition studies find that shock-driven turbulence is initially a strongly compressive form of driving \citep{Dhawalikar2022_shock_driving_parameter}. In supernova-remnant-motivated simulations of shocks propagating through inhomogeneous gas, however, the post-shock velocity field can become solenoidal-dominated and approach Kolmogorov-like scaling \citep{Hu2022_shock_amplification}. In simulations of a supernova-driven, multiphase ISM, initially compressive explosions can similarly feed both forward and backward turbulent cascades rather than producing a purely shock-compressive flow \citep{Beattie2025_so_long_kolmogorov}. A complementary analysis identifies small-scale baroclinic instabilities in structured post-shock gas as a mechanism by which supernovae generate large-scale incompressible motions \citep{Beattie2025_small_scale_instabilities}. The laboratory platform studied here isolates the same physical conversion in a controlled geometry.
    
    At the onset of ablation, the flow is initially irrotational, $|\bom|=0$. The compression and stretching terms in \autoref{eq:vorticity} therefore vanish identically at leading order, because both are proportional to $\bom$. Initial vorticity generation must then arise from the baroclinic term alone. In our setup, this occurs primarily during the ablation and hole-closure phase of the mesh targets, during the $t_{\rm rad}=30\,\rm ns$ radiative drive, rather than at the later interaction of the two counter-streaming flows (as has been previously assumed).

    During the X-ray drive, $\bnab\Ptot$ launches $\M > 1$ plasma streams from adjacent cell walls into each aperture. These streams collide within the mesh plane ($yz$ plane) and form oblique shocks. Because the aperture geometry produces strong local misalignment between $\bnab \Ptot$ and $\bnab \rho$, the baroclinic term generates out-of-plane vorticity, predominantly in the $\omega_x$ component. During this early hole-closure phase, the dominant $\bnab$ lie in the $yz$ plane $\bnab \approx \bnab_{\perp}$, so the local source of $\omega_x$ reduces to
    \begin{align}
       \frac{\d{\omega_x}}{\d{t}}
        &\approx \frac{1}{\rho^2}
        \left(
        \frac{\partial \rho}{\partial y}\frac{\partial \Ptot}{\partial z}
        -\frac{\partial \rho}{\partial z}\frac{\partial \Ptot}{\partial y}
        \right) = S_{{\rm baro},x}.
        \label{eq:omega_x_seed}
    \end{align}
    The first approximation keeps only the leading-order baroclinic source, $S_{{\rm baro},x}$, permitted by the initially irrotational ($|\bnab \times \u| = 0$) state; the second uses the local mesh geometry to neglect streamwise gradients during hole closure. Along a straight edge, one of the required cross-gradients is absent, so the strongest source is concentrated near the corners and at the intersections of the oblique shocks, where both $\bnab_{\perp} \Ptot$ and $\bnab_\perp \rho$ are non-zero.

    \begin{figure*}
        \centering
        \includegraphics[width=1\linewidth]{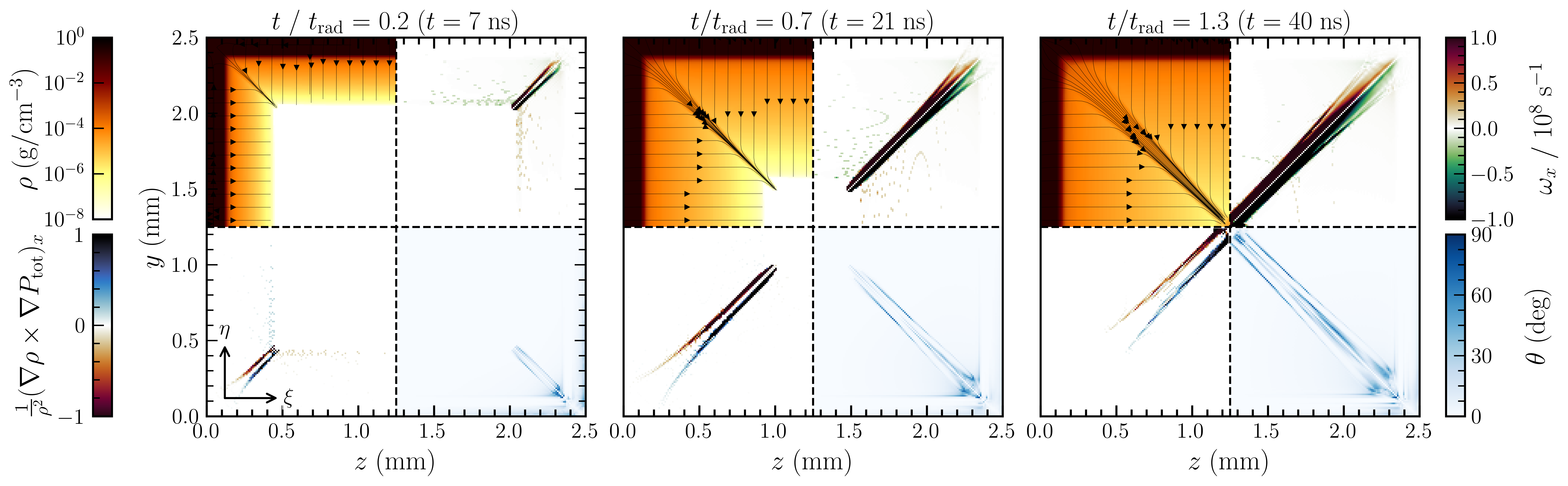}        
        \includegraphics[width=1\linewidth]{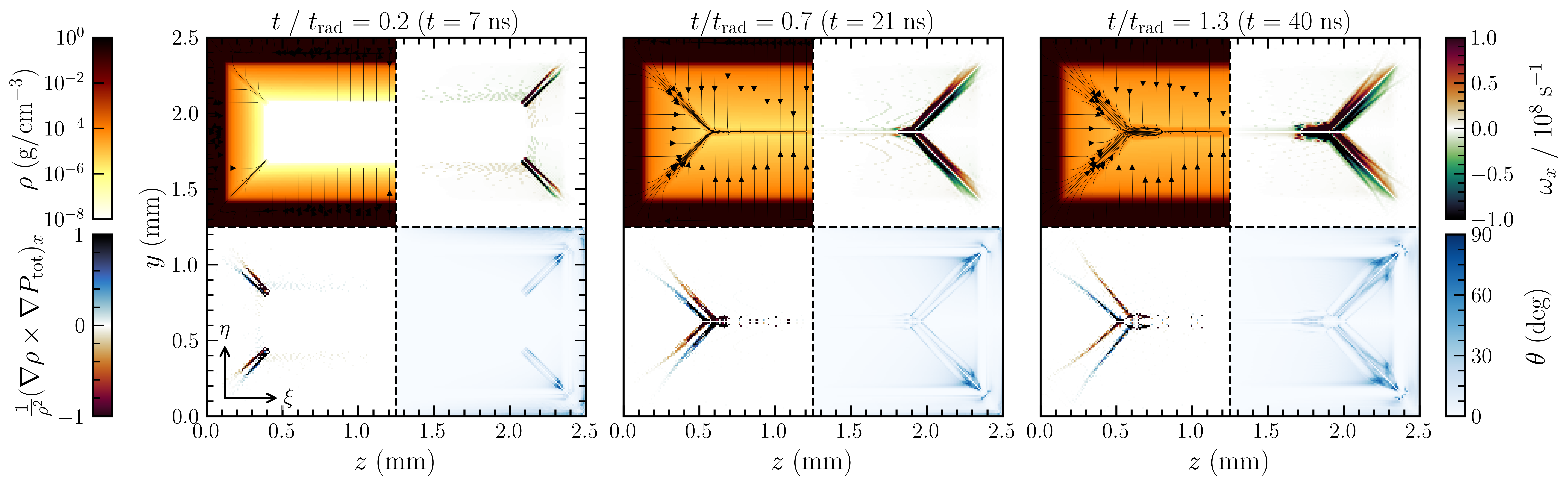}
        \caption{\textbf{Top row:} Time evolution ($t=7$-$40\,\rm{ns}$) of the plasma within a single cell of the front mesh, corresponding to the ablation of a $2\,\rm{mm}\times 2\,\rm{mm}$ square aperture. \textbf{Bottom row:} The same quantities for a single rear-mesh cell with a $2\,\rm{mm}\times 1\,\rm{mm}$ end-slot feature. In each sequence, the columns show the mass density, $\rho$, vorticity magnitude, $|\bom|$, baroclinic source term, $\baro$, and the angle $\theta$ between $\bnab \Ptot$ and $\bnab \rho$. Time is normalised by the duration of the X-ray drive pulse, $t_{\rm rad}=30\,\rm{ns}$. Baroclinicity is generated near the corners and along the internal oblique shocks, where $\bnab \Ptot$ and $\bnab \rho$ are misaligned (see \autoref{eq:vorticity} and \autoref{eq:omega_x_seed}). The resulting $\omega_x$ is then advected toward the centre of the cell by the converging in-plane flow, compressed into narrow channels, and embedded in the outgoing ablation stream. The symmetry properties of the corner source are discussed in \autoref{eq:baro_corner} and \autoref{eq:baro_antisym}. The different front- and rear-cell geometries produce distinct patterns of collimated vorticity injection.}
        \label{fig:FrontRearMeshVorticityGen}
    \end{figure*}

    This corner structure can be understood explicitly by introducing local coordinates $(\xi,\eta)$ within the cell, with $(\xi,\eta)=(0,0)$ at the lower-left corner and $\xi,\eta>0$ increasing away from the two walls. In these coordinates $S_{{\rm baro},x}$ is
    \begin{align}
        S_{{\rm baro},x}
        =
        \frac{1}{\rho^2}
        \left(
        \frac{\partial \rho}{\partial \xi}\frac{\partial \Ptot}{\partial \eta}
        -
        \frac{\partial \rho}{\partial \eta}\frac{\partial \Ptot}{\partial \xi}
        \right).
        \label{eq:baro_corner}
    \end{align}
    For a locally symmetric right-angle corner, the leading-order pressure and density fields are approximately invariant under exchange of the two wall-normal directions, $\xi \leftrightarrow \eta$. We may therefore write the Taylor expansions at the corners of the domain
    \begin{align}
        \Ptot(\xi,\eta)
        &= P_0 + a_1(\xi+\eta) + a_2(\xi^2+\eta^2) + a_3 \xi\eta + \hdots, \\
        \rho(\xi,\eta)
        &= \rho_0 + b_1(\xi+\eta) + b_2(\xi^2+\eta^2) + b_3 \xi\eta + \hdots,
    \end{align}
    where the coefficients $a_i$ and $b_i$ depend on time. Substituting these expansions into \autoref{eq:baro_corner} gives
    \begin{align}
        S_{{\rm baro},x}
        \propto (\xi-\eta) + \mathcal{O}\!\left[(\xi-\eta)(\xi+\eta)\right].
        \label{eq:baro_antisym}
    \end{align}
    Thus, the baroclinic source vanishes on the corner bisector, $\xi=\eta$, including the corner apex itself, and the leading non-zero contribution is antisymmetric $S_{{\rm baro},x}(\xi,\eta) = -S_{{\rm baro},x}(\eta,\xi)$ across the bisector. This explains the two adjacent strips of opposite sign seen in \autoref{fig:FrontRearMeshVorticityGen}: the corner itself is a symmetry point, while either side of the diagonal corresponds to opposite handedness in the local misalignment between $\bnab \Ptot$ and $\bnab \rho$. A more detailed derivation of these symmetry properties is provided in \autoref{eq:baro_corner_appendix} and \autoref{eq:baro_corner_factorised}.

    Once generated in these off-bisector ribbons, the resulting $\omega_x$ is transported by the converging in-plane flow toward the centre of the cell through the advective part of the material derivative, $\u\cdot\bnab\omega_x$. The figure therefore indicates that the baroclinicity first creates narrow bands of vorticity on either side of the bisector, and the subsequent in-plane velocity, $u_\perp$, sweeps this vorticity inward, collimating it into narrow channels. From the temporal sequence in \autoref{fig:FrontRearMeshVorticityGen}, this collimation occurs within the drive phase itself, over roughly $\Delta t_{\rm coll}\sim (0.7-0.2)t_{\rm rad}\sim 15\,\rm ns$. Taking the relevant transverse transport distance to be of order half of the shorter local aperture dimension, $\Delta \ell_\perp \sim \ell_{\rm min}/2$, gives $\Delta \ell_\perp \sim 1\,\rm mm$ for the $2\times2\,\rm mm$ front cell and $\Delta \ell_\perp \sim 0.5\,\rm mm$ for the $2\times1\,\rm mm$ rear cell. The implied in-plane advection speeds are therefore $u_\perp \sim \Delta \ell_\perp/\Delta t_{\rm coll} \sim 30-70\,\rm km\,s^{-1},$ comparable to the characteristic bulk outflow speed. 
    
    These channels are then embedded in the outgoing ablation stream and provide a natural route for injecting vorticity into the later mixing layer, seeding the turbulence from the off-diagonals of the mesh. The geometry of these channels is set by the cell itself, e.g., the location of the corners, the aperture aspect ratio, and the detailed shape of the converging compression layer together determine how the off-bisector vorticity is focused before ejection. The final collimation is therefore not universal, but depends directly on the mesh-cell geometry, which helps explain the distinct vorticity structures produced by the front and rear cells in \autoref{fig:FrontRearMeshVorticityGen} (and in our later $Q$-criterion analysis).

    \subsection{Amplification and transition in the mixing layer}

    Once the seeded vortical inflows collide, the flow is no longer irrotational and the kinematic terms, e.g., $-\bom\,(\bnab \cdot \u)$ and $\bom \cdot \bnab\otimes\u$ in \autoref{eq:vorticity} need not vanish. The problem therefore shifts from vorticity generation to vorticity amplification. The relevant question is which terms dominate the subsequent evolution of the inherited rotational content within the shocked interaction layer. In our platform, the collision acts primarily as an amplifier and re-organiser of pre-existing vorticity, rather than as the site of its initial creation. The resulting mixing layer remains mildly supersonic in the turbulent sense, with $\M \simeq 1.6$ in the interaction control volume (see \autoref{tab:plasma_inflows_params}).

    This behaviour differs from the ``self-generation'' picture of an unstable double-shock reverse system, in which vorticity is created primarily after the interaction of the counter-streaming flows \citep{Markwick_2021}. It also refines the interpretation of the TDYNO laser-driven turbulent-dynamo platform, where offset grids are used to corrugate the ablation fronts and the subsequent collision of interleaving fronts is argued to generate shear, Kelvin--Helmholtz instabilities, and turbulent mixing \citep{tzeferacos_numerical_2017,tzeferacos_laboratory_2018}. Our results suggest that this collision-stage interpretation is incomplete: in any gridded flow with strong pressure and density gradients around aperture corners (as is the case for both radiation-driven and flow-past-grid experiments), the same baroclinic mechanism should seed rotational structure before the counter-streaming flows collide. The later Kelvin--Helmholtz-like shear and turbulent mixing would then amplify and reorganise an already vortical upstream flow rather than generate turbulence from an initially laminar one. In this respect, the dynamics in our simulations are more closely analogous to shock-bounded interaction zones driven by non-uniform supersonic upstream structure, where downstream turbulence develops through the combined action of inherited vorticity, shear, compression, and shock curvature \citep{Folini_2006}. Indeed, baroclinic generation of vorticity is the key driver of supernova-driven turbulence \citep{Beattie2025_so_long_kolmogorov,Beattie2025_small_scale_instabilities}. To quantify the relative importance of these three contributions within the control volume (\autoref{fig:Setup}), we define time-dependent ratios that measure the relative weight of the terms in \autoref{eq:vorticity},

    \begin{equation}
        r_i(t) =
        \frac{\Exp{\|S_i(\x,t)\|}{\V}}
        {\sum_j \Exp{\|S_j(\x,t)\|}{\V}},
    \end{equation}
    where $j \in \{{\rm comp}, {\rm stretch}, {\rm baro}\}$, $\|\cdot\|$ denotes the Euclidean norm, and
    \begin{align}
        S_{\rm comp}   &= -\bom\,(\bnab \cdot \u), \\
        S_{\rm stretch} &= \bom \cdot \bnab\otimes\u, \\
        S_{\rm baro}    &= \frac{1}{\rho^2}\,\bnab \rho \times \bnab \Ptot.
    \end{align}
    
    \begin{figure}
        \centering
        \includegraphics[width=1\linewidth]{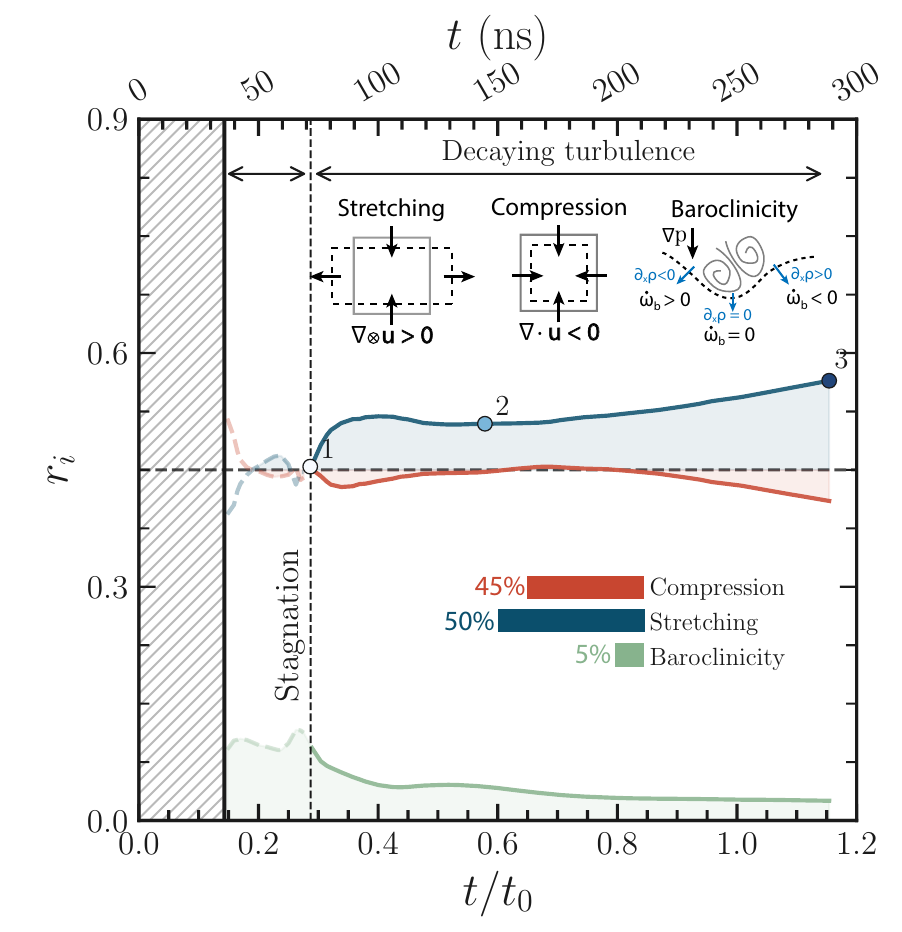}
        \caption{Time evolution of the volume-averaged vorticity-budget ratios, $r_{\rm stretch}$, $r_{\rm comp}$, and $r_{\rm baro}$, associated with the three terms in \autoref{eq:vorticity}, computed at each timestep within the interaction control volume shown in \autoref{fig:Setup}. The lower axis is normalised by the outer-scale eddy-turnover time, $t_0=261\,\rm{ns}$, and the upper axis gives the physical time in ns. The hatched region denotes the radiative drive phase ($t<t_{\rm rad}$), while the vertical dashed line marks the onset of stagnation and the formation of the turbulent mixing layer at $t/t_0 \simeq 0.3$ ($t \simeq 75\,\rm{ns}$). After stagnation, the vorticity budget is dominated by stretching and compression, which remain close to $50\%$ and $45\%$, respectively, whereas baroclinicity decays to a minor contribution of order $5\%$. The schematic insets illustrate the local flow deformations associated with the three terms: vortex stretching $\bom \cdot \bnab\otimes\u$, compressive amplification $-\bom\,(\bnab \cdot \u)$, and baroclinic generation, $\baro$.}
        \label{fig:LayerVorticityTimeResolved}
    \end{figure}
    
    The resulting ratios are shown in \autoref{fig:LayerVorticityTimeResolved}. Once the mixing layer forms, the vorticity budget is dominated by the two kinematic amplification terms, with $r_{\rm stretch}\simeq 0.5$ and $r_{\rm comp}\simeq 0.45$. Both are directly tied to the strong velocity gradients and compressible strain generated by the shock-processed converging flows. In contrast, baroclinicity remains subdominant after the collision, with $r_{\rm baro}\simeq 0.05$, although it reaches up to $\sim 0.1$ during the free-streaming inflow phase. \autoref{fig:LayerVorticityTimeResolved} therefore supports the physical picture established in \autoref{fig:FrontRearMeshVorticityGen}: baroclinicity seeds the inflows during launch, whereas compression and vortex stretching dominate the subsequent growth of vorticity in the shocked interaction zone.

    \begin{figure*}
        \centering
        \includegraphics[width=1\linewidth]{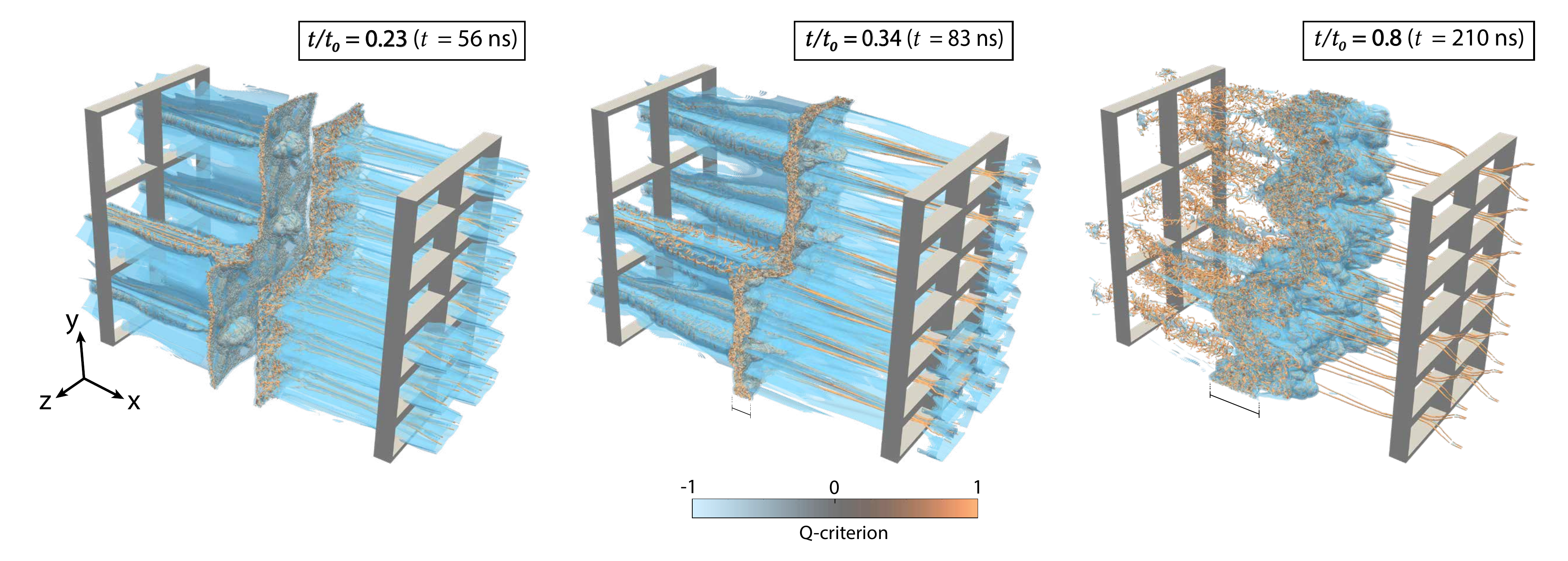}
        \caption{Three-dimensional visualisation of the flow using the Q criterion at three representative stages of the interaction. $Q > 0$ iso-contours identify vortex-dominated regions, whereas $Q < 0$ iso-contours identify strain-dominated regions associated with strong shear and compression. Before the main inflows collide, coherent vortical structures are already embedded in the outgoing streams as a consequence of the geometry-dependent baroclinic seeding shown in \autoref{fig:FrontRearMeshVorticityGen}. As the inflows stagnate and the mixing layer forms, these inherited vortical channels are amplified and wrapped by shock-bounded, strain-dominated structures. By approximately one outer-scale eddy-turnover time, $t_0 = 261\,\rm{ns}$, the layer is fully developed and contains a strongly intermixed network of vortical, shear, and compressive structures.}
        \label{fig:QCriteriaTimeSequence}
    \end{figure*}

    The $\omega$ budget identifies the dominant amplification mechanisms, but it does not show how the amplified vorticity is organised in space. To characterise the resulting flow topology within the mixing layer, we therefore employ the Q criterion, which is the second invariant of the velocity-gradient tensor $\partial u_i/\partial x_j$. Writing
    \begin{align}
        S^*_{ij} &= \frac{1}{2}\left(\partial_i u_j + \partial_j u_i\right), \\
        \Omega_{ij} &= \frac{1}{2}\left(\partial_i u_j - \partial_j u_i\right),
    \end{align}
    where $S^*_{ij}$ is the symmetric part (with trace) and $\Omega_{ij}$ the antisymmetric part of $\partial u_i/\partial x_j$, we define
    \begin{equation}
        Q = \frac{1}{2}\left(\Omega_{ij}\Omega_{ij} - S^*_{ij}S^*_{ij}\right).
    \end{equation}
    $Q > 0$ denotes regions where rotational motion dominates over strain, whereas $Q < 0$ highlights zones of strong shear or compression (since we leave the trace in $S^*_{ij}$), often associated with shocks. These strain-dominated regions correspond to locations of enhanced density and negative velocity divergence, and therefore naturally host large density gradients. As discussed later in \autoref{sec:DensityGrad}, such gradients play a central role in light dispersion and photon scattering within turbulent plasmas.
    
    \autoref{fig:QCriteriaTimeSequence} presents 3D iso-contours of $Q$ at three stages of the flow evolution, revealing the spatial interplay between vortices and shocklets in the turbulent layer. Before the main inflows collide, at $t=56\,\rm{ns}$, vortices are already embedded within both outflows as a consequence of the mesh-hole geometry. In the early interaction phase, once the inflows meet, these inherited vortical channels are amplified within the nascent mixing layer while shocks and strong shear layers wrap around them. The front mesh produces smaller-scale eddies, whereas the rear rectangular mesh generates more coherent and more tightly collimated elongated vortex streams. This difference is consistent with the geometry-dependent collimation of the in-cell vorticity channels established in \autoref{fig:FrontRearMeshVorticityGen}, where the shorter transverse dimension of the $2\times1\,\rm mm$ rear aperture imposes a more anisotropic focusing scale than the $2\times2\,\rm mm$ front cell. By approximately one outer-scale eddy-turnover time, $t_0 = 261\,\rm{ns}$ (see \autoref{tab:plasma_inflows_params}), nearly an order of magnitude longer than the $t_{\rm rad}=30\,\rm{ns}$ drive, the mixing layer is fully developed and contains a strongly intermixed network of both vortical and shock-dominated structures.

    \subsection{Summary of seeding chronology}
    The chronology of the vorticity seeding, transport, and amplification described in this section can be summarised as follows:
    \begin{enumerate}
      \item \textbf{Seeding at the mesh plane.} During the $t_{\rm rad}=30\,\rm ns$ radiative drive, oblique shocks formed during hole closure produce strong local misalignment between $\bnab \Ptot$ and $\bnab \rho$, so baroclinicity seeds $\omega_x$ near the hole boundaries and especially on the off-diagonals near the cell corners ($t/t_{\rm rad}\sim 0.2$), as shown in \autoref{fig:FrontRearMeshVorticityGen}.
    
      \item \textbf{Collimation within the cell.} The converging in-plane flow then advects the off-bisector vorticity toward the cell centre, where it is compressed into narrow geometry-dependent channels ($t/t_{\rm rad}\sim 0.7$), again visible in \autoref{fig:FrontRearMeshVorticityGen}. This implies a collimation time of order $\Delta t_{\rm coll}\sim (0.7-0.2)t_{\rm rad}\sim 15\,\rm ns$. Taking the relevant transverse transport distance to be of order half of the shorter local aperture dimension, $\Delta \ell_\perp \sim \ell_{\rm min}/2$, gives $\Delta \ell_\perp\sim 1\,\rm mm$ for the $2\times2\,\rm mm$ front cell and $\Delta \ell_\perp\sim 0.5\,\rm mm$ for the $2\times1\,\rm mm$ rear cell, implying $u_\perp\sim \Delta \ell_\perp/\Delta t_{\rm coll}\sim 30$--$70\,\rm km\,s^{-1}$. The differing aperture geometries of the front and rear meshes therefore produce distinct collimated vortical structures.
    
      \item \textbf{Injection into the inflows.} These collimated vortical channels are embedded in the outgoing ablation streams, so the counter-streaming inflows acquire non-zero vorticity before reaching the interaction region. The inflows propagate at a bulk speed of order $40\,\rm km\,s^{-1}$ through plasma with characteristic conditions $\rho \sim 5\times10^{-6}\,\rm g\,cm^{-3}$, $\Ti\simeq\Te\simeq 4\,\rm eV$, and $c_s\sim 9\,\rm km\,s^{-1}$, implying a bulk Mach number $\M\sim 4.5$ (see \autoref{tab:plasma_inflows_params}). The subsequent transition from baroclinic generation to kinematic amplification is quantified in \autoref{fig:LayerVorticityTimeResolved}.
    
      \item \textbf{Amplification in the mixing layer.} After the inflows collide at $t\simeq 75\,\rm ns$, compression and vortex stretching amplify and reorganise the inherited vorticity into the shock-rich turbulent structures seen in the Q-criterion analysis (\autoref{fig:LayerVorticityTimeResolved}; \autoref{fig:QCriteriaTimeSequence}). In this stage the interaction layer is heated to approximately $(\Ti,\Te)\simeq(24,10)\,\rm eV$, remains mildly supersonic in the turbulent sense with $\M\simeq 1.6$, and has a vorticity budget dominated by stretching and compression, with $r_{\rm stretch}\simeq 0.5$, $r_{\rm comp}\simeq 0.45$, and $r_{\rm baro}\simeq 0.05$.
    \end{enumerate}

    The analyses in this section establish the origin, transport, and amplification of the vorticity, as well as the resulting shock-vortex topology of the mixing layer. They do not, however, quantify how kinetic energy is partitioned across spatial scales or between compressive and solenoidal motions. We therefore turn next to a spectral and modal decomposition of the flow.

    \section{Fourier spectra and Helmholtz decomposition of the velocity}\label{sec:PowerSpectrum}

    We now quantify the scale-dependent velocity structure of the mixing layer. The previous sections showed that the collision produces a shock-rich turbulent region whose vorticity is seeded at the mesh and subsequently amplified by compression and stretching. The remaining questions are how the kinetic energy is distributed across spatial scales, how it partitions between compressive and solenoidal motions, and over what range of scales these statistics are resolved before numerical dissipation becomes important. We address these questions by analysing the velocity field in Fourier space within the interaction control volume, using Helmholtz decomposition to separate the compressive and solenoidal components and a strain-spectrum diagnostic to estimate the resolved dissipation scale. Because the present flow remains shock-rich and compressible, however, the total velocity spectrum alone is not sufficient: we must also separate the divergence-free and curl-free parts of the motion.

    These astrophysical connections motivate applying the same modal diagnostic to the laboratory layer. Measuring the compressive--solenoidal partition here is therefore not only a numerical characterization of the platform; it determines which part of the flow is analogous to the vortical turbulent cascade and which part remains directly tied to shocks, convergence, and acoustic response.
  
    \begin{figure*}
        \centering
        \includegraphics[width=1\linewidth]{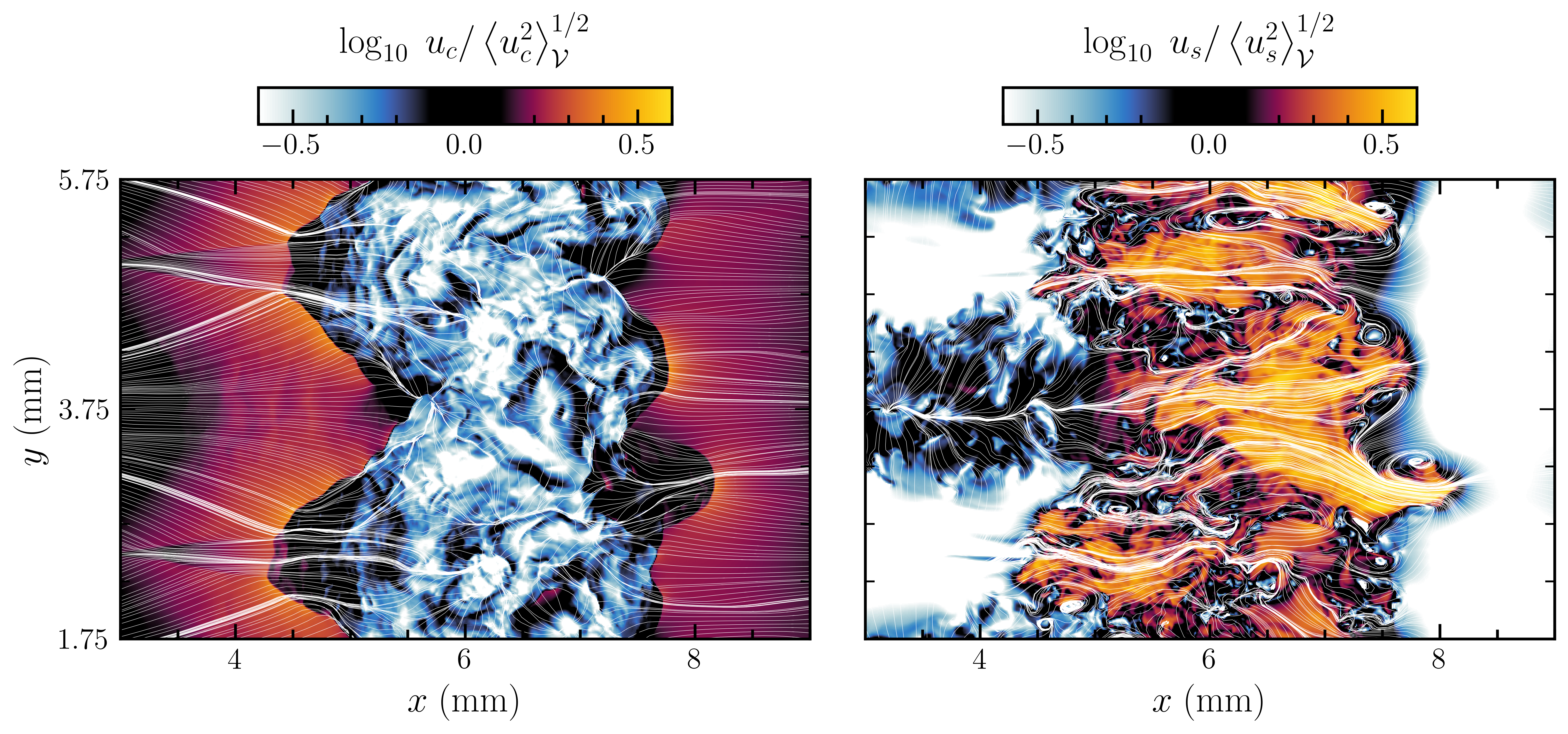}
        \caption{Helmholtz decomposition of the velocity field in an $xy$ slice through the mixing layer at $t = 261\,\rm{ns}$ $(t/t_0 = 1)$, with the compressive and solenoidal components, $\u_c$ and $\u_s$, defined in \autoref{eq:H_decomp}. \textbf{Left:} logarithm of $|\u_c|$ normalised by its root-mean-square amplitude, $\Exp{u_c^2}{\V}^{1/2}$. \textbf{Right:} logarithm of $|\u_s|$ normalised by its root-mean-square amplitude, $\Exp{u_s^2}{\V}^{1/2}$. White streamlines trace the in-plane morphology of each component. The strongest $\u_c$ is concentrated along the large-scale compressive structures at the shocked boundaries, whereas $\u_s$ dominates in the interior of the turbulent layer.}
        \label{fig:HelmholtzDec}
    \end{figure*}
    
        \subsection{Helmholtz Decomposition}
           We first decompose the velocity field into compressive and solenoidal parts, which provides the clearest measure of how shock-dominated and vortical motions coexist in the layer. We write
             \begin{align}\label{eq:H_decomp}
                \u = \u_c + \u_s, \qquad |\bnab \times \u_c| = 0, \qquad \bnab \cdot \u_s = 0.
            \end{align}
            In Fourier space, the longitudinal projection is
            \begin{align}
                \Tilde{\u}_c(\k) = \frac{\k \cdot \Tilde{\u}(\k)}{k^2}\,\k,
            \end{align}
            and the transverse remainder is
            \begin{align}\label{eq:Fourier}
                \Tilde{\u}_s(\k) = \Tilde{\u}(\k) - \Tilde{\u}_c(\k).
            \end{align}
            Here $k=|\k|$. The two components are orthogonal under the volume inner product, so their energies may be analysed separately. \autoref{fig:HelmholtzDec} shows both components at $t/t_0 = 1$, with $|\u_c|$ and $|\u_s|$ normalised by their respective root-mean-squared values. The strongest $\u_c$ is concentrated along the large-scale shocked boundaries and compressive interaction fronts, whereas $\u_s$ fills the interior of the layer. This is consistent with the physical picture developed in \autoref{sec:GenVorticity}: the collision directly injects compressive motions through shocks and convergence, while baroclinic seeding, shear, and nonlinear mode coupling populate the rotational component. The coherent structures in $\u_s$ also trace the vortices identified by the Q criterion.

\begin{figure}
                \centering
                \includegraphics[width=1\linewidth]{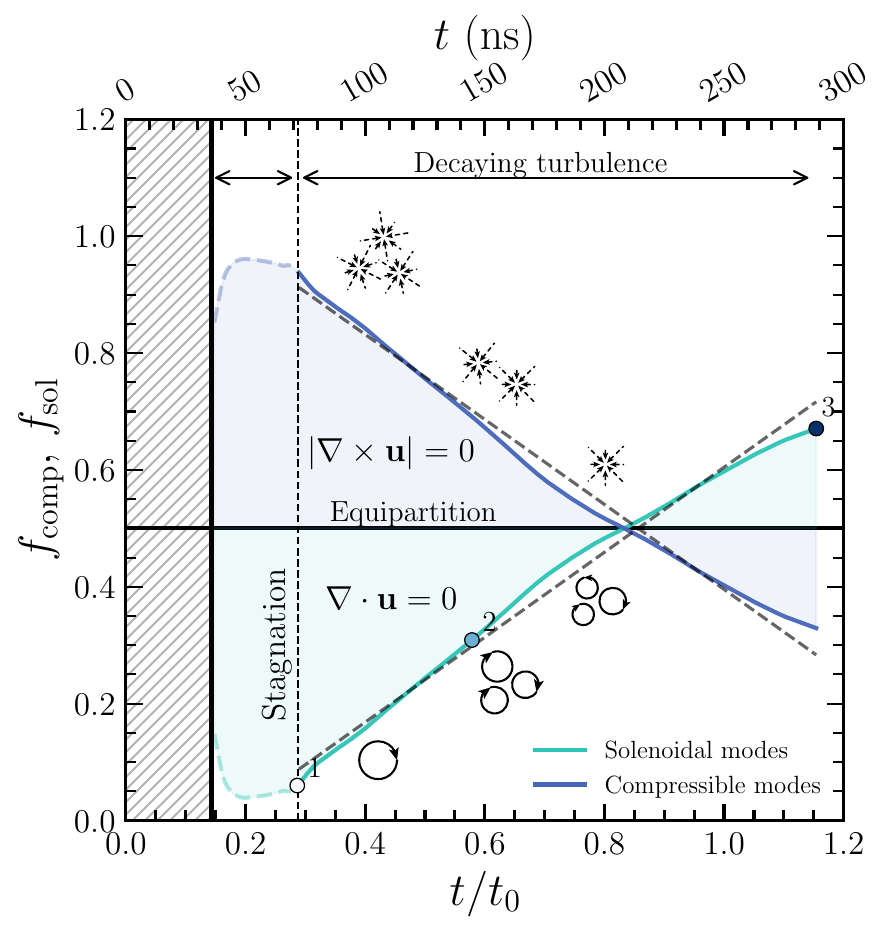}
                \caption{Time evolution of the volume-integrated Helmholtz kinetic-energy fractions, $f_{\rm sol}=\Exp{u_s^2}{\V}/\Exp{u^2}{\V}$ and $f_{\rm comp}=\Exp{u_c^2}{\V}/\Exp{u^2}{\V}$, within the control volume. The increase of $f_{\rm sol}$ and corresponding decrease of $f_{\rm comp}$ show that the layer becomes progressively more vortical as the interaction develops, although a non-negligible compressive component persists throughout the sampled interval.}
                \label{fig:LayerVelModesTimeResolved}
            \end{figure}

            To track the global mode content of the layer, we define the volume-integrated kinetic-energy fractions
            \begin{align}
                f_{\rm comp} &\equiv \frac{\Exp{u_c^2}{\V}}{\Exp{u^2}{\V}}, \qquad
                f_{\rm sol}  \equiv \frac{\Exp{u_s^2}{\V}}{\Exp{u^2}{\V}},
            \end{align}
            so that $f_{\rm comp}+f_{\rm sol}=1$ by the orthogonality of the Helmholtz decomposition. The resulting time evolution is shown in \autoref{fig:LayerVelModesTimeResolved}. Before stagnation, $f_{\rm sol}$ already rises as the mesh-modulated inflows enter the control volume, showing that the collision does not begin from a purely compressive state. After the layer forms, the evolution becomes more gradual, with solenoidal motions continuing to gain relative weight. By $t/t_0 = 1$, the volume-integrated kinetic energy is approximately $30\%$ compressive and $70\%$ solenoidal.

            This drift toward solenoidal dominance is consistent with the broader behaviour seen in numerical studies of compressible turbulence \citep{Federrath2010b,Kritsuk2011,beattie_taking_2025,Beattie2025_nature_astro,Connor2026_sn_turb}. A useful qualitative guide is that, for each non-zero Fourier mode in three dimensions, the divergence-free velocity field occupies two transverse polarisations whereas the compressive field occupies one longitudinal polarisation. In energy terms, a fully isotropised state would therefore approach $\Exp{u_s^2}{\V}:\Exp{u_c^2}{\V}\sim 2:1$. We invoke this only as qualitative context, not as a prediction for the measured fractions in our non-stationary, shock-imprinted flow. The mode fractions alone do not determine the scale-by-scale energy distribution, so we next turn to the velocity spectra.

            \subsection{Velocity power spectrum}
\begin{figure}
                \centering
                    \includegraphics[width=0.95\linewidth]{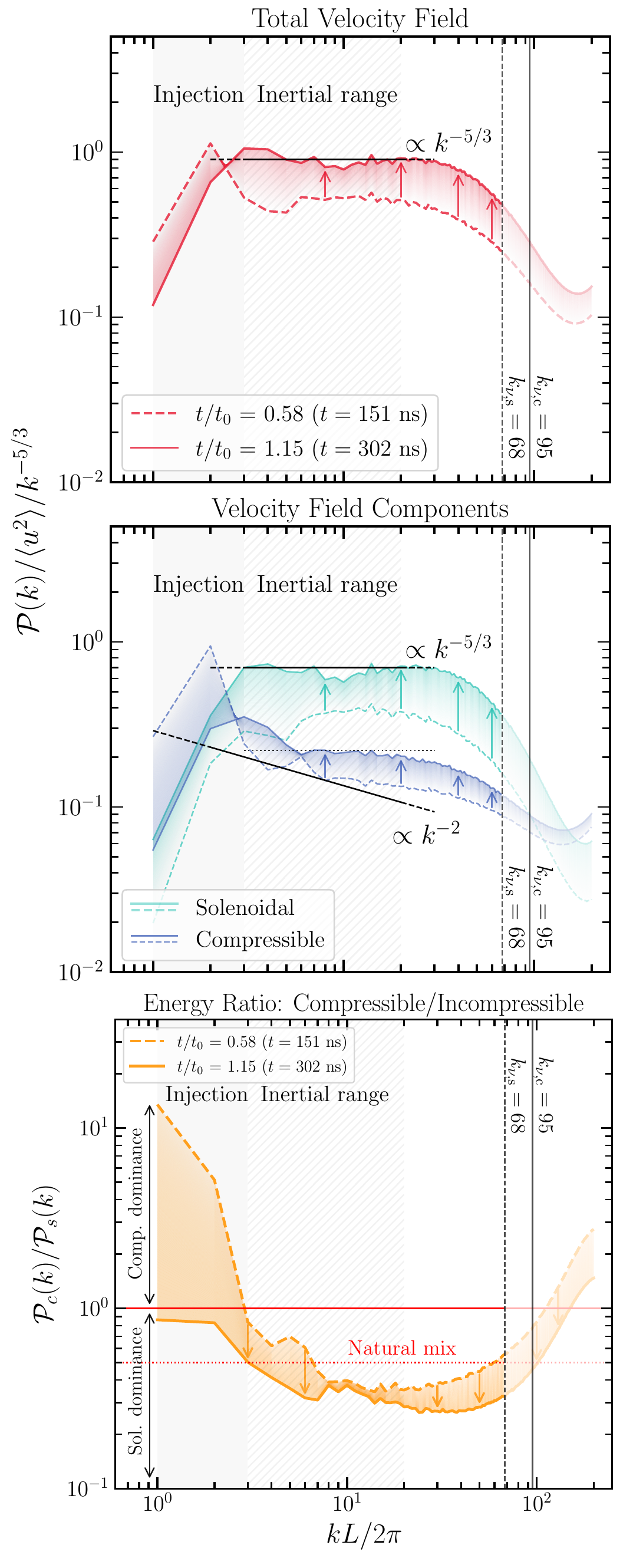}
                \caption{Compensated velocity spectra in the interaction control volume at $t/t_0=0.58$ ($t=151\,\rm ns$; dashed) and $t/t_0=1.15$ ($t=302\,\rm ns$; solid), after subtracting the volume-averaged mean flow. (a) Total spectrum, $\mathcal{P}_u(k)/\langle u^2\rangle\,k^{-5/3}$. (b) Solenoidal and compressive spectra, $\mathcal{P}_s(k)$ and $\mathcal{P}_c(k)$. (c) Scale-dependent spectral ratio, $\mathcal{P}_c(k)/\mathcal{P}_s(k)$. At late time, the solenoidal component is energetically dominant over most resolved scales, while the rise at the largest $k$ marks numerical dissipation.}
                \label{fig:PowerSpectrumVelocity}
            \end{figure}

            We now ask how the kinetic energy is distributed over $k$. Using the Helmholtz-decomposed fields, we compute the one-dimensional spectra of the total, solenoidal, and compressive velocities, $\mathcal{P}_u(k)$, $\mathcal{P}_s(k)$, and $\mathcal{P}_c(k)$, within the interaction control volume after subtracting the mean flow, $\u \rightarrow \u - \Exp{\u}{\V}$. The resulting spectra are shown in \autoref{fig:PowerSpectrumVelocity} at two representative times, $t/t_0=0.58$ and $1.15$. They are plotted in compensated form, so a flat segment corresponds to $\mathcal{P}(k)\propto k^{-5/3}$.

            Panel (a) shows that, at early time, $\mathcal{P}_u(k)$ peaks in the injection band, $2\pi/k \sim 1$-$3\,\rm mm$. By $t/t_0=1.15$, the compensated spectrum is flatter and extends to larger $k$, consistent with $\mathcal{P}_u(k)\propto k^{-5/3}$ over the shaded inertial interval. At sufficiently large $k$, both curves steepen and roll over as numerical dissipation becomes important; we quantify this scale in \autoref{sec:estimating_Reynolds}.

            Panel (b) sharpens the modal separation. The solenoidal spectrum, $\mathcal{P}_s(k)$, closely follows $\mathcal{P}_u(k)$ through the inertial interval and is roughly consistent with $k^{-5/3}$, whereas the compressive spectrum, $\mathcal{P}_c(k)$, is steeper, closer to $k^{-2}$. Thus, although shocks inject $\u_c$ on large scales, $\u_s$ is energetically dominant throughout the resolved inertial-type range in our simulations.
            
            Panel (c) makes this transition explicit. At $t/t_0=0.58$, $\mathcal{P}_c(k)/\mathcal{P}_s(k)>1$ only at the lowest $k$, then drops below unity near the end of the injection range and remains $\ll 1$ through the inertial interval. By $t/t_0=1.15$, the ratio is already $<1$ even at the largest scales shown and settles to $\sim 0.3$ through most of the cascade, i.e. clear solenoidal dominance. This is consistent with the picture developed in \autoref{sec:GenVorticity}, e.g., shocks and convergence inject $\u_c$ on large scales, while obliquity, shear, and mode coupling transfer part of that energy into $\u_s$ \citep{Federrath2010b,Kritsuk2011,beattie_taking_2025}. The upturn of $\mathcal{P}_c(k)/\mathcal{P}_s(k)$ near the largest resolved $k$ is likely a dissipation-range effect, where discretisation and effective viscosity bias the partitioning \citep{beattie_taking_2025}. This motivates the Reynolds-number estimate in the next section.

\begin{figure}
                \centering
                \includegraphics[width=\linewidth]{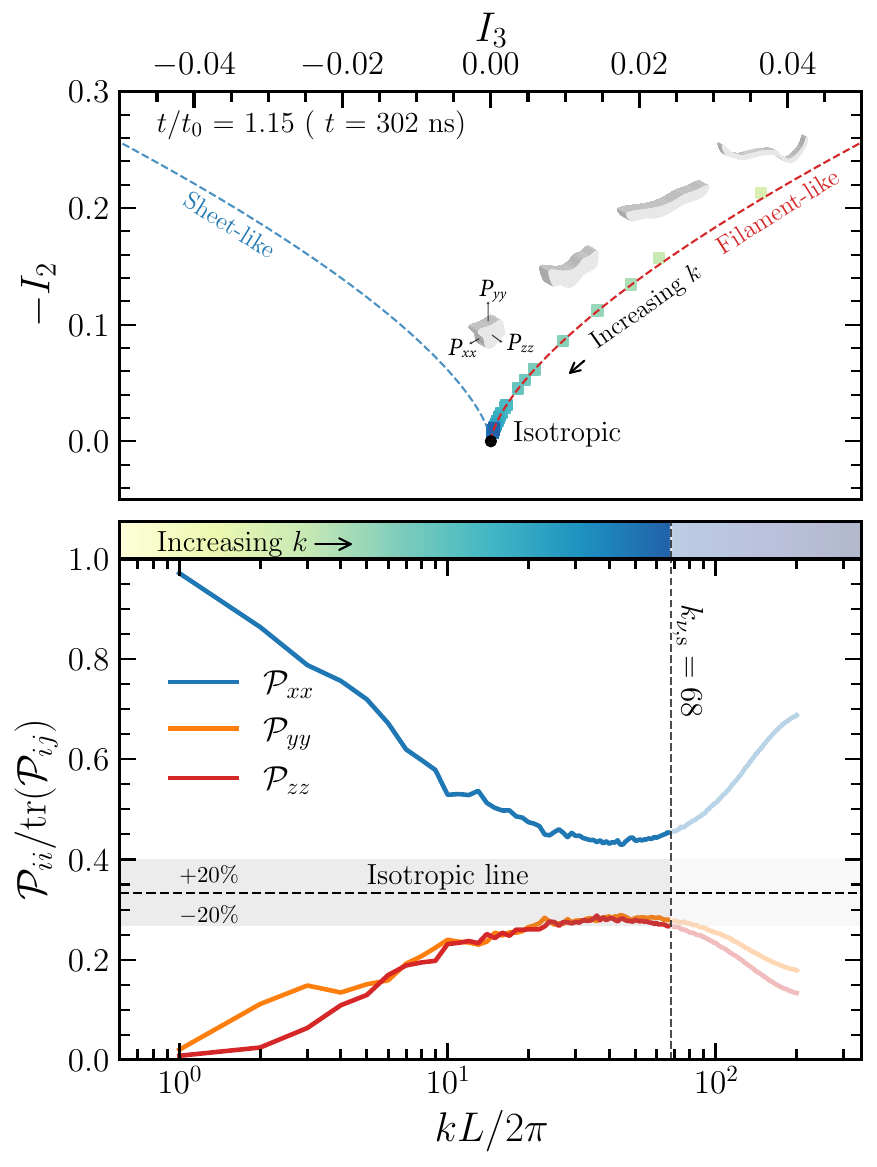}
                \caption{Scale-dependent Reynolds-stress anisotropy in the interaction control volume at $t/t_0=1.15$ ($t=302\,\rm ns$). Top panel: anisotropy-invariant map of the shell-wise Reynolds-stress tensor invariants, $I_2(k),\,I_3(k)$, defined in \autoref{eq:spectral_invariants_diag}. The coloured markers trace the tensor shape as shell wavenumber increases: the largest scales lie on the filament-like branch, while the stress moves back toward the isotropic point on higher-$k$ modes. Bottom panel: shell-integrated diagonal components of the spectral Reynolds-stress tensor, $\mathcal{P}_{ij}(k)$, defined in \autoref{eq:spectral_reynolds_stress}, shown here through $\mathcal{P}_{xx}(k)$, $\mathcal{P}_{yy}(k)$, and $\mathcal{P}_{zz}(k)$ and normalised by the shell trace ${\rm{tr}}(\mathcal{P}_{ij})$. The horizontal dashed line marks isotropy, $\mathcal{P}_{ii}/{\rm{tr}}(\mathcal{P}_{ij})=1/3$, and the shaded band marks $\pm20\%$ around that value. The vertical dashed line marks the solenoidal dissipation scale inferred from the strain-spectrum diagnostic in \autoref{eq:strain_spectrum}, $k_{\nu,\rm s}=68\,\rm mm^{-1}$. Most of the resolved inertial interval is substantially closer to isotropy than the outermost modes, while the highest-$k$ re-anisotropisation occurs only beyond the onset of numerical dissipation.}
                \label{fig:ScaleDependentAnisotropy}
            \end{figure}

            This shell-by-shell energy split does not by itself show whether the remaining motions are still directionally biased. To track that, we define the shell-wise spectral Reynolds-stress tensor using the same mean-subtracted velocity field, $\delta u_i=u_i-\Exp{u_i}{\V}$,
            \begin{align}\label{eq:spectral_reynolds_stress}
                \mathcal{P}_{ij}(k)
                \equiv
                \int_{\Omega_k}\widetilde{\delta u_i}(\k)\widetilde{\delta u_j}^{*}(\k)\,k^2\mathrm{d}\Omega_{\k},
            \end{align}
            whose diagonal components, $\mathcal{P}_{xx}(k)$, $\mathcal{P}_{yy}(k)$, and $\mathcal{P}_{zz}(k)$, give the component-wise power spectra. From these we form the shell-wise anisotropy tensor
            \begin{align}\label{eq:spectral_anisotropy_tensor}
                b_{ij}(k)
                =
                \frac{\mathcal{P}_{ij}(k)}{{\rm tr}[\mathcal{P}_{ij}(k)]}
                -\frac{1}{3}\delta_{ij},
                \qquad
                {\rm tr}[\mathcal{P}(k)]=\sum_i \mathcal{P}_{ii}(k),
            \end{align}
            and characterize its shape through the invariants ($I_2$, $I_3$). In the present analysis the off-diagonal shell contributions are neglected, so that $b_{ij}(k)$ is approximated as diagonal and
            \begin{align}\label{eq:spectral_invariants_diag}
                I_2(k)
                &= -\frac{1}{2}\left[b_{xx}^2(k)+b_{yy}^2(k)+b_{zz}^2(k)\right],\\
                I_3(k)
                &= \frac{1}{3}\left[b_{xx}^3(k)+b_{yy}^3(k)+b_{zz}^3(k)\right].
            \end{align}
            The $k$-dependent invariants $I_2(k)$ and $I_3(k)$ allow us to probe the scale-dependent anisotropy of the Reynolds stress in the same way that the eigen-structure of the tensor at the system scale does, e.g., for $I_2(k) = I_3(k) = 0$ the stress at $k$ is isotropic, for $I_2(k),\, I_3(k) < 0$ the stress is sheet-like, and for $I_2(k)< 0, \, I_3(k) > 0$, filament-like. The resulting scale-dependent anisotropy is shown in \autoref{fig:ScaleDependentAnisotropy}.

            \autoref{fig:ScaleDependentAnisotropy} refines the outer-scale picture from \autoref{ssec:large_scale_anisotropy}. At the very largest scales, nearly all of the spectral power resides in the streamwise component, with $\mathcal{P}_{xx}/{\rm{tr}}(\mathcal{P}_{ij}) \approx 1$ and $\mathcal{P}_{yy} \approx \mathcal{P}_{zz}\approx 0$ near $kL/2\pi \sim 1$. The corresponding stress therefore lies close to the filament-like morphology, as expected for motions still strongly aligned with the collision axis and the injected vortical jets.

            As $k$ increases, however, the anisotropy relaxes rapidly. Over roughly $20\lesssim kL/2\pi \lesssim 70$, the shell-wise power is much more nearly partitioned, with $\mathcal{P}_{xx}/{\rm{tr}}(\mathcal{P}_{ij}) \simeq 0.43$--$0.45$ and $\mathcal{P}_{yy}/{\rm{tr}}(\mathcal{P}_{ij}) \simeq \mathcal{P}_{zz}/{\rm{tr}}(\mathcal{P}_{ij}) \simeq 0.27$--$0.29$. The invariant trajectory simultaneously moves back toward the isotropic point, showing that most of the strong directional memory is confined to the first decade of modes. In other words, the residual anisotropy seen in the volume-integrated Reynolds stress is primarily an outer-scale property, whereas the resolved inertial-type range is much closer to axis-randomised turbulence.

            Beyond $k_{\nu,\rm s}$, the component spectra turn away from isotropy again, with $\mathcal{P}_{xx}(k)$ rising and $\mathcal{P}_{yy}(k)$, $\mathcal{P}_{zz}(k)$ falling. Because this occurs only after the onset of the dissipation turnover identified below, we do not interpret it as physical recovery of directional order. Rather, it is most naturally read as a numerical dissipation-range effect, consistent with the high-$k$ upturn already seen in $\mathcal{P}_c(k)/\mathcal{P}_s(k)$.

        \subsection{Estimating the Reynolds number}\label{sec:estimating_Reynolds}
        
            \begin{figure}
                \centering
                \includegraphics[width=\linewidth]{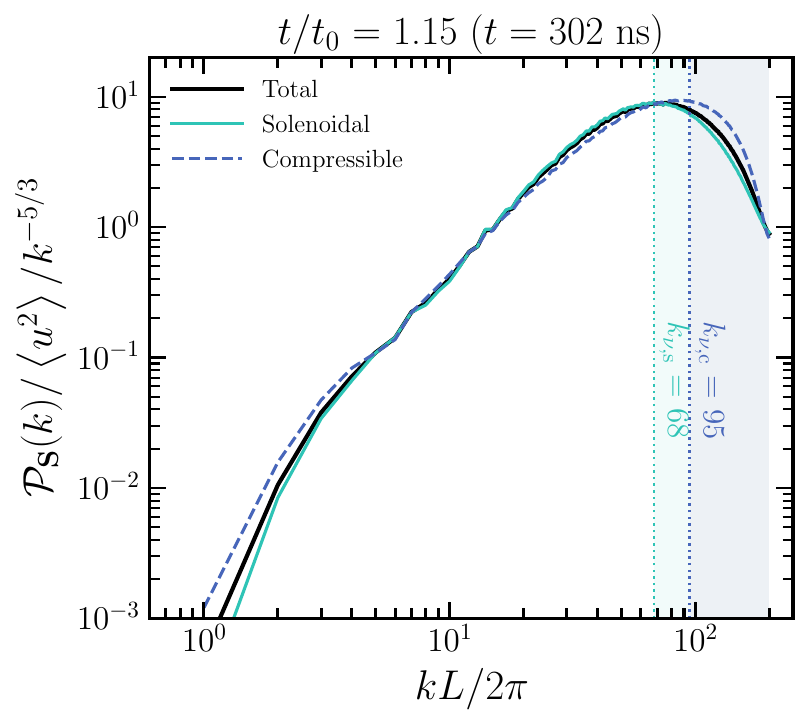}
                \caption{Compensated spectrum of the rate-of-strain tensor, $S_{ij}$, defined in \autoref{eq:rate_of_strain}, at $t/t_0=1.15$ ($t=302\,\rm ns$) for the total velocity field and for its Helmholtz-decomposed solenoidal and compressive parts (\autoref{eq:H_decomp}--\autoref{eq:Fourier}). The strain-spectrum diagnostic, $\mathcal{P}_{\mathbb S}(k)$, is defined in \autoref{eq:strain_spectrum}, so for spatially uniform viscosity $\nu_{\rm shear} \mathcal{P}_{\mathbb S}(k)$ is proportional to the scale-by-scale viscous dissipation rate. The spectra are plotted in compensated form, $\mathcal{P}_{\mathbb S}(k)/\langle u^2\rangle\,k^{-5/3}$, to highlight the turnover into the dissipation range. The vertical dotted lines mark the peak locations of the solenoidal and compressive strain spectra, $k_{\nu,\rm s}=68\,\rm mm^{-1}$ and $k_{\nu,\rm c}=95\,\rm mm^{-1}$, while the shaded region marks the onset of numerical dissipation.}
                \label{fig:PowerSpectrumStressTensor}
            \end{figure}

            We close by estimating the resolved dynamical range of the simulation, so that the spectral results above can be interpreted with a clear estimate of where numerical dissipation becomes important. Because the calculation is performed in the ILES regime, the Reynolds number inferred here should be interpreted as an effective numerical Reynolds number, i.e. a measure of the resolved cascade range rather than the microphysical plasma Reynolds number. To estimate it, we use the rate-of-strain tensor, since shear viscous dissipation is controlled by velocity gradients through $\varepsilon_{\rm visc} \sim \nu_{\rm shear} S_{ij}S_{ij}$, where $\nu_{\rm shear}$ is the kinematic shear viscosity coefficient \citep{Brandenburg2023_dissipation_MHD}, up to factors of order unity. The rate-of-strain tensor is
            \begin{align}\label{eq:rate_of_strain}
                S_{ij} = \frac{1}{2}\left( \partial_i u_j + \partial_j u_i \right) - \frac{1}{3}\delta_{ij} \partial_k u_k,
            \end{align}
            where the second term removes the trace, so $S_{ij}$ isolates the incompressible shear part of $\partial u_i/\partial x_j$ \citep{beattie_taking_2025}. The dissipation scale inferred below should therefore be understood as the dissipation scale of the incompressible cascade.
            
            The local viscous dissipation rate is $\varepsilon_{{\rm visc}} \sim \nu_{\rm shear} S_{ij}S_{ij}$, hence the power spectrum of the strain field is
            \begin{align}\label{eq:strain_spectrum}
                \frac{d\varepsilon_{{\rm visc}}(k)}{dk} \approx \mathcal{P}_{\mathbb{S}}(k) = \int_{\Omega_k} |S_{ij}(\k)S^{\dagger}_{ij}(\k)| k^2 \d{\Omega_k},
            \end{align}
            is the viscous dissipation rate spectrum, where $\Omega_k$ represents the spherical shell in Fourier space at wavenumber $k$. If the viscosity coefficient $\nu$ were spatially uniform, it would multiply the dissipation spectrum by a constant factor, i.e. rescale its amplitude, but it would not shift the characteristic wavenumber at which the dissipation power peaks. The dissipation wavenumber, $k_\nu$, is therefore identified as the mode where $\mathcal{P}_{\mathbb{S}}(k)$ peaks before sharply decaying, indicating the transition to dissipative dynamics and defining the shear-viscous scale of the resolved gradients. As shown in \autoref{fig:PowerSpectrumStressTensor}, the decomposed strain spectra peak at $k_{\nu,\rm s}=68\,\mathrm{mm}^{-1}$ for the solenoidal field and $k_{\nu,\rm c}=95\,\mathrm{mm}^{-1}$ for the compressive field. In what follows, we use the solenoidal peak to estimate $\Re$, because the Kolmogorov scaling invoked below pertains to the incompressible cascade; the corresponding compressive scale is discussed only as a comparison.
            
            Once $k_\nu$ is determined, the corresponding physical dissipation scale is $\ell_\nu = 2\pi/k_\nu$, where $k_\nu$ is understood here as the physical dissipation wavenumber. For a Kolmogorov-like cascade (as we show it is for the incompressible mode spectrum in \autoref{fig:PowerSpectrumVelocity}), the Kolmogorov scale is $\eta=(\nu^3/\varepsilon)^{1/4}$. Writing the dimensionless dissipation coefficient as $C_\epsilon\equiv \varepsilon \ell_0/u_0^3$ and the Reynolds number as $\Re\equiv u_0\ell_0/\nu$ gives
            \begin{align}
                \frac{\eta}{\ell_0} = C_\epsilon^{-1/4}\Re^{-3/4}.
            \end{align}
            If we identify the measured dissipation wavenumber with the Kolmogorov wavenumber, $k_\nu \approx k_\eta = 2\pi/\eta$, this becomes
            \begin{align}
                k_\nu = C_\nu \, k_0\, \Re^{3/4}, \qquad C_\nu \equiv C_\epsilon^{1/4},
            \end{align}
            where $k_0 = 2\pi/\ell_0$. For forced isotropic turbulence, \citet{McComb2015_forced_isotropic_dissipation} found $C_{\epsilon,\infty}=0.468\pm0.006$, which implies $C_\nu = C_\epsilon^{1/4}=0.827\pm0.003$\footnote{The literature more commonly reports $C_\epsilon$ or the Kolmogorov scale $\eta$ than the specific prefactor $C_\nu$ written here. \citet{Sreenivasan1998_update_dissipation} emphasised that the high-Reynolds-number dissipation coefficient is a constant of order unity but that its numerical value depends on the details of the low-$k$ forcing or large-scale structure. A representative DNS calibration for forced isotropic turbulence is the McComb et al. value used here. For comparison, \citet{Kriel2022_kinematic_dynamo_scales} calibrated $k_\nu=(0.025^{+0.005}_{-0.006})\,k_{\rm turb}\,\Re^{3/4}$ in \flash simulations with explicit dissipation, which would imply $\Re\simeq 2.2\times10^4$ for the present scale separation. That much larger value arises because their coefficient is a fitted exponential-cutoff parameter in a model spectrum, $\mathcal{P}_{\rm kin}(k)=A_{\rm kin}k^{\alpha_{\rm kin}}\exp(-k/k_\nu)$, rather than the Kolmogorov-scale prefactor inferred from $\eta/\ell_0=C_\epsilon^{-1/4}\Re^{-3/4}$. In the dissipation-range literature, \citet{PullinRogallo1994_pressure_higher_order_spectra} fitted the dissipation spectrum for $k\eta>0.2$, while \citet{BuariaSreenivasan2020_dissipation_range_spectrum} identified a stretched-exponential form over roughly $0.15\lesssim k\eta \lesssim 0.5$. We therefore regard the present McComb-based mapping as a conservative standard-turbulence estimate, while taking the directly measured quantity to be the scale separation $\ell_0/\ell_\nu$.}. Inverting the relation gives
            \begin{align}
                \Re
                =
                \left(
                \frac{k_\nu}{C_\nu k_0}
                \right)^{4/3}
                =
                C_\nu^{-4/3}
                \left(
                \frac{\ell_0}{\ell_\nu}
                \right)^{4/3}.
            \end{align}
            In the present simulation, the solenoidal strain spectrum peaks at $k_{\nu,\rm s} \simeq 68\,\mathrm{mm}^{-1}$, corresponding to $\ell_{\nu,\rm s} \simeq 2\pi/k_{\nu,\rm s} \simeq 0.092\,\mathrm{mm}=92\,\mu\mathrm{m}$. Using the outer scale measured at stagnation, $\ell_0 \simeq 4.5\,\mathrm{mm}$ (\autoref{sec:time_evol}), gives $k_0 \simeq 2\pi/(4.5\,\mathrm{mm}) \simeq 1.40\,\mathrm{mm}^{-1}$ and hence a resolved scale separation $\ell_0/\ell_{\nu,\rm s} \simeq 49$. The McComb calibration then gives
            \begin{align}
                \Re = \left(\frac{68}{0.827\times 1.40}\right)^{4/3} \simeq 2.3\times10^2.
            \end{align}
            For comparison, the compressive strain peak at $k_{\nu,\rm c}\simeq95\,\mathrm{mm}^{-1}$ corresponds to $\ell_{\nu,\rm c}\simeq 66\,\mu\mathrm{m}\simeq 6.6\,\delta$. The more robust statement is therefore that the resolved incompressible cascade extends over a scale separation of order $\ell_0/\ell_{\nu,\rm s}\sim 50$, with $\ell_{\nu,\rm s} \sim 9\,\delta$, where $\delta=10\,\mu\mathrm{m}$ is the grid spacing.
            
    \section{Mass and gradient density power spectra}\label{sec:DensityGrad}
\begin{figure}
            \centering
            \includegraphics[width=1\linewidth]{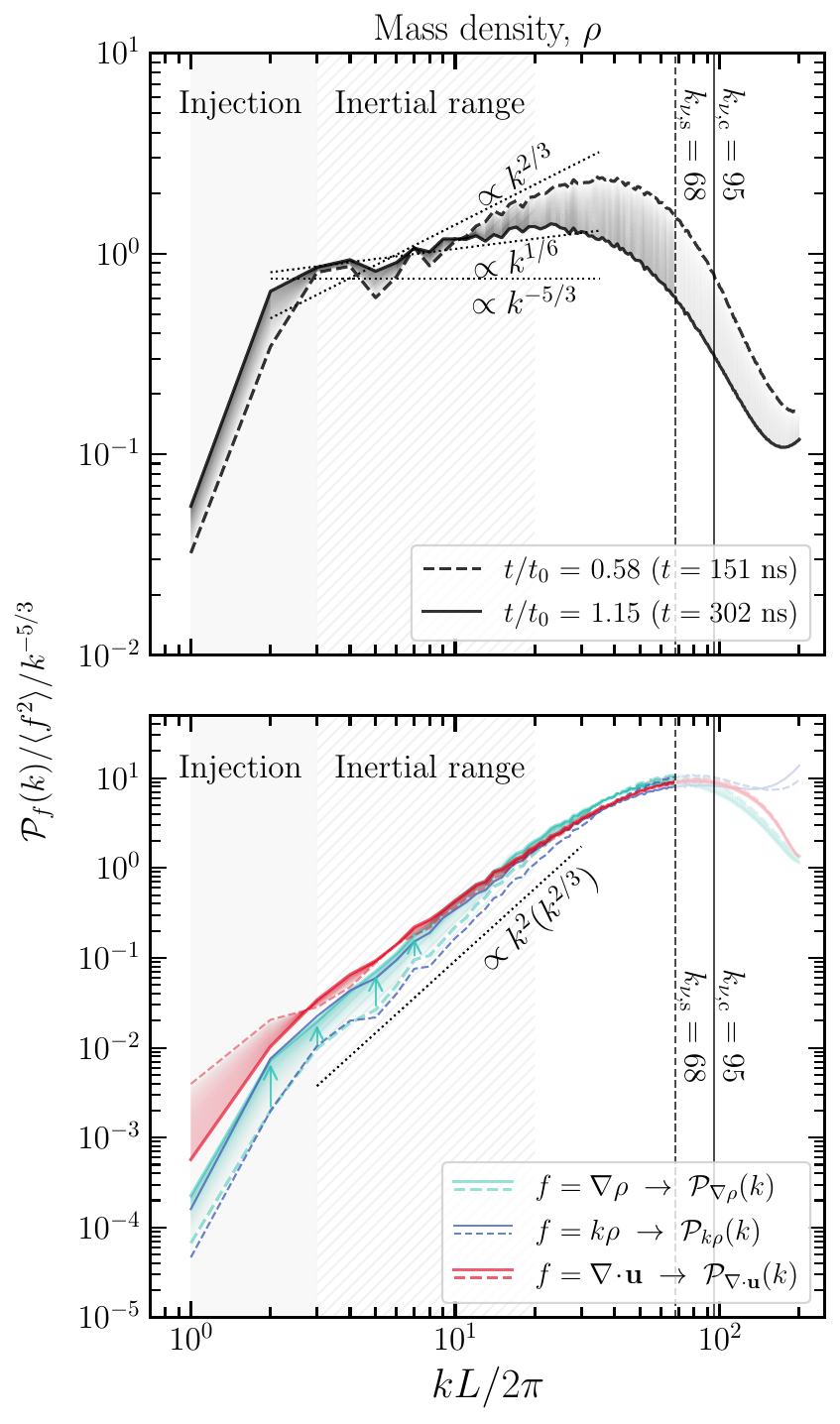}
            \caption{Compensated spectra of density diagnostics in the interaction control volume at $t/t_0=0.58$ ($t=151\,\rm ns$; dashed) and $t/t_0=1.15$ ($t=302\,\rm ns$; solid). Top panel: compensated density spectrum, $\mathcal{P}_\rho(k)/\langle \rho^2 \rangle \, k^{-5/3}$. The reference slopes shown in the panel correspond to the underlying density spectrum, with $\mathcal{P}_\rho(k)\propto k^{2/3}$ at early time and $\mathcal{P}_\rho(k)\propto k^{1/6}$ at late time. The density spectrum therefore becomes flatter in time; neither epoch follows a universal $k^{-5/3}$ law. Bottom panel: compensated spectra of $\mathcal{P}_{\bnab\rho}(k)$, $\mathcal{P}_{k^2\rho}(k)$, and $\mathcal{P}_{\bnab\cdot\u}(k)$. Their close tracking over most of the resolved range shows that density gradients are closely tied to compressive dynamics, making these spectra the simulated analogue of the density-gradient statistics probed by shadowgraphy-like laboratory diagnostics and radio scintillation observables. The common downturn near the vertical markers indicates the dissipation scales estimated with the strain-spectrum Reynolds-number diagnostic.}
            \label{fig:PowerSpectrumDensity}
        \end{figure}

        \autoref{sec:PowerSpectrum} established how the kinetic energy is partitioned between compressive and solenoidal motions. We now ask how that mode mixture imprints the scalar density field, $\rho$, and its gradients. \autoref{fig:PowerSpectrumDensity} shows the compensated spectra of $\mathcal{P}_\rho(k)$ (top) and of the compressive density-gradient diagnostics $\mathcal{P}_{\bnab\rho}(k)$, $\mathcal{P}_{k^2\rho}(k)$, and $\mathcal{P}_{\bnab\cdot\u}(k)$ (bottom), all measured in the interaction control volume.

        Density gradients have direct observational significance in ionized astrophysical plasmas. Classical radio-propagation measurements infer a broad power-law spectrum of electron-density fluctuations from scintillation and scattering observables \citep{Armstrong1995_power_law}, while scintillation arcs provide a more geometrical probe of small-scale scattering structure in the ISM \citep{Stinebring2006_scintillation_arcs}. Related analyses identify enhanced density-fluctuation amplitudes near stellar bow shocks, while noting that these enhancements may reflect either shock amplification or selection effects \citep{Ocker2021_bow_shock_turbulence}, and use FRB and quasar propagation to constrain density-fluctuation amplitudes and dissipation scales in circumgalactic turbulence \citep{Ocker2025_CGM_turbulence}. Plasma-lensing and extreme-scattering-event models further emphasize that compact refracting structures can dominate radio propagation signatures \citep{Pen2012MNRAS,Pen2014MNRAS,Jow2024MNRAS_cusp,Kempski2025ApJL}, and that wave-optics measurements can constrain the dimensionality of such lenses \citep{Jow2022MNRAS}. This motivates asking whether density-gradient statistics in the laboratory layer trace the nearly solenoidal velocity cascade, the compressive modes, or a separate thermodynamic source. The distinction matters for future synthetic scintillation diagnostics of HEDLA turbulence, and for interpreting astrophysical density structure inferred from propagation effects.

        \subsection{Density spectrum}
            The top panel of \autoref{fig:PowerSpectrumDensity} shows the compensated mass-density spectrum, $\mathcal{P}_\rho(k)/\langle \rho^2 \rangle \, k^{-5/3}$, at $t/t_0=0.58$ and $1.15$. The power-law guides shown in the figure correspond to the underlying density spectrum itself. At early time, $\mathcal{P}_\rho(k)\propto k^{2/3}$ over the resolved inertial interval, while by $t/t_0=1.15$ it is closer to $\mathcal{P}_\rho(k)\propto k^{1/6}$. The density spectrum therefore becomes flatter in time as the turbulence decays.

            This evolution is consistent with the physical picture developed in \autoref{sec:eos} and \autoref{sec:GenVorticity}. The largest density structures are set directly by shock compression and the geometry of the counter-streaming collision, so the low-$k$ spectrum remains shock-imprinted. At later times, however, the layer has already relaxed toward an effectively close-to-isothermal closure, and in-layer baroclinic generation is weak. The density field is then shaped less by fresh thermodynamic source terms and more by advection, mixing, and compressive transport within the developed layer. In that sense, $\mathcal{P}_\rho(k)$ records temporal redistribution of density structure by the turbulent flow, while still retaining a strong large-scale shock imprint and without implying a universal density cascade.

        \subsection{Density-gradient spectra and compressive coupling}

            The lower panel of \autoref{fig:PowerSpectrumDensity} compares three compressive diagnostics: $\mathcal{P}_{\bnab\rho}(k)$, $\mathcal{P}_{k^2\rho}(k)$, and $\mathcal{P}_{\bnab\cdot\u}(k)$. Over most of the injection and inertial ranges, these spectra track each other closely at both times and roll over near the same dissipation scale. This already shows that density gradients are primarily controlled by compressive dynamics. The close agreement between $\mathcal{P}_{\bnab\rho}(k)$ and $\mathcal{P}_{k^2\rho}(k)$ is also the expected Fourier derivative weighting, $\bnab \rightarrow i\k$.

            To interpret this result, we use a continuity-equation scaling model, with the full derivation deferred to \autoref{app:density_gradient_scaling}. Starting from the continuity equation,
            \begin{align}
                \partial_t \rho + \u\cdot\bnab\rho = -\rho\,\bnab\cdot\u.
            \end{align}
            and linearising about a slowly varying background, $\rho=\rho_0+\delta\rho$, gives
            \begin{align}\label{eq:small_dens_div}
                \bnab\cdot\u \approx -\frac{1}{\rho_0}
                \left(
                \partial_t\delta\rho + \u\cdot\bnab\delta\rho
                \right).
            \end{align}
            In Fourier space, with $\omega_{\rm tot}\equiv \omega_{c_s}+\omega_{\rm turb}$, $\omega_{c_s}=c_s k$, and $\omega_{\rm turb}\sim k u_c(k)$, this implies
            \begin{align}
                \mathcal{P}_{\bnab\cdot\u}(k)
                \sim
                \frac{\omega_{\rm tot}^2}{\rho_0^2}\,
                \mathcal{P}_{\rho}(k).
            \end{align}
            Since $\bnab\rho \rightarrow i\k\,\rho$, we also have $\mathcal{P}_{\bnab\rho}(k)\sim k^2 \mathcal{P}_{\rho}(k)$, and hence
            \begin{align}
                \mathcal{P}_{\bnab\cdot\u}(k)
                \sim
                \frac{\omega_{\rm tot}^2}{\rho_0^2 k^2}\,
                \mathcal{P}_{\bnab\rho}(k).
            \end{align}

\begin{figure}
                \centering
                \includegraphics[width=\linewidth]{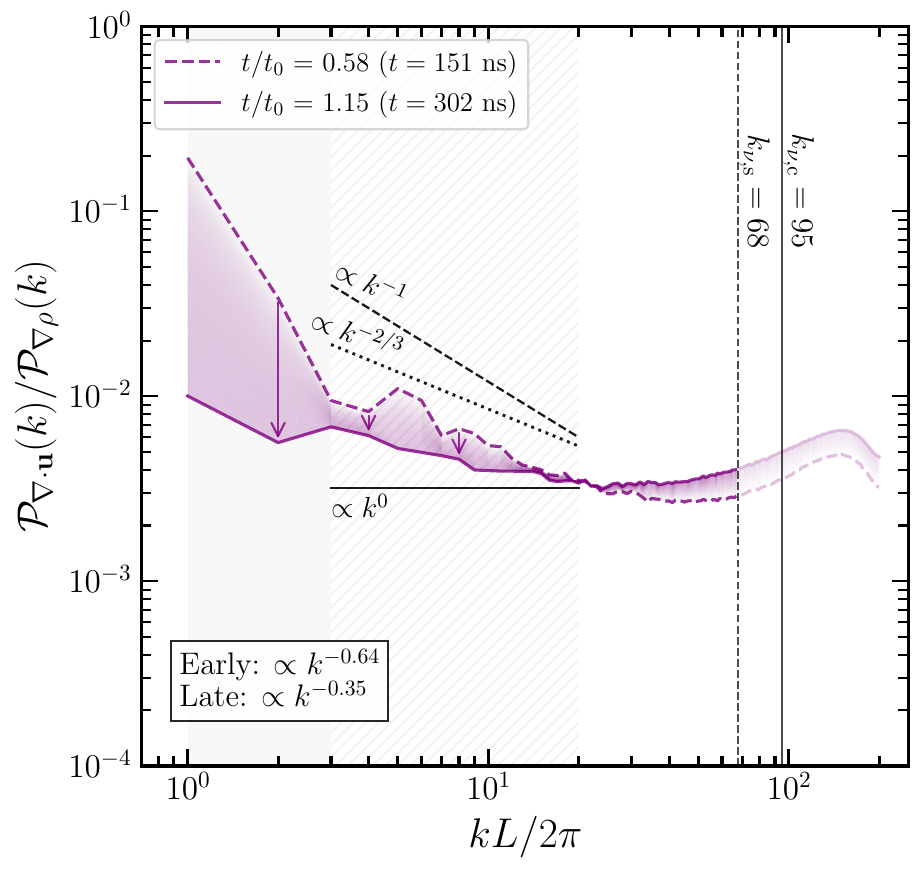}
                \caption{The ratio $\mathcal{P}_{\bnab\cdot\u}(k)/\mathcal{P}_{\bnab\rho}(k)$ at $t/t_0=0.58$ ($t=151\,\rm ns$; dashed) and $t/t_0=1.15$ ($t=302\,\rm ns$; solid). The shaded regions mark the injection and resolved inertial ranges, and the vertical lines indicate the solenoidal and compressive dissipation wavenumbers, $k_{\nu,\rm s}=68\,\rm mm^{-1}$ and $k_{\nu,\rm c}=95\,\rm mm^{-1}$. Reference slopes show the advective-limit expectations $\propto k^{-1}$ and $\propto k^{-2/3}$ for Burgers-like and Kolmogorov-like compressive spectra, respectively, while the horizontal guide marks the acoustic/source-dominated scaling $\mathcal{P}_{\bnab\cdot\u}(k)\propto\mathcal{P}_{\bnab\rho}(k)$. At early time, the ratio decreases with a slope close to the Kolmogorov-like prediction, indicating that density gradients retain a stronger imprint of cascade-mediated advection. By late time, the ratio is flatter across the resolved inertial interval, showing that density-gradient statistics are increasingly controlled by the local velocity-divergence source. This scale-dependent transition is the quantity most directly relevant for interpreting shadowgraphy, scintillation, and plasma-lensing diagnostics, which weight refractive density structure rather than the incompressible velocity cascade \citep{Armstrong1995_power_law,Stinebring2006_scintillation_arcs,Pen2012MNRAS,Jow2022MNRAS,Kempski2025ApJL}.}
                \label{fig:PowerSpectraHierarchy}
            \end{figure}

            Two limits are useful. In the acoustic limit, $\omega_{\rm tot}\approx \omega_{c_s}\approx c_s k$, so
            \begin{align}
                \mathcal{P}_{\bnab\cdot\u}(k) \sim \mathcal{P}_{\bnab\rho}(k).
            \end{align}
            In the advective-compressive limit, $\omega_{\rm tot}\approx \omega_{\rm turb}\approx k u_c(k)$, so
            \begin{align}
                \mathcal{P}_{\bnab\cdot\u}(k)
                \sim
                u_c^2(k)\,\mathcal{P}_{\bnab\rho}(k)
                \sim
                k \mathcal{P}_c(k)\,\mathcal{P}_{\bnab\rho}(k).
            \end{align}
            If the compressive velocity follows a Burgers-like spectrum, $\mathcal{P}_c(k)\propto k^{-2}$, then
            \begin{align}
                \mathcal{P}_{\bnab\cdot\u}(k) \sim k^{-1}\mathcal{P}_{\bnab\rho}(k).
            \end{align}
            If instead the relevant compressive velocity follows Kolmogorov scaling, $\mathcal{P}_c(k)\propto k^{-5/3}$, then
            \begin{align}
                \mathcal{P}_{\bnab\cdot\u}(k) \sim k^{-2/3}\mathcal{P}_{\bnab\rho}(k).
            \end{align}
            The ratio $\mathcal{P}_{\bnab\cdot\u}(k)/\mathcal{P}_{\bnab\rho}(k)$, shown in \autoref{fig:PowerSpectraHierarchy}, tests which of these limits controls the measured density-gradient structure. A flat ratio corresponds to the acoustic or local-source limit, in which velocity divergence directly sets the density-gradient spectrum. A decreasing ratio instead indicates an advective-compressive contribution, with the slope measuring how strongly the gradients inherit the scale dependence of the compressive velocity cascade. At $t/t_0=0.58$, soon after the layer has formed, the ratio decreases through the resolved inertial interval with an approximate slope close to the Kolmogorov-like prediction, $\propto k^{-2/3}$. This is consistent with the broader chronology above: the flow is still reorganising shock-imprinted density structure while the turbulent cascade is being established, so density gradients retain a stronger imprint of cascade-mediated advection.

            By $t/t_0=1.15$, the ratio is substantially flatter, closer to the acoustic/source-dominated limit than to either the Burgers-like or Kolmogorov-like advective scalings. This does not imply that the flow is globally weakly compressible; the layer remains shock-rich and the compressive velocity spectrum is still distinct from the solenoidal cascade. Rather, it indicates that once the post-stagnation turbulence has decayed and the interaction region has relaxed toward the nearly isothermal state described in \autoref{sec:eos}, the density-gradient statistics are increasingly controlled by the local velocity-divergence source. In Fourier space, $\widetilde{\bnab\cdot\u}(\k)=i\k\cdot\Tilde{\u}(\k)=i\k\cdot\Tilde{\u}_{c}(\k)$, because the solenoidal component satisfies $\k\cdot\Tilde{\u}_{s}(\k)=0$. Thus, up to shell-averaging convention, $\mathcal{P}_{\bnab\cdot\u}(k)\sim k^2 \mathcal{P}_c(k)$, so temporal evolution in the compressive velocity spectrum is inherited directly by the density and density-gradient statistics through the continuity equation. This explains why the near-universal behaviour seen in the incompressible velocity spectrum is not reproduced in $\mathcal{P}_\rho(k)$ or $\mathcal{P}_{\bnab\rho}(k)$: diagnostics sensitive to density gradients probe primarily the compressive part of the flow, transitioning from a more advective/cascade-imprinted regime at early time toward a more divergence-controlled regime at late time. The late-time hierarchy may therefore be summarised as
            \begin{align}\label{eq:density_gradient_hierarchy}
                \begin{aligned}
                    \mathcal{P}_{\bnab\rho}(k)
                    &\sim
                    k^2 \mathcal{P}_\rho(k)
                    \sim
                    \mathcal{P}_{\bnab\cdot\u}(k)
                    \sim
                    k^2 \mathcal{P}_c(k).
                \end{aligned}
            \end{align}

    \section{Summary and Conclusions}
    We have presented three-dimensional, three-temperature radiation-hydrodynamic simulations of a double-mesh laboratory platform for generating shock-driven turbulence (\autoref{sec:methods}, \autoref{fig:Setup}, \autoref{fig:ExampleFullcScaleSim}). A $t_{\rm rad}=30\,\rm ns$ X-ray pulse incident on two CH meshes separated by $9\,\rm mm$ launches counter-streaming ablation flows from apertures of size $2\times2\,\rm mm$ and $2\times1\,\rm mm$. These flows propagate toward the mid-plane at a bulk speed of order $40\,\rm km\,s^{-1}$, first collide at $t\simeq 75\,\rm ns$ near $x\simeq 6\,\rm mm$, and then form a shock-rich mixing layer that remains active to at least $t=302\,\rm ns \simeq 1.15\,t_0$, nearly an order of magnitude longer than the drive itself (\autoref{sec:time_evol}, \autoref{fig:TimeResolvedLayer}, \autoref{fig:TimeResolvedEddy}).

    The resulting chronology is quantitatively clear. Before collision, the inflows have characteristic conditions $\rho \simeq 5\times10^{-6}\,\rm g\,cm^{-3}$, $\Ti\simeq\Te\simeq 4\,\rm eV$, and $c_s\simeq 9\,\rm km\,s^{-1}$, implying a bulk Mach number $\M\approx 4.5$ and an electron-ion equilibration time of only $\tau_{\rm eq}\simeq 2\,\rm ns$ (\autoref{sec:time_evol}, \autoref{tab:plasma_inflows_params}). At first contact the outer scale reaches $\ell_0\simeq 4.5\,\rm mm$ inside a control volume of scale $L=0.7\,\rm cm$, while the largest eddy-turnover time is $t_0=261\,\rm ns$ and the ratio $t_0/t_{c_s}\simeq 0.2$ remains approximately constant during the subsequent decay. Over that same post-stagnation phase, the turbulent velocity amplitude decays approximately as $u_0(t)\propto t^{-1.1}$, consistent with freely decaying supersonic turbulence rather than sustained stationary forcing (\autoref{sec:time_evol}, \autoref{fig:TimeResolvedEddy}, \autoref{eq:u0_def_int}). The collision produces peak stagnation densities of order $10^{-4}\,\rm g\,cm^{-3}$ and heats the plasma from $(\Ti,\Te)\simeq(4,4)\,\rm eV$ to $(24,10)\,\rm eV$, but without forming a simple planar post-shock slab (\autoref{sec:time_evol}, \autoref{fig:PressureDensityCorrelation}).

    The anisotropy analysis shows that the interaction retains directional memory, but mainly on the largest scales. The volume-integrated Reynolds stress relaxes only partially toward isotropy: $R_{xx}/\mathrm{tr}(R)$ decreases from near unity at stagnation to $\simeq 0.72$ by $t/t_0=1.15$, while $R_{yy}/\mathrm{tr}(R)\simeq 0.16$ and $R_{zz}/\mathrm{tr}(R)\simeq 0.12$ remain unequal, demonstrating persistent memory of both the collision axis and the mesh geometry. This large-scale anisotropy is naturally tied to the pre-collision injection of narrow vortical jets by the mesh apertures discussed in \autoref{sec:GenVorticity}. The scale-dependent Reynolds-stress analysis then shows where that memory resides: the lowest modes remain close to a filament-like, streamwise state, but over $20\lesssim kL/2\pi\lesssim70$ the component-wise spectral power is much closer to isotropic, with only the highest-$k$ tail turning away again after the onset of numerical dissipation (\autoref{ssec:large_scale_anisotropy}, \autoref{fig:Anisotropy_reynolds_outer_scale}, \autoref{fig:ScaleDependentAnisotropy}). This scale dependence is important for proton-radiography and proton-tomography interpretations, because commonly used stochastic-field inversions infer magnetic spectra and $B_{\rm rms}$ under homogeneity and isotropy assumptions \citep{Bott2017_proton_imaging_stochastic_fields,tzeferacos_laboratory_2018,Bott2021_time_resolved_dynamo}. Our velocity anisotropy does not directly measure magnetic anisotropy, but it shows that isotropy must be demonstrated, not assumed, on the injection scales of gridded flow platforms.

    The interaction layer also relaxes toward an effectively close-to-isothermal closure. Between $t=131\,\rm ns$ and $261\,\rm ns$, the pressure-density relation evolves from $\gamma_{\rm eff}\simeq 1.16$ to $\gamma_{\rm eff}\simeq 1.07$ for the total pressure, while the radiation pressure remains nearly density-independent, $\gamma_{\rm eff}\ll 1$ (\autoref{sec:eos}, \autoref{fig:PressureDensityCorrelation}). This near-barotropic response is analogous to the close-to-isothermal closures often used for cold, rapidly cooling ISM turbulence \citep{Ferriere2020_reynolds_numbers_for_ism,beattie_taking_2025}, but here it emerges from the 3T radiation-hydrodynamic evolution rather than being imposed. It also explains why sustained in-layer baroclinic generation is weak even though the flow remains shock-rich.

    The key mechanism that generates vorticity, and therefore the incompressible component of the turbulence, occurs earlier, during the $30\,\rm ns$ ablation and hole-closure phase of each mesh cell. Oblique shocks inside the apertures create strong local misalignment between $\bnab\Ptot$ and $\bnab\rho$, so the baroclinic source in \autoref{eq:vorticity} generates predominantly $\omega_x$ near the cell corners and along the off-bisector ribbons identified in \autoref{eq:omega_x_seed} and \autoref{eq:baro_antisym}. The converging in-plane flow then advects this corner-generated vorticity toward the cell centre and compresses it into narrow, geometry-dependent channels on a timescale $\Delta t_{\rm coll}\sim 15\,\rm ns$, implying in-plane advection speeds of order $u_\perp\sim 30$--$70\,\rm km\,s^{-1}$ for the front and rear cell geometries. The outflows therefore emerge already carrying collimated vortical jets rather than featureless compressive streams (\autoref{sec:GenVorticity}, \autoref{fig:FrontRearMeshVorticityGen}). This mechanism should also operate in TDYNO-like gridded plasma flows, for which turbulence has been interpreted primarily as the product of shear and Kelvin--Helmholtz mixing when corrugated fronts interleave and collide \citep{tzeferacos_numerical_2017,tzeferacos_laboratory_2018}. The later collision in our platform therefore amplifies and reorganises pre-existing rotational structure, and analogous collision-stage instabilities in related gridded platforms may likewise act on already vortical upstream flows. After the inflows collide, the vorticity budget is dominated by stretching and compression, with $r_{\rm stretch}\simeq 0.5$ and $r_{\rm comp}\simeq 0.45$, while baroclinicity drops to $r_{\rm baro}\simeq 0.05$ (\autoref{fig:LayerVorticityTimeResolved}, \autoref{fig:QCriteriaTimeSequence}).

    The velocity spectra show that this shocked layer evolves toward a mixed but predominantly solenoidal cascade. By $t/t_0=1$, the volume-integrated kinetic energy is partitioned approximately $70\%$ in solenoidal motions and $30\%$ in compressive motions (\autoref{fig:LayerVelModesTimeResolved}), while the scale-by-scale spectral ratio settles to $\mathcal{P}_c(k)/\mathcal{P}_s(k)\sim 0.3$ through most of the resolved inertial interval (\autoref{fig:PowerSpectrumVelocity}). Over that same range, the total and solenoidal spectra are broadly consistent with $k^{-5/3}$, whereas the compressive spectrum is steeper, closer to $k^{-2}$ (\autoref{sec:PowerSpectrum}). The resolved dissipation scale inferred from the strain spectrum is $k_{\nu,\rm s}\simeq 68\,\rm mm^{-1}$, i.e. $\ell_{\nu,\rm s}\simeq 92\,\mu\rm m$, which together with $\ell_0\simeq 4.5\,\rm mm$ implies a resolved scale separation $\ell_0/\ell_{\nu,\rm s}\simeq 49$ and an effective numerical Reynolds number of order $\Re\sim 2\times10^2$ under standard Kolmogorov scaling (\autoref{sec:estimating_Reynolds}, \autoref{fig:PowerSpectrumStressTensor}, \autoref{eq:strain_spectrum}).

    The density statistics retain the imprint of that same shock-turbulence mixture. The mass-density spectrum becomes flatter, from $\mathcal{P}_\rho(k)\propto k^{2/3}$ at $t=151\,\rm ns$ to $\mathcal{P}_\rho(k)\propto k^{1/6}$ at $t=302\,\rm ns$, indicating temporal redistribution of density structure without evidence for a universal density cascade (\autoref{sec:DensityGrad}, \autoref{fig:PowerSpectrumDensity}). The spectra of $\mathcal{P}_{\bnab\rho}(k)$, $\mathcal{P}_{k^2\rho}(k)$, and $\mathcal{P}_{\bnab\cdot\u}(k)$ track one another over much of the resolved range, but the ratio $\mathcal{P}_{\bnab\cdot\u}(k)/\mathcal{P}_{\bnab\rho}(k)$ shows a clear temporal trend: at early time it decreases close to the Kolmogorov-like advective prediction, while by late time it flattens toward the acoustic/source-dominated limit in \autoref{eq:density_gradient_hierarchy} (\autoref{fig:PowerSpectraHierarchy}). Diagnostics sensitive to density gradients, including scintillation- and plasma-lensing-like observables, therefore probe the compressive part of the turbulence, transitioning from a more cascade-imprinted regime soon after stagnation toward a more local-divergence-controlled regime as the layer decays; they are not, by themselves, faithful tracers of the underlying incompressible velocity spectrum (\autoref{fig:PowerSpectrumDensity}, \autoref{fig:PowerSpectraHierarchy}, \autoref{app:density_gradient_scaling}).

    Taken together, these results establish the double-mesh platform as a quantitatively characterised laboratory realisation of sustained shock-driven turbulence: a $30\,\rm ns$ radiative drive seeds vorticity at the mesh, a collision at $75\,\rm ns$ sets a turbulence outer scale of $\ell_0\simeq 4.5\,\rm mm$, and the layer remains shock-rich and turbulently supersonic for at least one outer-scale eddy-turnover time. The velocity field evolves toward a predominantly solenoidal, Kolmogorov-like cascade with an effective resolved Reynolds number of order $2\times10^2$, while retaining geometry-imprinted anisotropy that is strongest at the outer scale but measurable over much of the resolved range. The scalar density field does not simply follow that incompressible cascade: its gradients remain tied to the compressive velocity divergence, evolving from a more cascade-imprinted regime soon after stagnation toward a flatter, source-dominated hierarchy at late time. This combination of persistent directional memory, increasingly isotropic inertial-range velocity structure, and compressively controlled density-gradient statistics provides a unique laboratory window into the nature of shock-driven interstellar turbulence (\autoref{sec:time_evol}, \autoref{sec:eos}, \autoref{sec:GenVorticity}, \autoref{sec:PowerSpectrum}, \autoref{sec:DensityGrad}). Future experiments on this platform, especially when combined with refraction-based diagnostics, such as shadowgraphy, can therefore connect the simulated density-gradient statistics to the refractive properties measured in the laboratory and to the density structures inferred from scintillation and plasma-lensing observations of the ionized ISM \citep{Armstrong1995_power_law,Stinebring2006_scintillation_arcs,Pen2012MNRAS,Pen2014MNRAS,Jow2022MNRAS,Kempski2025ApJL}.

    \section*{Acknowledgements}
    This work is supported by AFOSR under FA8655-23-1-7062, EPSRC and First Light Fusion under the AMPLIFI Prosperity Partnership, Sandia National Laboratories and the National Nuclear Security Administration (NNSA) under U.S. Department of Energy (DoE) DE-NA0004148. J.~R.~B. acknowledges compute allocations rrg-ripperda and rrg-essick from the Digital Research Alliance of Canada, funding from the Natural Sciences and Engineering Research Council of Canada (NSERC, funding reference number 568580), support from NSF Award 2206756, and high-performance computing resources provided by the Leibniz Rechenzentrum and the Gauss Center for Supercomputing (grants pn76gi, pr73fi, and pn76ga), which S.~M. and J.~R.~B. used for the running the simulations presented in this study.

    \software{We use \textsc{flash} for all simulations. Data analysis and visualization software used in this study includes \textsc{C++} \citep{Stroustrup2013}, \textsc{numpy} \citep{Oliphant2006,numpy2020}, \textsc{numba} \citep{numba:2015}, \textsc{matplotlib} \citep{Hunter2007}, \textsc{cython} \citep{Behnel2011}, \textsc{visit} \citep{Childs2012}, \textsc{scipy} \citep{Virtanen2020}, \textsc{scikit-image} \citep{vanderWalts2014}, \textsc{cmasher} \citep{Velden2020_cmasher}, \textsc{yt} \citep{yt}, \textsc{pandas} \citep{pandas}, \textsc{joblib} \citep{joblib}, and \textsc{pyfftw} \citep{2021ascl.soft09009G}. Our post-processing was performed using \textsc{plasmatools} \href{https://github.com/AstroJames/PLASMAtools}{https://github.com/AstroJames/PLASMAtools}. We acknowledge the use of \textsc{codex} \citep{OpenAI2025_Codex} for editing the manuscript.}

    \pagebreak
    
    \appendix

    \section{Numerical flux-order reconstruction study}\label{app:convergence}
\begin{figure*}
        \centering
        \includegraphics[width=1\linewidth]{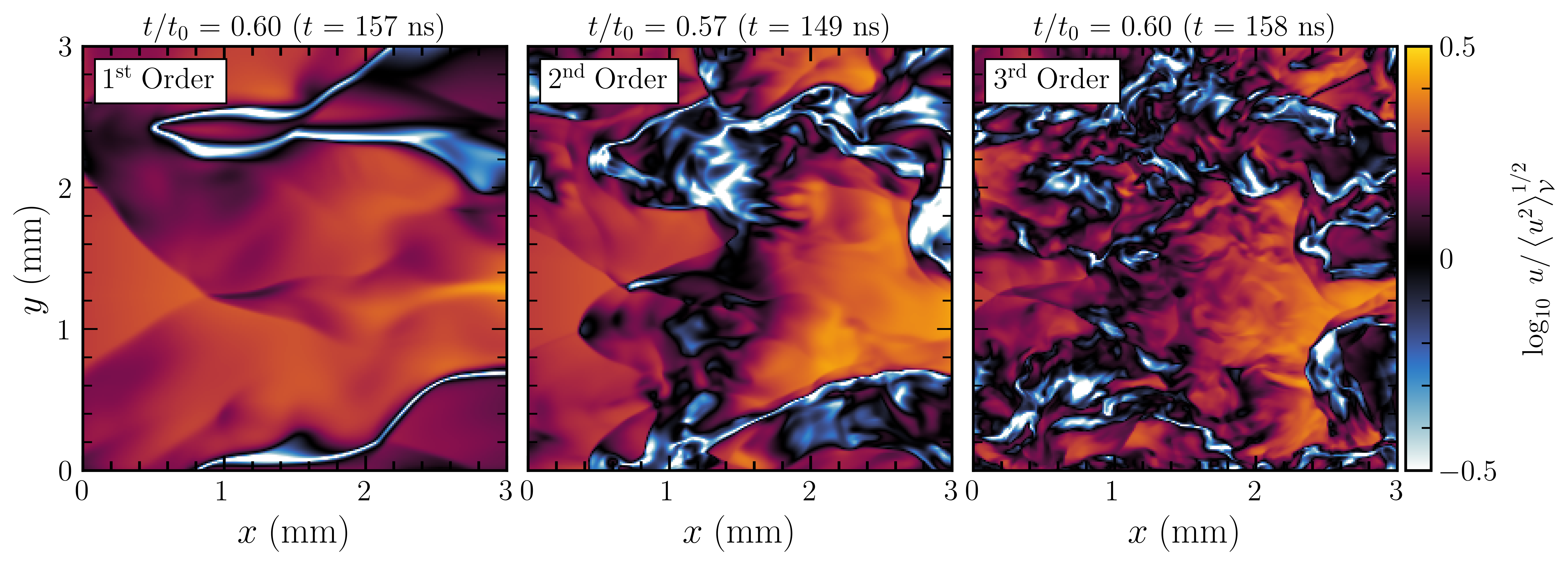}
        \caption{Flux-order comparison of the velocity-field morphology in a representative $3\,\rm mm\times3\,\rm mm$ $x$--$y$ subregion of the turbulent mixing layer. The panels show $1^{\rm st}$-, $2^{\rm nd}$-, and $3^{\rm rd}$-order hydrodynamic reconstruction from left to right, using comparable evolutionary times: $t/t_0=0.60$ ($t=157\,\rm ns$), $0.57$ ($t=149\,\rm ns$), and $0.60$ ($t=158\,\rm ns$). The colour scale is $\log_{10}(|\bm{u}|/\langle u^2\rangle_{\V}^{1/2})$, so each panel shows velocity contrast relative to the volume-averaged rms velocity of that run rather than absolute speed. The $1^{\rm st}$-order calculation preserves the large millimetre-scale shear geometry but smooths the interfaces and suppresses much of the sub-millimetre filamentation. The $2^{\rm nd}$- and $3^{\rm rd}$-order calculations retain sharper velocity jumps and thinner vortical structures across the same field of view, indicating that reconstruction order primarily controls the resolved small-scale content rather than the gross morphology set by the mesh-driven collision.}
        \label{fig:VelocityComparisonOrders}
    \end{figure*}

\begin{figure*}
        \centering
        \includegraphics[width=1\linewidth]{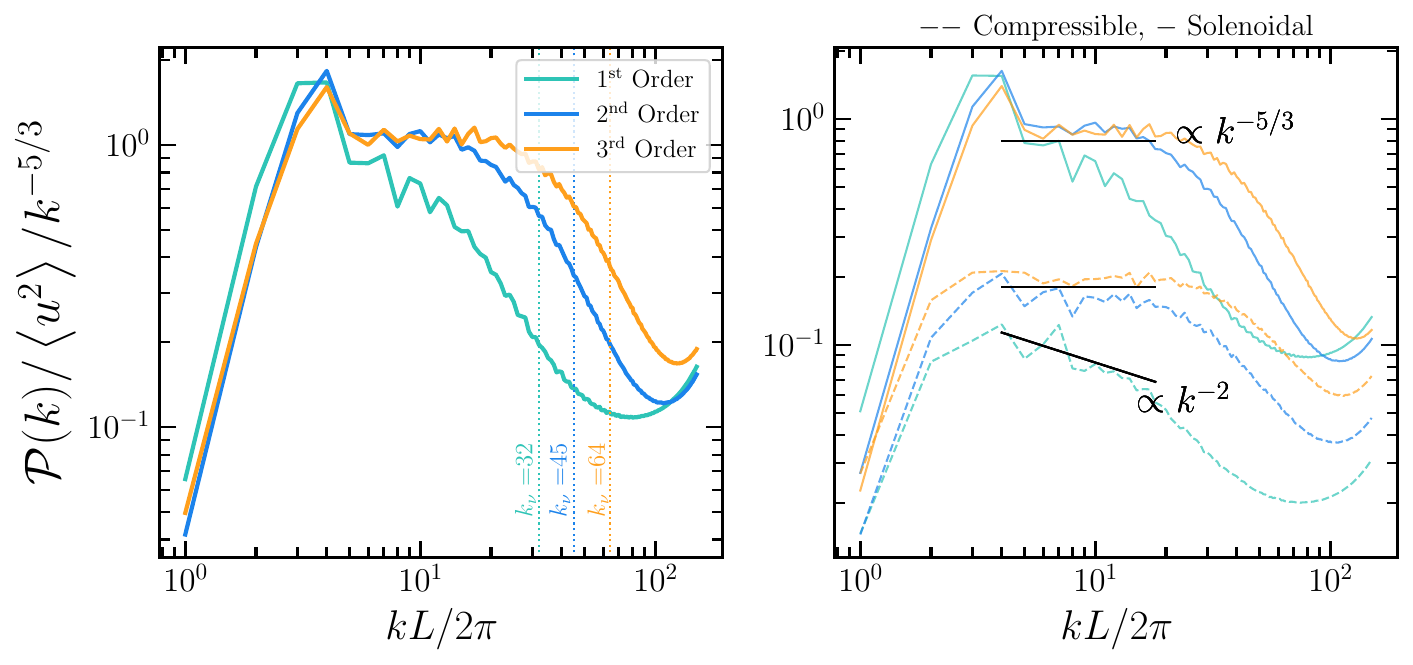}
        \caption{Flux-order comparison of the velocity spectra corresponding to \autoref{fig:VelocityComparisonOrders}. Left: compensated total velocity spectrum, $\mathcal{P}_u(k)/\langle u^2\rangle\,k^{-5/3}$, as a function of dimensionless shell wavenumber $kL/2\pi$. The black dashed line marks the injection scale, while the coloured dotted lines mark $k_\nu$, the numerical-dissipation wavenumber inferred using the same strain-spectrum diagnostic used for the Reynolds-number estimate in \autoref{sec:estimating_Reynolds} and \autoref{eq:strain_spectrum}: $k_\nu L/2\pi\simeq32$ for $1^{\rm st}$ order, $45$ for $2^{\rm nd}$ order, and $64$ for $3^{\rm rd}$ order. With $\ell_0\simeq4.5\,\rm mm$, $L=7\,\rm mm$, and $C_\nu=0.827$ in \autoref{sec:estimating_Reynolds}, these imply $\Re_{\rm eff}\simeq73$, $114$, and $183$, respectively. Right: Helmholtz-decomposed spectra in the same compensation, with dashed curves showing the compressive component, $\mathcal{P}_c(k)$, and solid curves showing the solenoidal component, $\mathcal{P}_s(k)$. The reference slopes indicate the approximate $k^{-2}$ compressive and $k^{-5/3}$ solenoidal scalings discussed in the main text. Across all three schemes, the large-scale peak and solenoidal-over-compressive ordering are similar, while $k_\nu$ shifts systematically to larger wavenumber with increasing reconstruction order.}
        \label{fig:PowerSpectrumOrderComparison}
    \end{figure*}

    To assess the robustness of the resolved turbulent structure, we compare simulations that vary the hydrodynamic flux reconstruction order. \autoref{fig:VelocityComparisonOrders} shows representative $3\,\rm mm\times3\,\rm mm$ velocity subregions from the mixing layer at comparable evolutionary stages, while \autoref{fig:PowerSpectrumOrderComparison} compares the corresponding compensated velocity spectra.
    
    The $1^{\rm st}$-order flux-reconstruction scheme is substantially more diffusive, producing smoother shear layers and suppressing much of the sub-millimetre filamentation visible in the higher-order calculations. The $2^{\rm nd}$- and $3^{\rm rd}$-order schemes retain sharper interfaces and smaller-scale vortical structure across the same $3\,\rm mm$ field of view. Quantitatively, the dissipation wavenumber inferred with the same strain-spectrum criterion used in \autoref{sec:estimating_Reynolds} moves from $k_\nu L/2\pi\simeq32$ for $1^{\rm st}$ order to $\simeq45$ for $2^{\rm nd}$ order and $\simeq64$ for $3^{\rm rd}$ order, increasing the resolved range by about a factor of two between the $1^{\rm st}$- and $3^{\rm rd}$-order calculations. Using the same mapping as \autoref{sec:estimating_Reynolds}, $\Re=(k_\nu/C_\nu k_0)^{4/3}$ with $C_\nu=0.827$, $k_0=2\pi/\ell_0$, $\ell_0\simeq4.5\,\rm mm$, and $L=7\,\rm mm$, these correspond to effective numerical Reynolds numbers $\Re_{\rm eff}\simeq73$, $114$, and $183$, respectively. The large-scale peak, the approximate total-spectrum slope, and the relative ordering of the compressive and solenoidal spectra are nevertheless consistent across schemes, suggesting that the mesh geometry and collision kinematics set the injection scale, while the reconstruction order chiefly controls how much of the downstream cascade remains resolved. We therefore adopt the $3^{\rm rd}$-order runs as the fiducial dataset in the main text, because they provide the broadest inertial interval and the highest inferred numerical-dissipation wavenumber.
    
    \section{Local corner expansion of the baroclinic source}\label{app:baroclinic_corner}
    In \autoref{sec:GenVorticity}, we argued that the baroclinic source term vanishes at a symmetric cell corner and changes sign across the corner bisector. Here we show this explicitly with a local Taylor expansion. We introduce local coordinates $(\xi,\eta)$ within the cell, with $(\xi,\eta)=(0,0)$ at the lower-left corner and $\xi,\eta>0$ increasing away from the two solid walls. The $x$-component of the baroclinic source term is then
    \begin{align}
        S_{{\rm baro},x}
        =
        \frac{1}{\rho^2}
        \left(
        \frac{\partial \rho}{\partial \xi}\frac{\partial \Ptot}{\partial \eta}
        -
        \frac{\partial \rho}{\partial \eta}\frac{\partial \Ptot}{\partial \xi}
        \right).
        \label{eq:baro_corner_appendix}
    \end{align}
    The derivation assumes a locally symmetric right-angle corner, so that $(\xi,\eta)$ provide a natural basis aligned with the two wall-normal directions. We further assume that the wall material and radiative drive are locally symmetric under exchange of these directions, $\xi \leftrightarrow \eta$, and that the pressure and density fields are smooth enough near the corner to admit Taylor expansions over the region of interest. Under this exchange symmetry, the local fields satisfy
    \begin{align}
        \Ptot(\xi,\eta) &= \Ptot(\eta,\xi), \\
        \rho(\xi,\eta)  &= \rho(\eta,\xi).
    \end{align}
    To make the symmetry reduction explicit, begin with the most general quadratic Taylor expansion about the corner,
    \begin{align}
        \Ptot(\xi,\eta)
        &= P_0 + c_1 \xi + c_2 \eta + c_3 \xi^2 + c_4 \eta^2 + c_5\xi\eta
        + \mathcal{O}(\xi^3,\xi^2\eta,\xi\eta^2,\eta^3), \\
        \rho(\xi,\eta)
        &= \rho_0 + d_1 \xi + d_2 \eta + d_3 \xi^2 + d_4 \eta^2 + d_5\xi\eta
        + \mathcal{O}(\xi^3,\xi^2\eta,\xi\eta^2,\eta^3),
    \end{align}
    where $P_0$ and $\rho_0$ are the corner values, and the remaining coefficients are proportional to derivatives of $\Ptot$ and $\rho$ evaluated at the corner. Exchange symmetry, $\xi \leftrightarrow \eta$, then requires
    \begin{align}
        c_1=c_2,\qquad c_3=c_4,\qquad d_1=d_2,\qquad d_3=d_4.
    \end{align}
    The expansions therefore reduce to
    \begin{align}
        \Ptot(\xi,\eta)
        &= P_0 + a_1(\xi+\eta) + a_2(\xi^2+\eta^2) + a_3\xi\eta
        + \mathcal{O}(\xi^3,\xi^2\eta,\xi\eta^2,\eta^3), \\
        \rho(\xi,\eta)
        &= \rho_0 + b_1(\xi+\eta) + b_2(\xi^2+\eta^2) + b_3\xi\eta
        + \mathcal{O}(\xi^3,\xi^2\eta,\xi\eta^2,\eta^3),
    \end{align}
    where the coefficients $a_i$ and $b_i$ depend on time. Differentiating gives
    \begin{align}
        \frac{\partial \Ptot}{\partial \eta}
        &= a_1 + 2a_2\eta + a_3\xi + \mathcal{O}(\xi^2,\xi\eta,\eta^2), \\
        \frac{\partial \Ptot}{\partial \xi}
        &= a_1 + 2a_2\xi + a_3\eta + \mathcal{O}(\xi^2,\xi\eta,\eta^2), \\
        \frac{\partial \rho}{\partial \xi}
        &= b_1 + 2b_2\xi + b_3\eta + \mathcal{O}(\xi^2,\xi\eta,\eta^2), \\
        \frac{\partial \rho}{\partial \eta}
        &= b_1 + 2b_2\eta + b_3\xi + \mathcal{O}(\xi^2,\xi\eta,\eta^2).
    \end{align}
    Substituting these into \autoref{eq:baro_corner_appendix} yields
    \begin{align}
        S_{{\rm baro},x}
        &=
        \frac{1}{\rho^2}
        \Big[
        \left(b_1 + 2b_2\xi + b_3\eta\right)\left(a_1 + 2a_2\eta + a_3\xi\right)
        \nonumber \\
        &\hspace{2em}
        -
        \left(b_1 + 2b_2\eta + b_3\xi\right)\left(a_1 + 2a_2\xi + a_3\eta\right)
        \Big]
        + \mathcal{O}(\xi^2,\xi\eta,\eta^2).
    \end{align}
    Expanding and collecting terms gives
    \begin{align}
        S_{{\rm baro},x}
        =
        \frac{\xi-\eta}{\rho^2}
        \left[
        a_1(2b_2-b_3)
        -
        b_1(2a_2-a_3)
        \right]
        + \mathcal{O}\!\left[(\xi-\eta)(\xi+\eta)\right].
        \label{eq:baro_corner_factorised}
    \end{align}
    The baroclinic source therefore contains an explicit factor of $(\xi-\eta)$ at leading order, while the denominator remains symmetric and finite for non-zero density. It follows immediately that $S_{{\rm baro},x}=0$ on the corner bisector, $\xi=\eta$, and hence also at the corner apex, $(\xi,\eta)=(0,0)$. Moreover, $S_{{\rm baro},x}$ changes sign across the bisector because $(\xi-\eta)$ changes sign. Thus, the two adjacent strips of opposite sign seen in \autoref{fig:FrontRearMeshVorticityGen} are the generic first non-zero departure from exact corner symmetry: the corner itself is a symmetry point, while either side of the diagonal corresponds to opposite handedness in the local misalignment between $\bnab \Ptot$ and $\bnab \rho$.

    \section{Continuity-equation scaling model for density-gradient spectra}\label{app:density_gradient_scaling}
    In \autoref{sec:DensityGrad}, we argued that the density-gradient spectra are closely tied to the velocity-divergence spectrum through the continuity equation. Here we state the simple scaling derivation more explicitly. Start from
    \begin{align}
        \partial_t \rho + \u\cdot\bnab\rho = -\rho\,\bnab\cdot\u, \qquad
        \rho = \rho_0+\delta\rho.
    \end{align}
    For a slowly varying background density $\rho_0$ and $|\delta\rho|/\rho_0<1$, this gives
    \begin{align}\label{eq:app_small_dens_div}
        \bnab\cdot\u \approx -\frac{1}{\rho_0}\left(\partial_t\delta\rho + \u\cdot\bnab\delta\rho\right).
    \end{align}
    In Fourier space, retaining only characteristic magnitudes,
    \begin{align}
        \partial_t\delta\rho &\sim \omega_{c_s}\,\delta\rho(k), &
        \u\cdot\bnab\delta\rho &\sim \omega_{\rm turb}\,\delta\rho(k), &
        \bnab\delta\rho &\sim k\,\delta\rho(k),\\
        \omega_{c_s} &= c_s k, &
        \omega_{\rm turb} &\sim k u_c(k), &
        \omega_{\rm tot} &\equiv \omega_{c_s}+\omega_{\rm turb}.
    \end{align}
    Here $\omega_{\rm tot}$ is the total characteristic frequency, $\omega_{c_s}$ is the acoustic frequency, and $\omega_{\rm turb}$ is the reciprocal turbulent turnover time. Then \autoref{eq:app_small_dens_div} implies
    \begin{align}
        (\bnab\cdot\u)(k)\sim \frac{\omega_{\rm tot}}{\rho_0}\,\delta\rho(k), \qquad
        \mathcal{P}_{\bnab\cdot\u}(k)\sim \frac{\omega_{\rm tot}^2}{\rho_0^2}\,\mathcal{P}_{\rho}(k).
        \label{eq:app_divrho}
    \end{align}
    Since $\bnab\rho \rightarrow i\k\,\rho$,
    \begin{align}
        \mathcal{P}_{\bnab\rho}(k)\sim k^2 \mathcal{P}_{\rho}(k), \qquad \text{so} \qquad
        \mathcal{P}_{\bnab\cdot\u}(k)\sim \frac{\omega_{\rm tot}^2}{\rho_0^2 k^2}\,\mathcal{P}_{\bnab\rho}(k).
        \label{eq:app_density_gradient_scaling}
    \end{align}
    These relations are intended to describe scale dependence, so we suppress scale-independent normalisation factors such as $c_s^2/\rho_0^2$. The useful limits are immediate. In the acoustic limit,
    \begin{align}
        \omega_{\rm tot}\approx \omega_{c_s}\approx c_s k, \qquad
        \mathcal{P}_{\bnab\cdot\u}(k)\propto \mathcal{P}_{\bnab\rho}(k).
    \end{align}
    In the advective limit, using the shell convention $u_c^2(k)\sim k\mathcal{P}_c(k)$,
    \begin{align}
        \omega_{\rm tot}\approx \omega_{\rm turb}\approx k u_{c}(k), \qquad
        \mathcal{P}_{\bnab\cdot\u}(k)\sim u_{c}^2(k)\,\mathcal{P}_{\bnab\rho}(k)\sim k \mathcal{P}_c(k)\,\mathcal{P}_{\bnab\rho}(k).
    \end{align}
    Thus, for $\mathcal{P}_c(k)\propto k^{-n}$,
    \begin{align}
        \mathcal{P}_{\bnab\cdot\u}(k)\sim k^{1-n}\mathcal{P}_{\bnab\rho}(k),
    \end{align}
    so Burgers scaling ($n=2$) gives
    \begin{align}
        \mathcal{P}_{\bnab\cdot\u}(k)\sim k^{-1}\mathcal{P}_{\bnab\rho}(k), \qquad
        \text{while Kolmogorov scaling } (n=5/3) \text{ gives } \mathcal{P}_{\bnab\cdot\u}(k)\sim k^{-2/3}\mathcal{P}_{\bnab\rho}(k).
    \end{align}
    This derivation should be interpreted as an interpretive scaling model, not a formal closure theory. Its purpose is to identify the spectral relation expected when density gradients are controlled primarily by the local velocity-divergence source, or instead retain a stronger imprint of advective cascade dynamics.

    \bibliography{Beattie,Merlini}{}

@ARTICLE{Mohapatra2019_turbulent_heat_flux_ICM,
       author = {{Mohapatra}, Rajsekhar and {Sharma}, Prateek},
        title = "{Turbulence in the intracluster medium: simulations, observables, and thermodynamics}",
      journal = {The Monthly Notices of The Royal Astronomical Society},
         year = 2019,
        month = apr,
       volume = {484},
       number = {4},
        pages = {4881-4896},
          doi = {10.1093/mnras/stz328},
archivePrefix = {arXiv},
       eprint = {1810.00018},
 primaryClass = {astro-ph.GA},
       adsurl = {https://ui.adsabs.harvard.edu/abs/2019MNRAS.484.4881M}
}

@ARTICLE{Zhang2026_Perseus_merger_turbulence,
       author = {{Zhang}, Congyao and {Zhuravleva}, Irina and {Heinrich}, Annie and {Bellomi}, Elena and {Truong}, Nhut and {Zuhone}, John and {Churazov}, Eugene and {Eckart}, Megan E. and {Fujita}, Yutaka and {Hlavacek-Larrondo}, Julie and {Ichinohe}, Yuto and {Markevitch}, Maxim and {Matsushita}, Kyoko and {Mernier}, Fran{\c c}ois and {Miller}, Eric D. and {Mori}, Koji and {Nakajima}, Hiroshi and {Ogorzalek}, Anna and {Porter}, Frederick S. and {T{\"u}mer}, Ay{\c s}eg{\"u}l and {Ueda}, Shutaro and {Werner}, Norbert},
        title = "{Mapping the Perseus galaxy cluster with XRISM: gas kinematic features and their implications for turbulence}",
      journal = {\aap},
         year = 2026,
        month = mar,
       volume = {707},
          eid = {A109},
        pages = {A109},
          doi = {10.1051/0004-6361/202557660},
       adsurl = {https://ui.adsabs.harvard.edu/abs/2026A&A...707A.109Z}
}

@ARTICLE{Appleton2023_Stephans_Quintet,
       author = {{Appleton}, P.~N. and {Guillard}, P. and {Emonts}, Bjorn and {Boulanger}, Francois and {Togi}, Aditya and {Reach}, William T. and {Alatalo}, Katherine and {Cluver}, M. and {Diaz Santos}, T. and {Duc}, P.-A. and {Gallagher}, S. and {Ogle}, P. and {O'Sullivan}, E. and {Voggel}, K. and {Xu}, C.~K.},
        title = "{Multiphase Gas Interactions on Subarcsec Scales in the Shocked Intergalactic Medium of Stephan's Quintet with JWST and ALMA}",
      journal = {\apj},
         year = 2023,
        month = jul,
       volume = {951},
       number = {2},
          eid = {104},
        pages = {104},
          doi = {10.3847/1538-4357/accc2a},
archivePrefix = {arXiv},
       eprint = {2301.02928},
 primaryClass = {astro-ph.GA},
       adsurl = {https://ui.adsabs.harvard.edu/abs/2023ApJ...951..104A}
}

@ARTICLE{Sharda2022_driving_parameter,
       author = {{Sharda}, Piyush and {Menon}, Shyam H. and {Federrath}, Christoph and {Krumholz}, Mark R. and {Beattie}, James R. and {Jameson}, Katherine E. and {Tokuda}, Kazuki and {Burkhart}, Blakesley and {Crocker}, Roland M. and {Law}, Charles J. and {Seta}, Amit and {Gaetz}, Terrance J. and {Pingel}, Nickolas M. and {Seitenzahl}, Ivo R. and {Sano}, Hidetoshi and {Fukui}, Yasuo},
        title = "{First extragalactic measurement of the turbulence driving parameter: ALMA observations of the star-forming region N159E in the Large Magellanic Cloud}",
      journal = {\mnras},
         year = 2022,
        month = jan,
       volume = {509},
       number = {2},
        pages = {2180-2193},
          doi = {10.1093/mnras/stab3048},
archivePrefix = {arXiv},
       eprint = {2109.03983},
 primaryClass = {astro-ph.GA},
       adsurl = {https://ui.adsabs.harvard.edu/abs/2022MNRAS.509.2180S}
}

@ARTICLE{Gerrard2023_driving_parameter,
       author = {{Gerrard}, Isabella A. and {Federrath}, Christoph and {Pingel}, Nickolas M. and {McClure-Griffiths}, Naomi M. and {Marchal}, Antoine and {Joncas}, Gilles and {Clark}, Susan E. and {Stanimirovi{\'c}}, Sne{\v{z}}ana and {Lee}, Min-Young and {van Loon}, Jacco Th and {Dickey}, John and {D{\'e}nes}, Helga and {Ma}, Yik Ki and {Dempsey}, James and {Lynn}, Callum},
        title = "{A new method for spatially resolving the turbulence-driving mixture in the ISM with application to the Small Magellanic Cloud}",
      journal = {\mnras},
         year = 2023,
        month = nov,
       volume = {526},
       number = {1},
        pages = {982-999},
          doi = {10.1093/mnras/stad2718},
archivePrefix = {arXiv},
       eprint = {2309.10755},
 primaryClass = {astro-ph.GA},
       adsurl = {https://ui.adsabs.harvard.edu/abs/2023MNRAS.526..982G}
}

@misc{2021ascl.soft09009G,
       author = {{Gomersall}, Henry},
        title = "{pyFFTW: Python wrapper around FFTW}",
 howpublished = {Astrophysics Source Code Library, record ascl:2109.009},
         year = 2021,
        month = sep,
          eid = {ascl:2109.009},
       adsurl = {https://ui.adsabs.harvard.edu/abs/2021ascl.soft09009G}
}

@article{Kolmogorov1941,
author = {Kolmogorov, A. N.},
doi = {10.1098/rspa.1991.0075},
isbn = {0962-8444},
issn = {13645021},
journal = {Doklady Akademii Nauk Sssr},
number = {1890},
pages = {301--305},
title = {{The local structure of turbulence in incompressible viscous fluid for very large Reynolds numbers}},
url = {http://rspa.royalsocietypublishing.org/cgi/doi/10.1098/rspa.1991.0075},
volume = {30},
year = {1941}
}

@article{Federrath2009,
archivePrefix = {arXiv},
arxivId = {0808.0605},
author = {Federrath, Christoph and Klessen, Ralf S. and Schmidt, Wolfram},
doi = {10.1086/595280},
eprint = {0808.0605},
isbn = {5223105001},
issn = {0004-637X},
journal = {The Astrophysical Journal},
number = {1},
pages = {364--374},
title = {{The Fractal Density Structure in Supersonic Isothermal Turbulence: Solenoidal Versus Compressive Energy Injection}},
url = {http://arxiv.org/abs/0808.0605{\%}0Ahttp://dx.doi.org/10.1086/595280},
volume = {692},
year = {2009}
}

@ARTICLE{Federrath2018,
       author = {{Federrath}, Christoph},
        title = "{The turbulent formation of stars}",
      journal = {Physics Today},
         year = 2018,
        month = jun,
       volume = {71},
       number = {6},
        pages = {38-42},
          doi = {10.1063/PT.3.3947},
archivePrefix = {arXiv},
       eprint = {1806.05312},
 primaryClass = {astro-ph.SR},
       adsurl = {https://ui.adsabs.harvard.edu/abs/2018PhT....71f..38F}
}

@article{scikit-image,
 title = {scikit-image: image processing in {P}ython},
 author = {van der Walt, {S}t\'efan and {S}ch\"onberger, {J}ohannes {L}. and
           {Nunez-Iglesias}, {J}uan and {B}oulogne, {F}ran\c{c}ois and {W}arner,
           {J}oshua {D}. and {Y}ager, {N}eil and {G}ouillart, {E}mmanuelle and
           {Y}u, {T}ony and the scikit-image contributors},
 year = {2014},
 month = {6},
 volume = {2},
 pages = {453},
 journal = {PeerJ},
 issn = {2167-8359},
 url = {http://dx.doi.org/10.7717/peerj.453},
 doi = {10.7717/peerj.453}
}

@article{Tzeferacos2018_dynamo_in_the_lab,
	Author = {Tzeferacos, P. and Rigby, A. and Bott, A. F. A. and Bell, A. R. and Bingham, R. and Casner, A. and Cattaneo, F. and Churazov, E. M. and Emig, J. and Fiuza, F. and Forest, C. B. and Foster, J. and Graziani, C. and Katz, J. and Koenig, M. and Li, C. -K. and Meinecke, J. and Petrasso, R. and Park, H. -S. and Remington, B. A. and Ross, J. S. and Ryu, D. and Ryutov, D. and White, T. G. and Reville, B. and Miniati, F. and Schekochihin, A. A. and Lamb, D. Q. and Froula, D. H. and Gregori, G.},
	Da = {2018/02/09},
	Doi = {10.1038/s41467-018-02953-2},
	Id = {Tzeferacos2018},
	Isbn = {2041-1723},
	Journal = {Nature Communications},
	Number = {1},
	Pages = {591},
	Title = {Laboratory evidence of dynamo amplification of magnetic fields in a turbulent plasma},
	Ty = {JOUR},
	Url = {https://doi.org/10.1038/s41467-018-02953-2},
	Volume = {9},
	Year = {2018}}

@article{Federrath2013_universality,
archivePrefix = {arXiv},
arxivId = {1306.3989},
author = {Federrath, Christoph},
doi = {10.1093/mnras/stt1644},
eprint = {1306.3989},
issn = {00358711},
journal = {The Monthly Notices of The Royal Astronomical Society},
number = {2},
pages = {1245--1257},
title = {{On the universality of supersonic turbulence}},
volume = {436},
year = {2013}
}

@article{McKee2007,
archivePrefix = {arXiv},
arxivId = {0707.3514},
author = {McKee, Christopher F. and Ostriker, Eve C.},
doi = {10.1146/annurev.astro.45.051806.110602},
eprint = {0707.3514},
isbn = {0066-4146},
issn = {0066-4146},
journal = {Annu. Rev. Astron. Astrophys.},
number = {1},
pages = {565--687},
pmid = {23887428},
title = {{Theory of Star Formation}},
url = {http://arxiv.org/abs/0707.3514{\%}0Ahttp://dx.doi.org/10.1146/annurev.astro.45.051806.110602},
volume = {45},
year = {2007}
}

@article{MacLow2004,
archivePrefix = {arXiv},
arxivId = {astro-ph/0301093},
author = {Mac Low, Mordecai Mark and Klessen, Ralf S.},
doi = {10.1103/RevModPhys.76.125},
eprint = {0301093},
isbn = {0034-6861},
issn = {00346861},
journal = {Reviews of Modern Physics},
number = {1},
pages = {125--194},
primaryClass = {astro-ph},
title = {{Control of star formation by supersonic turbulence}},
volume = {76},
year = {2004}
}

@article{Federrath2010_solendoidal_versus_compressive,
archivePrefix = {arXiv},
arxivId = {0905.1060},
author = {Federrath, C. and Roman-Duval, J. and Klessen, R. and Schmidt, W. and {Mac Low}, M. -M.},
doi = {10.1051/0004-6361/200912437},
eprint = {0905.1060},
issn = {0004-6361},
journal = {Astronomy and Astrophysics},
number = {A81},
title = {{Comparing the statistics of interstellar turbulence in simulations and observations: Solenoidal versus compressive turbulence forcing}},
url = {http://arxiv.org/abs/0905.1060{\%}0Ahttp://dx.doi.org/10.1051/0004-6361/200912437},
volume = {512},
year = {2010}
}

@ARTICLE{Ferriere2020_reynolds_numbers_for_ism,
       author = {{Ferri{\`e}re}, K.},
        title = "{Plasma turbulence in the interstellar medium}",
      journal = {Plasma Physics and Controlled Fusion},
         year = 2020,
        month = jan,
       volume = {62},
       number = {1},
        pages = {014014},
          doi = {10.1088/1361-6587/ab49eb},
archivePrefix = {arXiv},
       eprint = {1912.08237},
 primaryClass = {astro-ph.GA},
       adsurl = {https://ui.adsabs.harvard.edu/abs/2020PPCF...62a4014F}
}

@ARTICLE{Howes2024_fundamental_parameters,
       author = {{Howes}, Gregory G.},
        title = "{The fundamental parameters of astrophysical plasma turbulence and its dissipation: non-relativistic limit}",
      journal = {Journal of Plasma Physics},
         year = 2024,
        month = oct,
       volume = {90},
       number = {5},
          eid = {905900504},
        pages = {905900504},
          doi = {10.1017/S0022377824001090},
archivePrefix = {arXiv},
       eprint = {2402.12829},
 primaryClass = {physics.plasm-ph}
}

@ARTICLE{Jin2017,
       author = {{Jin}, Keitaro and {Salim}, Diane M. and {Federrath}, Christoph and
         {Tasker}, Elizabeth J. and {Habe}, Asao and {Kainulainen}, Jouni T.},
        title = "{On the effective turbulence driving mode of molecular clouds formed in disc galaxies}",
      journal = {The Monthly Notices of The Royal Astronomical Society},
         year = 2017,
        month = jul,
       volume = {469},
       number = {1},
        pages = {383-393},
          doi = {10.1093/mnras/stx737},
archivePrefix = {arXiv},
       eprint = {1703.09709},
 primaryClass = {astro-ph.GA},
       adsurl = {https://ui.adsabs.harvard.edu/abs/2017MNRAS.469..383J}
}

@ARTICLE{Kortgen2017,
       author = {{K{\"o}rtgen}, Bastian and {Federrath}, Christoph and {Banerjee}, Robi},
        title = "{The driving of turbulence in simulations of molecular cloud formation and evolution}",
      journal = {The Monthly Notices of The Royal Astronomical Society},
         year = 2017,
        month = dec,
       volume = {472},
       number = {2},
        pages = {2496-2503},
          doi = {10.1093/mnras/stx2208},
archivePrefix = {arXiv},
       eprint = {1703.07232},
 primaryClass = {astro-ph.GA},
       adsurl = {https://ui.adsabs.harvard.edu/abs/2017MNRAS.472.2496K}
}

@ARTICLE{Lu2020_supernova_driving,
       author = {{Lu}, Zu-Jia and {Pelkonen}, Veli-Matti and {Padoan}, Paolo and {Pan}, Liubin and {Haugb{\o}lle}, Troels and {Nordlund}, {\r{A}}ke},
        title = "{The Effect of Supernovae on the Turbulence and Dispersal of Molecular Clouds}",
      journal = {The Astrophysical Journal},
         year = 2020,
        month = nov,
       volume = {904},
       number = {1},
          eid = {58},
        pages = {58},
          doi = {10.3847/1538-4357/abbd8f},
archivePrefix = {arXiv},
       eprint = {2007.09518},
 primaryClass = {astro-ph.GA},
       adsurl = {https://ui.adsabs.harvard.edu/abs/2020ApJ...904...58L}
}

@ARTICLE{Klessen2000,
       author = {{Klessen}, Ralf S. and {Heitsch}, Fabian and {Mac Low}, Mordecai-Mark},
        title = "{Gravitational Collapse in Turbulent Molecular Clouds. I. Gasdynamical Turbulence}",
      journal = {The Astrophysical Journal},
         year = "2000",
        month = "Jun",
       volume = {535},
        pages = {887-906},
          doi = {10.1086/308891},
archivePrefix = {arXiv},
       eprint = {astro-ph/9911068},
 primaryClass = {astro-ph},
       adsurl = {https://ui.adsabs.harvard.edu/\#abs/2000ApJ...535..887K}
}

@ARTICLE{Wolfire1995_isothermal_ISM,
       author = {{Wolfire}, M.~G. and {Hollenbach}, D. and {McKee}, C.~F. and {Tielens}, A.~G.~G.~M. and {Bakes}, E.~L.~O.},
        title = "{The Neutral Atomic Phases of the Interstellar Medium}",
      journal = {The Astrophysical Journal},
         year = 1995,
        month = apr,
       volume = {443},
        pages = {152},
          doi = {10.1086/175510},
       adsurl = {https://ui.adsabs.harvard.edu/abs/1995ApJ...443..152W}
}

@ARTICLE{Federrath2010b,
       author = {{Federrath}, C. and {Duval}, J. and {Klessen}, R.~S. and {Schmidt}, W. and
         {Low}, M. -M. Mac},
        title = "{Solenoidal versus compressive turbulence forcing}",
      journal = {Highlights of Astronomy},
         year = "2010",
        month = "Nov",
       volume = {15},
        pages = {404-404},
          doi = {10.1017/S1743921310009944},
archivePrefix = {arXiv},
       eprint = {0910.5469},
 primaryClass = {astro-ph.SR},
       adsurl = {https://ui.adsabs.harvard.edu/\#abs/2010HiA....15..404F}
}

@ARTICLE{Fryxell2000,
   author = {{Fryxell}, B. and {Olson}, K. and {Ricker}, P. and {Timmes}, F.~X. and 
	{Zingale}, M. and {Lamb}, D.~Q. and {MacNeice}, P. and {Rosner}, R. and 
	{Truran}, J.~W. and {Tufo}, H.},
    title = "{FLASH: An Adaptive Mesh Hydrodynamics Code for Modeling Astrophysical Thermonuclear Flashes}",
  journal = {The Astrophysical Journal Supplement },
     year = 2000,
    month = nov,
   volume = 131,
    pages = {273-334},
      doi = {10.1086/317361},
   adsurl = {https://ui.adsabs.harvard.edu/abs/2000ApJS..131..273F}
}

@ARTICLE{Bacchini2020_supernova_drives_turb,
       author = {{Bacchini}, Cecilia and {Fraternali}, Filippo and {Iorio}, Giuliano and {Pezzulli}, Gabriele and {Marasco}, Antonino and {Nipoti}, Carlo},
        title = "{Evidence for supernova feedback sustaining gas turbulence in nearby star-forming galaxies}",
      journal = {\aap},
         year = 2020,
        month = sep,
       volume = {641},
          eid = {A70},
        pages = {A70},
          doi = {10.1051/0004-6361/202038223},
archivePrefix = {arXiv},
       eprint = {2006.10764},
 primaryClass = {astro-ph.GA},
       adsurl = {https://ui.adsabs.harvard.edu/abs/2020A&A...641A..70B}
}

@INPROCEEDINGS{numba:2015,
       author = {{Lam}, Siu Kwan and {Pitrou}, Antoine and {Seibert}, Stanley},
        title = "{Numba: A LLVM-based Python JIT Compiler}",
    booktitle = {Proc. Second Workshop on the LLVM Compiler Infrastructure in HPC},
         year = 2015,
        month = nov,
        pages = {1-6},
          doi = {10.1145/2833157.2833162},
       adsurl = {https://ui.adsabs.harvard.edu/abs/2015llvm.confE...1L}
}

@ARTICLE{Velden2020_cmasher,
       author = {{van der Velden}, Ellert},
        title = "{CMasher: Scientific colormaps for making accessible, informative and 'cmashing' plots}",
      journal = {The Journal of Open Source Software},
         year = 2020,
        month = feb,
       volume = {5},
       number = {46},
          eid = {2004},
        pages = {2004},
          doi = {10.21105/joss.02004},
archivePrefix = {arXiv},
       eprint = {2003.01069},
 primaryClass = {eess.IV},
       adsurl = {https://ui.adsabs.harvard.edu/abs/2020JOSS....5.2004V}
}

@article{Mandal2020,
    author = {Mandal, Ankush and Federrath, Christoph and Körtgen, Bastian},
    title = "{Molecular cloud formation by compression of magnetized turbulent gas subjected to radiative cooling}",
    journal = {Monthly Notices of the Royal Astronomical Society},
    volume = {493},
    number = {3},
    pages = {3098-3113},
    year = {2020},
    month = {02},
    issn = {0035-8711},
    doi = {10.1093/mnras/staa468},
    url = {https://doi.org/10.1093/mnras/staa468},
    eprint = {https://academic.oup.com/mnras/article-pdf/493/3/3098/32848623/staa468.pdf},
}

@ARTICLE{Kriel2022_kinematic_dynamo_scales,
       author = {{Kriel}, Neco and {Beattie}, James R. and {Seta}, Amit and {Federrath}, Christoph},
        title = "{Fundamental scales in the kinematic phase of the turbulent dynamo}",
      journal = {The Monthly Notices of The Royal Astronomical Society},
         year = 2022,
        month = jun,
       volume = {513},
       number = {2},
        pages = {2457-2470},
          doi = {10.1093/mnras/stac969},
archivePrefix = {arXiv},
       eprint = {2204.00828},
 primaryClass = {astro-ph.SR},
       adsurl = {https://ui.adsabs.harvard.edu/abs/2022MNRAS.513.2457K}
}

@Article{Hunter2007,
  Author    = {Hunter, J. D.},
  Title     = {Matplotlib: A 2D graphics environment},
  Journal   = {Computing in Science \& Engineering},
  Volume    = {9},
  Number    = {3},
  Pages     = {90--95},
  publisher = {IEEE COMPUTER SOC},
  doi       = {10.1109/MCSE.2007.55},
  year      = 2007
}

@ARTICLE{Virtanen2020,
       author = {{Virtanen}, Pauli and {Gommers}, Ralf and {Oliphant},
         Travis E. and {Haberland}, Matt and {Reddy}, Tyler and
         {Cournapeau}, David and {Burovski}, Evgeni and {Peterson}, Pearu
         and {Weckesser}, Warren and {Bright}, Jonathan and {van der Walt},
         St{\'e}fan J.  and {Brett}, Matthew and {Wilson}, Joshua and
         {Jarrod Millman}, K.  and {Mayorov}, Nikolay and {Nelson}, Andrew
         R.~J. and {Jones}, Eric and {Kern}, Robert and {Larson}, Eric and
         {Carey}, CJ and {Polat}, {\.I}lhan and {Feng}, Yu and {Moore},
         Eric W. and {Vand erPlas}, Jake and {Laxalde}, Denis and
         {Perktold}, Josef and {Cimrman}, Robert and {Henriksen}, Ian and
         {Quintero}, E.~A. and {Harris}, Charles R and {Archibald}, Anne M.
         and {Ribeiro}, Ant{\^o}nio H. and {Pedregosa}, Fabian and
         {van Mulbregt}, Paul and {Contributors}, SciPy 1. 0},
        title = "{SciPy 1.0: Fundamental Algorithms for Scientific
                  Computing in Python}",
      journal = {Nature Methods},
      year = "2020",
      volume={17},
      pages={261--272},
      adsurl = {https://rdcu.be/b08Wh},
      doi = {https://doi.org/10.1038/s41592-019-0686-2},
}

@Misc{Oliphant2006,
  author =    {Travis Oliphant},
  title =     {{NumPy}: A guide to {NumPy}},
  year =      {2006},
  howpublished = {USA: Trelgol Publishing},
  url = "http://www.numpy.org/",
  note = {[Online; accessed <today>]}
 }

@InCollection{Childs2012,
    author = {Hank Childs and Eric Brugger and Brad Whitlock and Jeremy Meredith and Sean Ahern and David Pugmire and Kathleen Biagas and Mark Miller and Cyrus Harrison and Gunther H. Weber and Hari Krishnan and Thomas Fogal and Allen Sanderson and Christoph Garth and E. Wes Bethel and David Camp and Oliver R\"{u}bel and Marc Durant and Jean M. Favre and Paul Navr\'{a}til},
    title = {{VisIt: An End-User Tool For Visualizing and Analyzing Very Large Data}},
    year = "2012",
    pages = "357-372",
    month = "Oct",
    booktitle = {{High Performance Visualization--Enabling Extreme-Scale Scientific Insight}},
    publisher = {Taylor \& Francis}
}

@article{Behnel2011, 
  title={Cython: The best of both worlds}, 
  author={Behnel, Stefan and Bradshaw, Robert and Citro, Craig and Dalcin, Lisandro and Seljebotn, Dag Sverre and Smith, Kurt}, 
  journal={Computing in Science \& Engineering}, 
  volume={13}, 
  number={2}, 
  pages={31--39}, 
  year={2011}, 
  publisher={IEEE} 
}

@ARTICLE{Dhawalikar2022_shock_driving_parameter,
       author = {{Dhawalikar}, Saee and {Federrath}, Christoph and {Davidovits}, Seth and {Teyssier}, Romain and {Nagel}, Sabrina R. and {Remington}, Bruce A. and {Collins}, David C.},
        title = "{The driving mode of shock-driven turbulence}",
      journal = {The Monthly Notices of The Royal Astronomical Society},
         year = 2022,
        month = aug,
       volume = {514},
       number = {2},
        pages = {1782-1800},
          doi = {10.1093/mnras/stac1480},
archivePrefix = {arXiv},
       eprint = {2205.14417},
 primaryClass = {astro-ph.GA},
       adsurl = {https://ui.adsabs.harvard.edu/abs/2022MNRAS.514.1782D}
}

@ARTICLE{Padoan2016_supernova_driving,
       author = {{Padoan}, Paolo and {Pan}, Liubin and {Haugb{\o}lle}, Troels and {Nordlund}, {\r{A}}ke},
        title = "{Supernova Driving. I. The Origin of Molecular Cloud Turbulence}",
      journal = {The Astrophysical Journal},
         year = 2016,
        month = may,
       volume = {822},
       number = {1},
          eid = {11},
        pages = {11},
          doi = {10.3847/0004-637X/822/1/11},
archivePrefix = {arXiv},
       eprint = {1509.04663},
 primaryClass = {astro-ph.GA},
       adsurl = {https://ui.adsabs.harvard.edu/abs/2016ApJ...822...11P}
}

@ARTICLE{Mohapatra2020,
       author = {{Mohapatra}, Rajsekhar and {Federrath}, Christoph and {Sharma}, Prateek},
        title = "{Turbulence in stratified atmospheres: implications for the intracluster medium}",
      journal = {The Monthly Notices of The Royal Astronomical Society},
         year = 2020,
        month = mar,
       volume = {493},
       number = {4},
        pages = {5838-5853},
          doi = {10.1093/mnras/staa711},
archivePrefix = {arXiv},
       eprint = {2001.06494},
 primaryClass = {astro-ph.GA},
       adsurl = {https://ui.adsabs.harvard.edu/abs/2020MNRAS.493.5838M}
}

@article{vanderWalts2014,
 title = {scikit-image: image processing in {P}ython},
 author = {van der Walt, {S}t\'efan and {S}ch\"onberger, {J}ohannes {L}. and
           {Nunez-Iglesias}, {J}uan and {B}oulogne, {F}ran\c{c}ois and {W}arner,
           {J}oshua {D}. and {Y}ager, {N}eil and {G}ouillart, {E}mmanuelle and
           {Y}u, {T}ony and the scikit-image contributors},
 year = {2014},
 month = {6},
 volume = {2},
 pages = {e453},
 journal = {PeerJ},
 issn = {2167-8359},
 url = {https://doi.org/10.7717/peerj.453},
 doi = {10.7717/peerj.453}
}

@book{Stroustrup2013, 
author = {Stroustrup, Bjarne}, 
title = {The C++ Programming Language}, 
year = {2013}, 
isbn = {0321563840}, 
publisher = {Addison-Wesley Professional}, 
edition = {4th} 
}

@ARTICLE{Hew2023_lagrangian_stats,
       author = {{Hew}, Justin Kin Jun and {Federrath}, Christoph},
        title = "{Lagrangian statistics of a shock-driven turbulent dynamo in decaying turbulence}",
      journal = {The Monthly Notices of The Royal Astronomical Society},
         year = 2023,
        month = apr,
       volume = {520},
       number = {4},
        pages = {6268-6282},
          doi = {10.1093/mnras/stad545},
archivePrefix = {arXiv},
       eprint = {2301.06033},
 primaryClass = {astro-ph.GA},
       adsurl = {https://ui.adsabs.harvard.edu/abs/2023MNRAS.520.6268H}
}

@ARTICLE{Bott2021_time_resolved_dynamo,
       author = {{Bott}, Archie F.~A. and {Tzeferacos}, Petros and {Chen}, Laura and {Palmer}, Charlotte A.~J. and {Rigby}, Alexandra and {Bell}, Anthony R. and {Bingham}, Robert and {Birkel}, Andrew and {Graziani}, Carlo and {Froula}, Dustin H. and {Katz}, Joseph and {Koenig}, Michel and {Kunz}, Matthew W. and {Li}, Chikang and {Meinecke}, Jena and {Miniati}, Francesco and {Petrasso}, Richard and {Park}, Hye-Sook and {Remington}, Bruce A. and {Reville}, Brian and {Ross}, J. Steven and {Ryu}, Dongsu and {Ryutov}, Dmitri and {S{\'e}guin}, Fredrick H. and {White}, Thomas G. and {Schekochihin}, Alexander A. and {Lamb}, Donald Q. and {Gregori}, Gianluca},
        title = "{Time-resolved turbulent dynamo in a laser plasma}",
      journal = {Proceedings of the National Academy of Science},
         year = 2021,
        month = mar,
       volume = {118},
       number = {11},
          eid = {e2015729118},
        pages = {e2015729118},
          doi = {10.1073/pnas.2015729118},
archivePrefix = {arXiv},
       eprint = {2007.12837},
 primaryClass = {physics.plasm-ph},
       adsurl = {https://ui.adsabs.harvard.edu/abs/2021PNAS..11815729B}
}

@ARTICLE{Bott2017_proton_imaging_stochastic_fields,
       author = {{Bott}, Archie F.~A. and {Graziani}, Carlo and {Tzeferacos}, Petros and {White}, Thomas G. and {Lamb}, Donald Q. and {Gregori}, Gianluca and {Schekochihin}, Alexander A.},
        title = "{Proton imaging of stochastic magnetic fields}",
      journal = {Journal of Plasma Physics},
         year = 2017,
        month = dec,
       volume = {83},
       number = {6},
          eid = {905830614},
        pages = {905830614},
          doi = {10.1017/S0022377817000939},
archivePrefix = {arXiv},
       eprint = {1708.01738},
 primaryClass = {physics.plasm-ph}
}

@article{numpy2020,
	Author = {Harris, Charles R. and Millman, K. Jarrod and van der Walt, St{\'e}fan J. and Gommers, Ralf and Virtanen, Pauli and Cournapeau, David and Wieser, Eric and Taylor, Julian and Berg, Sebastian and Smith, Nathaniel J. and Kern, Robert and Picus, Matti and Hoyer, Stephan and van Kerkwijk, Marten H. and Brett, Matthew and Haldane, Allan and del R{\'\i}o, Jaime Fern{\'a}ndez and Wiebe, Mark and Peterson, Pearu and G{\'e}rard-Marchant, Pierre and Sheppard, Kevin and Reddy, Tyler and Weckesser, Warren and Abbasi, Hameer and Gohlke, Christoph and Oliphant, Travis E.},
	Da = {2020/09/01},
	Doi = {10.1038/s41586-020-2649-2},
	Id = {Harris2020},
	Isbn = {1476-4687},
	Journal = {Nature},
	Number = {7825},
	Pages = {357--362},
	Title = {Array programming with NumPy},
	Ty = {JOUR},
	Url = {https://doi.org/10.1038/s41586-020-2649-2},
	Volume = {585},
	Year = {2020}
}

@ARTICLE{Menon2020,
       author = {{Menon}, Shyam H. and {Federrath}, Christoph and {Klaassen}, Pamela and {Kuiper}, Rolf and {Reiter}, Megan},
        title = "{On the compressive nature of turbulence driven by ionizing feedback in the pillars of the Carina Nebula}",
      journal = {The Monthly Notices of The Royal Astronomical Society},
         year = 2021,
        month = jan,
       volume = {500},
       number = {2},
        pages = {1721-1740},
          doi = {10.1093/mnras/staa3271},
archivePrefix = {arXiv},
       eprint = {2010.09861},
 primaryClass = {astro-ph.GA},
       adsurl = {https://ui.adsabs.harvard.edu/abs/2021MNRAS.500.1721M}
}

@ARTICLE{Menon2020b,
       author = {{Menon}, Shyam H. and {Federrath}, Christoph and {Kuiper}, Rolf},
        title = "{On the turbulence driving mode of expanding H II regions}",
      journal = {The Monthly Notices of The Royal Astronomical Society},
         year = 2020,
        month = apr,
       volume = {493},
       number = {4},
        pages = {4643-4656},
          doi = {10.1093/mnras/staa580},
archivePrefix = {arXiv},
       eprint = {2002.08707},
 primaryClass = {astro-ph.GA},
       adsurl = {https://ui.adsabs.harvard.edu/abs/2020MNRAS.493.4643M}
}

@ARTICLE{Mohapatra2021,
       author = {{Mohapatra}, Rajsekhar and {Federrath}, Christoph and {Sharma}, Prateek},
        title = "{Turbulent density and pressure fluctuations in the stratified intracluster medium}",
      journal = {The Monthly Notices of The Royal Astronomical Society},
         year = 2021,
        month = jan,
       volume = {500},
       number = {4},
        pages = {5072-5087},
          doi = {10.1093/mnras/staa3564},
archivePrefix = {arXiv},
       eprint = {2010.12602},
 primaryClass = {astro-ph.GA},
       adsurl = {https://ui.adsabs.harvard.edu/abs/2021MNRAS.500.5072M}
}

@ARTICLE{McComb2015_forced_isotropic_dissipation,
       author = {{McComb}, W.~D. and {Berera}, A. and {Yoffe}, S.~R. and {Linkmann}, M.~F.},
        title = "{Energy transfer and dissipation in forced isotropic turbulence}",
      journal = {Physical Review E},
         year = 2015,
        month = apr,
       volume = {91},
       number = {4},
          eid = {043013},
        pages = {043013},
          doi = {10.1103/PhysRevE.91.043013}
}

@ARTICLE{Sreenivasan1998_update_dissipation,
       author = {{Sreenivasan}, K.~R.},
        title = "{An update on the energy dissipation rate in isotropic turbulence}",
      journal = {Physics of Fluids},
         year = 1998,
        month = feb,
       volume = {10},
       number = {2},
        pages = {528-529},
          doi = {10.1063/1.869575}
}

@ARTICLE{BuariaSreenivasan2020_dissipation_range_spectrum,
       author = {{Buaria}, Dhawal and {Sreenivasan}, Katepalli R.},
        title = "{Dissipation range of the energy spectrum in high Reynolds number turbulence}",
      journal = {Physical Review Fluids},
         year = 2020,
        month = sep,
       volume = {5},
       number = {9},
          eid = {092601},
        pages = {092601},
          doi = {10.1103/PhysRevFluids.5.092601}
}

@INPROCEEDINGS{PullinRogallo1994_pressure_higher_order_spectra,
       author = {{Pullin}, D.~I. and {Rogallo}, R.~S.},
        title = "{Pressure and higher-order spectra for homogeneous isotropic turbulence}",
    booktitle = {Studying Turbulence Using Numerical Simulation Databases V: Proceedings of the 1994 Summer Program},
         year = 1994,
        month = dec,
       publisher = {Center for Turbulence Research, Stanford University},
          note = {NASA Technical Reports Server document 19950014628}
}

@ARTICLE{Schmidt2021_CGM_turbulence,
       author = {{Schmidt}, W. and {Schmidt}, J.~P. and {Grete}, P.},
        title = "{Turbulence in the intragroup and circumgalactic medium}",
      journal = {Astronomy and Astrophysics},
         year = 2021,
        month = oct,
       volume = {654},
          eid = {A115},
        pages = {A115},
          doi = {10.1051/0004-6361/202140920},
archivePrefix = {arXiv},
       eprint = {2107.12125},
 primaryClass = {astro-ph.GA},
       adsurl = {https://ui.adsabs.harvard.edu/abs/2021A&A...654A.115S}
}

@ARTICLE{Chen2023_QSO_CGM_turbulence,
       author = {{Chen}, Mandy C. and {Chen}, Hsiao-Wen and {Rauch}, Michael and {Qu}, Zhijie and {Johnson}, Sean D. and {Li}, Jean I.~H. and {Schaye}, Joop and {Rudie}, Gwen C. and {Zahedy}, Fakhri S. and {Boettcher}, Erin and {Cooksey}, Kathy L. and {Cantalupo}, Sebastiano},
        title = "{Empirical constraints on the turbulence in QSO host nebulae from velocity structure function measurements}",
      journal = {\mnras},
         year = 2023,
        month = jan,
       volume = {518},
       number = {2},
        pages = {2354-2372},
          doi = {10.1093/mnras/stac3193},
archivePrefix = {arXiv},
       eprint = {2210.10057},
 primaryClass = {astro-ph.GA},
       adsurl = {https://ui.adsabs.harvard.edu/abs/2023MNRAS.518.2354C}
}

@ARTICLE{Kritsuk2011,
       author = {{Kritsuk}, Alexei G. and {Nordlund}, {\r{A}}ke and {Collins}, David and {Padoan}, Paolo and {Norman}, Michael L. and {Abel}, Tom and {Banerjee}, Robi and {Federrath}, Christoph and {Flock}, Mario and {Lee}, Dongwook and {Li}, Pak Shing and {M{\"u}ller}, Wolf-Christian and {Teyssier}, Romain and {Ustyugov}, Sergey D. and {Vogel}, Christian and {Xu}, Hao},
        title = "{Comparing Numerical Methods for Isothermal Magnetized Supersonic Turbulence}",
      journal = {The Astrophysical Journal},
         year = 2011,
        month = aug,
       volume = {737},
       number = {1},
          eid = {13},
        pages = {13},
          doi = {10.1088/0004-637X/737/1/13},
archivePrefix = {arXiv},
       eprint = {1103.5525},
 primaryClass = {astro-ph.SR},
       adsurl = {https://ui.adsabs.harvard.edu/abs/2011ApJ...737...13K}
}

@ARTICLE{Armstrong1995_power_law,
       author = {{Armstrong}, J.~W. and {Rickett}, B.~J. and {Spangler}, S.~R.},
        title = "{Electron Density Power Spectrum in the Local Interstellar Medium}",
      journal = {The Astrophysical Journal},
         year = 1995,
        month = apr,
       volume = {443},
        pages = {209},
          doi = {10.1086/175515},
       adsurl = {https://ui.adsabs.harvard.edu/abs/1995ApJ...443..209A}
}

@ARTICLE{Grete2023_as_a_matter_of_dynamical_range,
       author = {{Grete}, Philipp and {O'Shea}, Brian W. and {Beckwith}, Kris},
        title = "{As a Matter of Dynamical Range - Scale Dependent Energy Dynamics in MHD Turbulence}",
      journal = {The Astrophysical Journal Letters},
         year = 2023,
        month = jan,
       volume = {942},
       number = {2},
          eid = {L34},
        pages = {L34},
          doi = {10.3847/2041-8213/acaea7},
archivePrefix = {arXiv},
       eprint = {2211.09750},
 primaryClass = {astro-ph.GA},
       adsurl = {https://ui.adsabs.harvard.edu/abs/2023ApJ...942L..34G}
}

@ARTICLE{Beattie2022_spdf,
       author = {{Beattie}, James R. and {Mocz}, Philip and {Federrath}, Christoph and {Klessen}, Ralf S.},
        title = "{The density distribution and physical origins of intermittency in supersonic, highly magnetized turbulence with diverse modes of driving}",
      journal = {The Monthly Notices of The Royal Astronomical Society},
         year = 2022,
        month = dec,
       volume = {517},
       number = {4},
        pages = {5003-5031},
          doi = {10.1093/mnras/stac3005},
archivePrefix = {arXiv},
       eprint = {2109.10470},
 primaryClass = {astro-ph.GA},
       adsurl = {https://ui.adsabs.harvard.edu/abs/2022MNRAS.517.5003B}
}

@ARTICLE{Field1969,
       author = {{Field}, G.~B. and {Goldsmith}, D.~W. and {Habing}, H.~J.},
        title = "{Cosmic-Ray Heating of the Interstellar Gas}",
      journal = {The Astrophysical Journal Letters},
         year = 1969,
        month = mar,
       volume = {155},
        pages = {L149},
          doi = {10.1086/180324},
       adsurl = {https://ui.adsabs.harvard.edu/abs/1969ApJ...155L.149F}
}

@article{Hosking2021_reconnection_controlled_decay,
  title = {Reconnection-Controlled Decay of Magnetohydrodynamic Turbulence and the Role of Invariants},
  author = {Hosking, David N. and Schekochihin, Alexander A.},
  journal = {Phys. Rev. X},
  volume = {11},
  issue = {4},
  pages = {041005},
  numpages = {31},
  year = {2021},
  month = {Oct},
  publisher = {American Physical Society},
  doi = {10.1103/PhysRevX.11.041005},
  url = {https://link.aps.org/doi/10.1103/PhysRevX.11.041005}
}

@ARTICLE{Korpi1999_supernova_regulated_ISM,
       author = {{Korpi}, M.~J. and {Brandenburg}, A. and {Shukurov}, A. and {Tuominen}, I. and {Nordlund}, {\r{A}}.},
        title = "{A Supernova-regulated Interstellar Medium: Simulations of the Turbulent Multiphase Medium}",
      journal = {The Astrophysical Journal Letters},
         year = 1999,
        month = apr,
       volume = {514},
       number = {2},
        pages = {L99-L102},
          doi = {10.1086/311954},
       adsurl = {https://ui.adsabs.harvard.edu/abs/1999ApJ...514L..99K}
}

@ARTICLE{Hu2022_shock_amplification,
       author = {{Hu}, Yue and {Xu}, Siyao and {Stone}, James M. and {Lazarian}, Alex},
        title = "{Turbulent Magnetic Field Amplification by the Interaction of a Shock Wave and Inhomogeneous Medium}",
      journal = {The Astrophysical Journal},
         year = 2022,
        month = dec,
       volume = {941},
       number = {2},
          eid = {133},
        pages = {133},
          doi = {10.3847/1538-4357/ac9ebc},
archivePrefix = {arXiv},
       eprint = {2207.06941},
 primaryClass = {astro-ph.GA},
       adsurl = {https://ui.adsabs.harvard.edu/abs/2022ApJ...941..133H}
}

@ARTICLE{Gaesnsler_2011_trans_ISM,
       author = {{Gaensler}, B.~M. and {Haverkorn}, M. and {Burkhart}, B. and {Newton-McGee}, K.~J. and {Ekers}, R.~D. and {Lazarian}, A. and {McClure-Griffiths}, N.~M. and {Robishaw}, T. and {Dickey}, J.~M. and {Green}, A.~J.},
        title = "{Low-Mach-number turbulence in interstellar gas revealed by radio polarization gradients}",
      journal = {\nat},
         year = 2011,
        month = oct,
       volume = {478},
       number = {7368},
        pages = {214-217},
          doi = {10.1038/nature10446},
archivePrefix = {arXiv},
       eprint = {1110.2896},
 primaryClass = {astro-ph.GA},
       adsurl = {https://ui.adsabs.harvard.edu/abs/2011Natur.478..214G}
}

@ARTICLE{Beattie2025_so_long_kolmogorov,
       author = {{Beattie}, James R. and {Noer Kolborg}, Anne and {Ramirez-Ruiz}, Enrico and {Federrath}, Christoph},
        title = "{So long Kolmogorov: the forward and backward turbulence cascades in a supernovae-driven, multiphase interstellar medium}",
      journal = {arXiv e-prints},
         year = 2025,
        month = jan,
          eid = {arXiv:2501.09855},
        pages = {arXiv:2501.09855},
          doi = {10.48550/arXiv.2501.09855},
archivePrefix = {arXiv},
       eprint = {2501.09855},
 primaryClass = {astro-ph.GA},
       adsurl = {https://ui.adsabs.harvard.edu/abs/2025arXiv250109855B}
}

@ARTICLE{Semenov2025_subgrid_turbulence,
       author = {{Semenov}, Vadim A.},
        title = "{Capturing Turbulence with Numerical Dissipation: A Simple Dynamical Model for Unresolved Turbulence in Hydrodynamic Simulations}",
      journal = {\apjs},
         year = 2025,
        month = dec,
       volume = {281},
       number = {2},
          eid = {37},
        pages = {37},
          doi = {10.3847/1538-4365/ae0cc6},
archivePrefix = {arXiv},
       eprint = {2410.23339},
 primaryClass = {astro-ph.GA},
       adsurl = {https://ui.adsabs.harvard.edu/abs/2025ApJS..281...37S}
}

@ARTICLE{Teissier2024_higher_order_schemes,
       author = {{Teissier}, Jean-Mathieu and {M{\"a}usle}, Raquel and {M{\"u}ller}, Wolf-Christian},
        title = "{Cost-efficient finite-volume high-order schemes for compressible magnetohydrodynamics}",
      journal = {Journal of Computational Physics},
         year = 2024,
        month = oct,
       volume = {515},
          eid = {113287},
        pages = {113287},
          doi = {10.1016/j.jcp.2024.113287},
archivePrefix = {arXiv},
       eprint = {2306.09856},
 primaryClass = {math.NA},
       adsurl = {https://ui.adsabs.harvard.edu/abs/2024JCoPh.51513287T}
}

@ARTICLE{Grete2015_supersonic_closures,
       author = {{Grete}, Philipp and {Vlaykov}, Dimitar G. and {Schmidt}, Wolfram and {Schleicher}, Dominik R.~G. and {Federrath}, Christoph},
        title = "{Nonlinear closures for scale separation in supersonic magnetohydrodynamic turbulence}",
      journal = {New Journal of Physics},
         year = 2015,
        month = feb,
       volume = {17},
       number = {2},
          eid = {023070},
        pages = {023070},
          doi = {10.1088/1367-2630/17/2/023070},
archivePrefix = {arXiv},
       eprint = {1501.07170},
 primaryClass = {physics.flu-dyn},
       adsurl = {https://ui.adsabs.harvard.edu/abs/2015NJPh...17b3070G}
}

@ARTICLE{Grehan2025_numerical_reconnection,
       author = {{Grehan}, Michael P. and {Ghosal}, Tanisha and {Beattie}, James R. and {Ripperda}, Bart and {Porth}, Oliver and {Bacchini}, Fabio},
        title = "{Comparison of magnetic diffusion and reconnection in ideal and resistive relativistic magnetohydrodynamics, ideal magnetodynamics and resistive force-free electrodynamics}",
      journal = {arXiv e-prints},
         year = 2025,
        month = mar,
          eid = {arXiv:2503.20013},
        pages = {arXiv:2503.20013},
          doi = {10.48550/arXiv.2503.20013},
archivePrefix = {arXiv},
       eprint = {2503.20013},
 primaryClass = {astro-ph.HE},
       adsurl = {https://ui.adsabs.harvard.edu/abs/2025arXiv250320013G}
}

@ARTICLE{Shivakumar2025_numerical_diffusion,
       author = {{Shivakumar}, Lakshmi Malvadi and {Federrath}, Christoph},
        title = "{Numerical viscosity and resistivity in MHD turbulence simulations}",
      journal = {\mnras},
         year = 2025,
        month = mar,
       volume = {537},
       number = {4},
        pages = {2961-2986},
          doi = {10.1093/mnras/staf160},
archivePrefix = {arXiv},
       eprint = {2311.10350},
 primaryClass = {astro-ph.SR},
       adsurl = {https://ui.adsabs.harvard.edu/abs/2025MNRAS.537.2961S}
}

@misc{joblib,
title = {Joblib: running Python functions as pipeline jobs},
author = {{Joblib Development Team}},
year = {2020},
url = {https://joblib.readthedocs.io/},
}

@misc{pandas,
    author       = {The pandas development team},
    title        = {pandas-dev/pandas: Pandas},
    month        = dec,
    year         = 2023,
    publisher    = {Zenodo},
    version      = {2.1.4},
    doi          = {10.5281/zenodo.10304236},
    url          = {https://doi.org/10.5281/zenodo.10304236}
}

@ARTICLE{yt,
       author = {{Turk}, Matthew J. and {Smith}, Britton D. and {Oishi}, Jeffrey S. and {Skory}, Stephen and {Skillman}, Samuel W. and {Abel}, Tom and {Norman}, Michael L.},
        title = "{yt: A Multi-code Analysis Toolkit for Astrophysical Simulation Data}",
      journal = {\apjs},
         year = 2011,
        month = jan,
       volume = {192},
       number = {1},
          eid = {9},
        pages = {9},
          doi = {10.1088/0067-0049/192/1/9},
archivePrefix = {arXiv},
       eprint = {1011.3514},
 primaryClass = {astro-ph.IM},
       adsurl = {https://ui.adsabs.harvard.edu/abs/2011ApJS..192....9T}
}

@ARTICLE{Beattie2025_small_scale_instabilities,
       author = {{Beattie}, James R.},
        title = "{Supernovae drive large-scale, incompressible turbulence through small-scale instabilities}",
      journal = {arXiv e-prints},
         year = 2025,
        month = sep,
          eid = {arXiv:2509.07354},
        pages = {arXiv:2509.07354},
          doi = {10.48550/arXiv.2509.07354},
archivePrefix = {arXiv},
       eprint = {2509.07354},
 primaryClass = {astro-ph.GA},
       adsurl = {https://ui.adsabs.harvard.edu/abs/2025arXiv250907354B}
}

@misc{OpenAI2025_Codex,
  author = {{OpenAI}},
  title = {Introducing Codex},
  year = {2025},
  month = may,
  url = {https://openai.com/index/introducing-codex/},
  note = {Accessed 25 April 2026}
}

@ARTICLE{Draine1993_ISM_shocks,
       author = {{Draine}, Bruce T. and {McKee}, Christopher F.},
        title = "{Theory of interstellar shocks.}",
      journal = {\araa},
         year = 1993,
        month = jan,
       volume = {31},
        pages = {373-432},
          doi = {10.1146/annurev.aa.31.090193.002105},
       adsurl = {https://ui.adsabs.harvard.edu/abs/1993ARA&A..31..373D}
}

@ARTICLE{Beattie2025_nature_astro,
       author = {{Beattie}, James R. and {Federrath}, Christoph and {Klessen}, Ralf S. and {Cielo}, Salvatore and {Bhattacharjee}, Amitava},
        title = "{The spectrum of magnetized turbulence in the interstellar medium}",
      journal = {Nature Astronomy},
         year = 2025,
        month = aug,
       volume = {9},
        pages = {1195-1205},
          doi = {10.1038/s41550-025-02551-5},
archivePrefix = {arXiv},
       eprint = {2504.07136},
 primaryClass = {astro-ph.GA},
       adsurl = {https://ui.adsabs.harvard.edu/abs/2025NatAs...9.1195B}
}

@ARTICLE{Gent2021_supernova_turbulence_and_dynamo,
       author = {{Gent}, Frederick A. and {Mac Low}, Mordecai-Mark and {K{\"a}pyl{\"a}}, Maarit J. and {Singh}, Nishant K.},
        title = "{Small-scale Dynamo in Supernova-driven Interstellar Turbulence}",
      journal = {\apjl},
         year = 2021,
        month = apr,
       volume = {910},
       number = {2},
          eid = {L15},
        pages = {L15},
          doi = {10.3847/2041-8213/abed59},
archivePrefix = {arXiv},
       eprint = {2010.01833},
 primaryClass = {astro-ph.GA},
       adsurl = {https://ui.adsabs.harvard.edu/abs/2021ApJ...910L..15G}
}

@misc{halliday_investigating_2022,
    title = {Investigating radiatively driven, magnetised plasmas with a university scale pulsed-power generator},
    url = {http://arxiv.org/abs/2203.11881},
    language = {en},
    urldate = {2024-01-24},
    publisher = {arXiv},
    author = {Halliday, Jack W. D. and Crilly, Aidan and Chittenden, Jeremy and Mancini, Roberto C. and Merlini, Stefano and Rose, Steven and Russell, Danny R. and Suttle, Lee G. and Valenzuela-Villaseca, Vicente and Bland, Simon N. and Lebedev, Sergey V.},
    month = mar,
    year = {2022},
    note = {arXiv:2203.11881 [physics]},
}

@article{Ishida2006_decay_isotropic,
  author = {Ishida, T. and Davidson, P. A. and Kaneda, Y.},
  title = {On the decay of isotropic turbulence},
  journal = {Journal of Fluid Mechanics},
  volume = {564},
  pages = {455--475},
  year = {2006},
  doi = {10.1017/S0022112006001625}
}

@article{KrogstadDavidson2010_grid,
  author = {Krogstad, P.-{\AA}. and Davidson, P. A.},
  title = {Is grid turbulence {Saffman} turbulence?},
  journal = {Journal of Fluid Mechanics},
  volume = {642},
  pages = {373--394},
  year = {2010},
  doi = {10.1017/S0022112009991807}
}

@article{GorceFalcon2024_saffman,
  author = {Gorce, Jean-Baptiste and Falcon, Eric},
  title = {Freely Decaying {Saffman} Turbulence Experimentally Generated by Magnetic Stirrers},
  journal = {Physical Review Letters},
  volume = {132},
  pages = {264001},
  year = {2024},
  doi = {10.1103/PhysRevLett.132.264001}
}

@article{MacLow1998_decay_supersonic,
  author = {Mac Low, Mordecai-Mark and Klessen, Ralf S. and Burkert, Andreas and Smith, Michael D.},
  title = {Kinetic Energy Decay Rates of Supersonic and Super-Alfv\'enic Turbulence in Star-Forming Clouds},
  journal = {Physical Review Letters},
  volume = {80},
  number = {13},
  pages = {2754--2757},
  year = {1998},
  doi = {10.1103/PhysRevLett.80.2754}
}

@article{Samtaney2001_decaying_compressible,
  author = {Samtaney, Ravi and Pullin, D. I. and Kosovi{\'c}, Branko},
  title = {Direct numerical simulation of decaying compressible turbulence and shocklet statistics},
  journal = {Physics of Fluids},
  volume = {13},
  number = {5},
  pages = {1415--1430},
  year = {2001},
  doi = {10.1063/1.1355682}
}

@ARTICLE{Brandenburg2023_dissipation_MHD,
       author = {{Brandenburg}, Axel and {Rogachevskii}, Igor and {Schober}, Jennifer},
        title = "{Dissipative magnetic structures and scales in small-scale dynamos}",
      journal = {\mnras},
         year = 2023,
        month = feb,
       volume = {518},
       number = {4},
        pages = {6367-6375},
          doi = {10.1093/mnras/stac3555},
archivePrefix = {arXiv},
       eprint = {2209.08717},
 primaryClass = {astro-ph.GA},
       adsurl = {https://ui.adsabs.harvard.edu/abs/2023MNRAS.518.6367B}
}

@ARTICLE{Connor2026_sn_turb,
       author = {{Connor}, Isabelle and {Beattie}, James R. and {Kolborg}, Anne Noer and {Ramirez-Ruiz}, Enrico},
        title = "{Cascading from the Winds to the Disk: The Universality of Supernovae-driven Turbulence in Different Galactic Interstellar Media}",
      journal = {\apj},
         year = 2026,
        month = jan,
       volume = {997},
       number = {1},
          eid = {33},
        pages = {33},
          doi = {10.3847/1538-4357/ae17b1},
archivePrefix = {arXiv},
       eprint = {2509.01653},
 primaryClass = {astro-ph.GA},
      adsurl = {https://ui.adsabs.harvard.edu/abs/2026ApJ...997...33C}
}

@ARTICLE{Stinebring2006_scintillation_arcs,
  author = {{Stinebring}, Daniel R.},
  title = "{Scintillation Arcs: Probing Turbulence and Structure in the ISM}",
  journal = {Chinese Journal of Astronomy and Astrophysics Supplement},
  year = {2006},
  volume = {6},
  pages = {204--214}
}

@ARTICLE{Ocker2021_bow_shock_turbulence,
  author = {{Ocker}, Stella Koch and {Cordes}, James M. and {Chatterjee}, Shami and {Dolch}, Timothy},
  title = "{An In Situ Study of Turbulence near Stellar Bow Shocks}",
  journal = {\apj},
  year = {2021},
  month = dec,
  volume = {922},
  number = {2},
  pages = {233},
  doi = {10.3847/1538-4357/ac2b28}
}

@ARTICLE{Ocker2025_CGM_turbulence,
  author = {{Ocker}, Stella Koch and {Chen}, Mandy C. and {Oh}, S. Peng and {Sharma}, Prateek},
  title = "{Microphysics of Circumgalactic Turbulence Probed by Fast Radio Bursts and Quasars}",
  journal = {\apj},
  year = {2025},
  volume = {988},
  pages = {69},
  doi = {10.3847/1538-4357/ade0bc},
  archivePrefix = {arXiv},
  eprint = {2503.02329},
  primaryClass = {astro-ph.GA}
}

@ARTICLE{Kempski2025ApJL,
  author = {{Kempski}, Philipp and {Li}, Dongzi and {Fielding}, Drummond B. and {Quataert}, Eliot and {Phinney}, E. Sterl and {Kunz}, Matthew W. and {Jow}, Dylan L. and {Philippov}, Alexander A.},
  title = "{A Unified Model of Cosmic-Ray Propagation and Radio Extreme Scattering Events from Intermittent Interstellar Structures}",
  journal = {\apjl},
  year = {2025},
  volume = {990},
  number = {1},
  pages = {L18},
  doi = {10.3847/2041-8213/adde55}
}

@ARTICLE{Jow2022MNRAS,
  author = {{Jow}, Dylan L. and {Pen}, Ue-Li},
  title = "{Measuring Lens Dimensionality in Extreme Scattering Events through Wave Optics}",
  journal = {\mnras},
  year = {2022},
  volume = {514},
  number = {3},
  pages = {4069--4077},
  doi = {10.1093/mnras/stac1652},
  archivePrefix = {arXiv},
  eprint = {2110.07119},
  primaryClass = {astro-ph.HE}
}

@ARTICLE{Jow2024MNRAS_cusp,
  author = {{Jow}, Dylan L. and {Pen}, Ue-Li and {Baker}, Daniel},
  title = "{On the Cusp of Cusps: A Universal Model for Extreme Scattering Events in the ISM}",
  journal = {\mnras},
  year = {2024},
  month = mar,
  volume = {528},
  number = {4},
  pages = {6292--6301},
  doi = {10.1093/mnras/stae300},
  archivePrefix = {arXiv},
  eprint = {2301.08344},
  primaryClass = {astro-ph.HE}
}

@ARTICLE{Pen2014MNRAS,
  author = {{Pen}, Ue-Li and {Levin}, Yuri},
  title = "{Pulsar Scintillations from Corrugated Reconnection Sheets in the Interstellar Medium}",
  journal = {\mnras},
  year = {2014},
  volume = {442},
  number = {4},
  pages = {3338--3346},
  doi = {10.1093/mnras/stu1020},
  archivePrefix = {arXiv},
  eprint = {1302.1897},
  primaryClass = {astro-ph.GA}
}

@ARTICLE{Pen2012MNRAS,
  author = {{Pen}, Ue-Li and {King}, Lindsay},
  title = "{Refractive Convergent Plasma Lenses Explain Extreme Scattering Events and Pulsar Scintillation}",
  journal = {Monthly Notices of the Royal Astronomical Society: Letters},
  year = {2012},
  volume = {421},
  number = {1},
  pages = {L132--L136},
  doi = {10.1111/j.1745-3933.2012.01223.x}
}

@article{Folini_2006,
  title = {Supersonic Turbulence in Shock-Bound Interaction Zones: {{I}}. {{Symmetric}} Settings},
  shorttitle = {Supersonic Turbulence in Shock-Bound Interaction Zones},
  author = {Folini, D. and Walder, R.},
  year = 2006,
  month = nov,
  journal = {Astronomy \& Astrophysics},
  volume = {459},
  number = {1},
  pages = {1--19},
  issn = {0004-6361, 1432-0746},
  doi = {10.1051/0004-6361:20053898},
  urldate = {2026-04-01},
  langid = {english}
}

@article{Markwick_2021,
  title = {Cooling and Instabilities in Colliding Flows},
  author = {Markwick, R N and Frank, A and {Carroll-Nellenback}, J and Liu, B and Blackman, E G and Lebedev, S V and Hartigan, P M},
  year = 2021,
  month = oct,
  journal = {Monthly Notices of the Royal Astronomical Society},
  volume = {508},
  number = {2},
  pages = {2266--2278},
  issn = {0035-8711, 1365-2966},
  doi = {10.1093/mnras/stab2577},
  urldate = {2026-04-01},
  copyright = {https://academic.oup.com/journals/pages/open\_access/funder\_policies/chorus/standard\_publication\_model},
  langid = {english}
}

@misc{Merlini_2023, 
    title={Structure of accretion shocks and radiative cooling effects in high energy density plasma experiments},
    url={http://hdl.handle.net/10044/1/110790}, 
    DOI={https://doi.org/10.25560/110790}, 
    abstractNote={Accretion shocks are ubiquitous phenomena in many astrophysical systems which can be significantly affected by radiative cooling effects, leading to instability and turbulence. This thesis presents a study of accretion shock experiments at the MAGPIE pulsed power facility. Two methods are used to produce reverse shocks in the laboratory. The first investigates the structure of reverse shocks resulting from the collision of supersonic, magnetised plasma flows generated by an inverse wire array interacting with a planar conducting obstacle, whereas the second approach consists of the ablation by X-ray radiation from a Z-pinch wire array of solid targets. In the planar obstacle experiments, variations in the reverse shock structure are observed depending on the wire material used, despite similar upstream flow velocities and mass densities. Specifically, when aluminium wire arrays are employed, a well-defined, sharp shock is formed that aligns with magneto-hydrodynamic theory. However, in the case of tungsten wires, a distinct stand-off shock is not observed, instead, a broad region ahead of the obstacle is characterised by density fluctuations spanning a wide range of spatial scales. These two contrasty interactions are diagnosed with laser interferometry, Thomson scattering, shadowgraphy, and a newly developed imaging refractometer, enabling to characterise the small-scale density perturbations by detecting deflections of the probing laser. These measurements suggest that the differences in shock structure are most likely due to radiative cooling effects which give rise to density perturbations elongated along magnetic field lines. In the second experimental campaign, reverse shocks formed from the collision of counter-streaming supersonic plasma flows are investigated. The properties of the shocked layer were determined using various laser diagnostics and optical self-emission images, revealing overall consistency with a 1-D accretion shock model for a value of γ ≤ 1.2. In addition, the capability of the X-ray-driven platform to easily change the B-field orientation allowed us to explore the case for colliding flows in a transverse magnetic field, revealing promising results for investigating new astrophysical systems in the laboratory.}, 
    author={Merlini, Stefano}, 
    year={2023}, 
    month={Oct}
}

@article{MerliniRadiativeCoolingEffects_2023,
	title = {Radiative cooling effects on reverse shocks formed by magnetized supersonic plasma flows},
	volume = {30},
	issn = {1070-664X, 1089-7674},
	url = {https://pubs.aip.org/pop/article/30/9/092102/2909321/Radiative-cooling-effects-on-reverse-shocks-formed},
	doi = {10.1063/5.0160809},
	language = {en},
	number = {9},
	urldate = {2024-01-24},
	journal = {Physics of Plasmas},
	author = {Merlini, S. and Hare, J. D. and Burdiak, G. C. and Halliday, J. W. D. and Ciardi, A. and Chittenden, J. P. and Clayson, T. and Crilly, A. J. and Eardley, S. J. and Marrow, K. E. and Russell, D. R. and Smith, R. A. and Stuart, N. and Suttle, L. G. and Tubman, E. R. and Valenzuela-Villaseca, V. and Varnish, T. W. O. and Lebedev, S. V.},
	month = sep,
	year = {2023},
	pages = {092102},
    }

@article{Lebedev_2001_wire_array,
  author = {Lebedev, S. V. and Beg, F. N. and Bland, S. N. and Chittenden, J. P. and Dangor, A. E. and Haines, M. G. and Kwek, K. H. and Pikuz, S. A. and Shelkovenko, T. A.},
  title = {Effect of discrete wires on the implosion dynamics of wire array {Z} pinches},
  journal = {Physics of Plasmas},
  volume = {8},
  number = {8},
  pages = {3734--3747},
  year = {2001},
  doi = {10.1063/1.1385373}
}

@article{Bland_2006_diagnostics,
  author = {Bland, S. N. and Bott, S. C. and Hall, G. N. and Lebedev, S. V. and Suzuki, F. and Ampleford, D. J. and Palmer, J. B. A. and Pikuz, S. A. and Shelkovenko, T. A.},
  title = {Diagnostics for studying the dynamics of wire array {Z} pinches},
  journal = {Review of Scientific Instruments},
  volume = {77},
  pages = {10F326},
  year = {2006}
}

@article{Marrow_2026,
	title = {Measuring dynamics of differentially rotating, unmagnetized, free-boundary plasma produced by soft x-ray irradiation},
	volume = {68},
	issn = {0741-3335, 1361-6587},
	url = {https://iopscience.iop.org/article/10.1088/1361-6587/ae44cc},
	doi = {10.1088/1361-6587/ae44cc},
	language = {en},
	number = {2},
	urldate = {2026-03-28},
	journal = {Plasma Physics and Controlled Fusion},
	author = {Marrow, K E and Strucka, J and Merlini, S and Suttle, L G and Mundy, T R and Uras, A and Bland, S N and Valenzuela-Villaseca, V and Lebedev, S V},
	month = feb,
	year = {2026},
	pages = {025027},
}

@article{Braginskii_1965,
       author = {{Braginskii}, S.~I.},
        title = "{Transport Processes in a Plasma}",
      journal = {Reviews of Plasma Physics},
         year = 1965,
        month = jan,
       volume = {1},
        pages = {205},
       adsurl = {https://ui.adsabs.harvard.edu/abs/1965RvPP....1..205B}
}

@article{Ryutov_1999,
  title = {Similarity Criteria for the Laboratory Simulation of Supernova Hydrodynamics},
  author = {Ryutov, D. and Drake, R. P. and Kane, J. and Liang, E. and Remington, B. A. and {Wood-Vasey}, W. M.},
  year = 1999,
  month = jun,
  journal = {The Astrophysical Journal},
  volume = {518},
  number = {2},
  pages = {821},
  doi = {10.1086/307293}
}

@article{gittings_RAGE_Code_2008,
  title = {The {{RAGE}} Radiation-Hydrodynamic Code},
  author = {Gittings, Michael and Weaver, Robert and Clover, Michael and Betlach, Thomas and Byrne, Nelson and Coker, Robert and Dendy, Edward and Hueckstaedt, Robert and New, Kim and Oakes, W Rob and Ranta, Dale and Stefan, Ryan},
  year = 2008,
  month = nov,
  journal = {Computational Science \& Discovery},
  volume = {1},
  number = {1},
  pages = {015005},
  issn = {1749-4699},
  doi = {10.1088/1749-4699/1/1/015005},
  urldate = {2026-01-21},
  langid = {english}
}

@article{tzeferacos_FLASHMHDSimulations_2015,
  title = {{{FLASH MHD}} Simulations of Experiments That Study Shock-Generated Magnetic Fields},
  author = {Tzeferacos, P. and Fatenejad, M. and Flocke, N. and Graziani, C. and Gregori, G. and Lamb, D.Q. and Lee, D. and Meinecke, J. and Scopatz, A. and Weide, K.},
  year = {2015},
  month = dec,
  journal = {High Energy Density Physics},
  volume = {17},
  pages = {24--31},
  issn = {15741818},
  doi = {10.1016/j.hedp.2014.11.003},
  urldate = {2026-01-20},
  langid = {english}
}

@article{macfarlane_HELIOSCR1DRadiation_2006,
  title = {{{HELIOS-CR}} – {{A}} 1-{{D}} Radiation-Magnetohydrodynamics Code with Inline Atomic Kinetics Modeling},
  author = {MacFarlane, J.J. and Golovkin, I.E. and Woodruff, P.R.},
  year = {2006},
  journal = {Journal of Quantitative Spectroscopy and Radiative Transfer},
  shortjournal = {Journal of Quantitative Spectroscopy and Radiative Transfer},
  volume = {99},
  number = {1--3},
  pages = {381--397},
  issn = {00224073},
  doi = {10.1016/j.jqsrt.2005.05.031},
  url = {https://linkinghub.elsevier.com/retrieve/pii/S0022407305001627},
  urldate = {2026-01-20},
  langid = {english}
}

@article{tzeferacos_laboratory_2018,
	title = {Laboratory evidence of dynamo amplification of magnetic fields in a turbulent plasma},
	volume = {9},
	issn = {2041-1723},
	url = {https://www.nature.com/articles/s41467-018-02953-2},
	doi = {10.1038/s41467-018-02953-2},
	language = {en},
	number = {1},
	urldate = {2026-01-16},
	journal = {Nature Communications},
	author = {Tzeferacos, P. and Rigby, A. and Bott, A. F. A. and Bell, A. R. and Bingham, R. and Casner, A. and Cattaneo, F. and Churazov, E. M. and Emig, J. and Fiuza, F. and Forest, C. B. and Foster, J. and Graziani, C. and Katz, J. and Koenig, M. and Li, C.-K. and Meinecke, J. and Petrasso, R. and Park, H.-S. and Remington, B. A. and Ross, J. S. and Ryu, D. and Ryutov, D. and White, T. G. and Reville, B. and Miniati, F. and Schekochihin, A. A. and Lamb, D. Q. and Froula, D. H. and Gregori, G.},
	month = feb,
	year = {2018},
	pages = {591},
}

@article{tzeferacos_numerical_2017,
	title = {Numerical modeling of laser-driven experiments aiming to demonstrate magnetic field amplification via turbulent dynamo},
	volume = {24},
	issn = {1070-664X, 1089-7674},
	url = {https://pubs.aip.org/pop/article/24/4/041404/110295/Numerical-modeling-of-laser-driven-experiments},
	doi = {10.1063/1.4978628},
	language = {en},
	number = {4},
	urldate = {2026-01-16},
	journal = {Physics of Plasmas},
	author = {Tzeferacos, P. and Rigby, A. and Bott, A. and Bell, A. R. and Bingham, R. and Casner, A. and Cattaneo, F. and Churazov, E. M. and Emig, J. and Flocke, N. and Fiuza, F. and Forest, C. B. and Foster, J. and Graziani, C. and Katz, J. and Koenig, M. and Li, C.-K. and Meinecke, J. and Petrasso, R. and Park, H.-S. and Remington, B. A. and Ross, J. S. and Ryu, D. and Ryutov, D. and Weide, K. and White, T. G. and Reville, B. and Miniati, F. and Schekochihin, A. A. and Froula, D. H. and Gregori, G. and Lamb, D. Q.},
	month = apr,
	year = {2017},
	pages = {041404},
}

@article{meinecke_developed_2015,
	title = {Developed turbulence and nonlinear amplification of magnetic fields in laboratory and astrophysical plasmas},
	volume = {112},
	issn = {0027-8424, 1091-6490},
	url = {https://pnas.org/doi/full/10.1073/pnas.1502079112},
	doi = {10.1073/pnas.1502079112},
	language = {en},
	number = {27},
	urldate = {2026-01-16},
	journal = {Proceedings of the National Academy of Sciences},
	author = {Meinecke, Jena and Tzeferacos, Petros and Bell, Anthony and Bingham, Robert and Clarke, Robert and Churazov, Eugene and Crowston, Robert and Doyle, Hugo and Drake, R. Paul and Heathcote, Robert and Koenig, Michel and Kuramitsu, Yasuhiro and Kuranz, Carolyn and Lee, Dongwook and MacDonald, Michael and Murphy, Christopher and Notley, Margaret and Park, Hye-Sook and Pelka, Alexander and Ravasio, Alessandra and Reville, Brian and Sakawa, Youichi and Wan, Willow and Woolsey, Nigel and Yurchak, Roman and Miniati, Francesco and Schekochihin, Alexander and Lamb, Don and Gregori, Gianluca},
	month = jul,
	year = {2015},
	pages = {8211--8215},
}

@article{casner_turbulent_2018,
	title = {Turbulent hydrodynamics experiments in high energy density plasmas: scientific case and preliminary results of the {TurboHEDP} project},
	volume = {6},
	copyright = {http://creativecommons.org/licenses/by/4.0/},
	issn = {2095-4719, 2052-3289},
	shorttitle = {Turbulent hydrodynamics experiments in high energy density plasmas},
	url = {https://www.cambridge.org/core/product/identifier/S2095471918000348/type/journal_article},
	doi = {10.1017/hpl.2018.34},
	language = {en},
	urldate = {2026-01-19},
	journal = {High Power Laser Science and Engineering},
	author = {Casner, A. and Rigon, G. and Albertazzi, B. and Michel, Th. and Pikuz, T. and Faenov, A. and Mabey, P. and Ozaki, N. and Sakawa, Y. and Sano, T. and Ballet, J. and Tzeferacos, P. and Lamb, D. and Falize, E. and Gregori, G. and Koenig, M.},
	year = {2018},
	pages = {e44},
}

@article{collins_role_2020,
	title = {Role of collisionality and radiative cooling in supersonic plasma jet collisions of different materials},
	volume = {101},
	issn = {2470-0045, 2470-0053},
	url = {https://link.aps.org/doi/10.1103/PhysRevE.101.023205},
	doi = {10.1103/PhysRevE.101.023205},
	language = {en},
	number = {2},
	urldate = {2026-01-19},
	journal = {Physical Review E},
	author = {Collins, G. W. and Valenzuela, J. C. and Speliotopoulos, C. A. and Aybar, N. and Conti, F. and Beg, F. N. and Tzeferacos, P. and Khiar, B. and Bott, A. F. A. and Gregori, G.},
	month = feb,
	year = {2020},
	pages = {023205},
}

@article{bott_insensitivity_2022,
	title = {Insensitivity of a turbulent laser-plasma dynamo to initial conditions},
	volume = {7},
	issn = {2468-2047, 2468-080X},
	url = {http://arxiv.org/abs/2201.01705},
	doi = {10.1063/5.0084345},
	language = {en},
	number = {4},
	urldate = {2026-01-19},
	journal = {Matter and Radiation at Extremes},
	author = {Bott, A. F. A. and Chen, L. and Tzeferacos, P. and Palmer, C. A. J. and Bell, A. R. and Bingham, R. and Birkel, A. and Froula, D. H. and Katz, J. and Kunz, M. W. and Li, C.-K. and Park, H.-S. and Petrasso, R. and Ross, J. S. and Reville, B. and Ryu, D. and Séguin, F. H. and White, T. G. and Schekochihin, A. A. and Lamb, D. Q. and Gregori, G.},
	month = jul,
	year = {2022},
	note = {arXiv:2201.01705 [physics]},
	pages = {046901},
}

@article{davidovits_turbulence_2022,
	title = {Turbulence generation by shock interaction with a highly nonuniform medium},
	volume = {105},
	issn = {2470-0045, 2470-0053},
	url = {https://link.aps.org/doi/10.1103/PhysRevE.105.065206},
	doi = {10.1103/PhysRevE.105.065206},
	language = {en},
	number = {6},
	urldate = {2026-01-19},
	journal = {Physical Review E},
	author = {Davidovits, Seth and Federrath, Christoph and Teyssier, Romain and Raman, Kumar S. and Collins, David C. and Nagel, Sabrina R.},
	month = jun,
	year = {2022},
	pages = {065206},
}

@article{beattie_taking_2025,
	title = {Taking control of compressible modes: bulk viscosity and the turbulent dynamo},
	volume = {542},
	issn = {0035-8711, 1365-2966},
	shorttitle = {Taking control of compressible modes},
	url = {http://arxiv.org/abs/2312.03984},
	doi = {10.1093/mnras/staf1318},
	language = {en},
	number = {4},
	urldate = {2026-01-19},
	journal = {Monthly Notices of the Royal Astronomical Society},
	author = {Beattie, James R. and Federrath, Christoph and Kriel, Neco and Hew, Justin Kin Jun and Bhattacharjee, Amitava},
	month = sep,
	year = {2025},
	note = {arXiv:2312.03984 [astro-ph]},
	pages = {2669--2697},
}
    \bibliographystyle{aasjournal}

\end{document}